\documentclass[aps, 10pt, prd,
               notitlepage, twocolumn, superscriptaddress,
               longbibliography, tightenlines,
               nofootinbib, floatfix]{revtex4-2}
\usepackage{colortbl}
\usepackage{tabularx}
\usepackage{graphicx}
\usepackage{booktabs}
\usepackage{threeparttable}
\usepackage{epsfig}
\usepackage{amssymb}
\usepackage{amsmath}
\usepackage{tensor}
\usepackage{textcomp}
\usepackage[varg]{txfonts}
\usepackage{enumerate}
\usepackage{mathtools}
\usepackage[normalem]{ulem}
\usepackage{bm}
\usepackage[OT1]{fontenc}
\usepackage{multirow}
\usepackage{makecell}
\usepackage{soul}
\usepackage{anyfontsize}
\usepackage{microtype}
\usepackage[dvipsnames]{xcolor}
\usepackage[colorlinks,linktocpage,breaklinks]{hyperref}
\usepackage[capitalize]{cleveref}
\usepackage{booktabs}

\hypersetup{colorlinks=true,
            citecolor=NavyBlue,
            linkcolor=NavyBlue,
            urlcolor=NavyBlue}
            
\newcommand{\dd}{{\rm d}}
\newcommand{\ii}{{\rm i}}

\newcommand{\reply}[1]{{#1}}

\newcommand{\aei}{\affiliation{Max Planck Institute for Gravitational Physics (Albert Einstein Institute),
D-14476 Potsdam, Germany}}
\newcommand{\uiuc}{\affiliation{Department of Physics and Illinois Center for Advanced Studies of the Universe,\\University of Illinois Urbana-Champaign, Urbana, Illinois 61801, USA}}
\newcommand{\yz}{\affiliation{Center for Gravitation and Cosmology, College of Physical Scienceand Technology,\\Yangzhou University, Yangzhou 225009, China}}
\newcommand{\yukawa}{\affiliation{Center of Gravitational Physics and Quantum Information, Yukawa Institute for Theoretical Physics, \\Kyoto University, Kyoto, 606-8502, Japan}}

\begin{document}

\title{Nonradial oscillations of stratified neutron stars with solid crusts:\\
Mode characterization and tidal resonances in coalescing binaries}

\date{\today}

\author{Yong Gao} \email{yong.gao@aei.mpg.de}     
\aei 

\author{Hao-Jui Kuan} 
\aei

\author{Cheng-Jun Xia} 
\yz

\author{Hector O. Silva} 
\uiuc \aei   

\author{Masaru Shibata} 
\aei \yukawa

\begin{abstract}

Dynamical tides of neutron stars in the late stages of binary inspirals provide a viable probe into dense matter through gravitational waves, and potentially trigger electromagnetic precursors. We model the tidal response as a set of driven harmonic oscillators, where the natural frequencies are given by the quasinormal modes of a nonrotating neutron star. These modes are calculated in general relativity by applying linear perturbation theory to stellar models that include a solid crust and compositional stratification. For the mode spectrum, we find that the canonical interface mode associated with crust-core boundary vanishes in stratified neutron stars and is replaced by compositional gravity modes with mixed gravity–interfacial character, driven primarily by strong buoyancy in the outer core. We also find that fluid modes such as the core gravity mode and the fundamental mode can penetrate the crust, and we establish a criterion for such penetration. Regarding the tidal interaction, we find that transfer of binding energy to oscillations is dominated by the fundamental mode despite its frequency being too high to resonate with the tidal forcing. In general, we find that lower-frequency modes induce gravitational-wave phase shifts smaller than $\sim 10^{-3}\,\rm rad$ for the equation of state we consider. We discover that nonresonant fundamental and crustal shear modes can trigger crust breaking already near the first gravity-mode resonance, while gravity-mode resonance concentrates strain at the base of the crust and may marginally crack it. These results suggest that both resonant and nonresonant excitations can overstress the crust and may channel energy into the magnetosphere prior to merger, potentially powering electromagnetic precursors. Our work represents an important step toward realistic modeling of dynamical tides of neutron stars in multimessenger observations.

\end{abstract}
\maketitle

\section{Introduction}

As the densest stable form of matter known in the Universe, neutron stars (NSs) are expected to have complex interiors dominated by strong interactions inaccessible through terrestrial experiments.
Despite extensive experimental and theoretical efforts in probing the nuclear equation of state (EOS), our current knowledge about the EOS at the densities most relevant for determining the internal structure of NSs remains poor~\cite{Lattimer:2000nx,Ozel:2016oaf,Baym:2017whm}. 

Compact binary mergers involving NSs provide multiple pathways to probe the dense matter, from subtle finite-size effects in the late inspiral~\cite{Hinderer:2007mb,Flanagan:2007ix,LIGOScientific:2017vwq,Hinderer:2009ca,Hotokezaka:2016bzh,Read:2009yp,Shibata:1993qc,Lai:1993di,Kokkotas:1995xe,Hinderer:2016eia,Steinhoff:2016rfi} to the dramatic dynamics of merger remnants~\cite{Shibata:2005ss,Takami:2014zpa,Bauswein:2018bma,Bernuzzi:2014kca,Dietrich:2020efo,Hotokezaka:2011dh}, tidal disruption~\cite{Lackey:2011vz,Lackey:2013axa,Kyutoku:2011vz}, and mass ejection~\cite{Hotokezaka:2012ze,Kiuchi:2019lls,Shibata:2019wef}. 
The landmark detection of gravitational waves (GWs) from the binary NS inspiral GW170817 by the LIGO-Virgo-KAGRA Collaboration~\cite{LIGOScientific:2017vwq}, along with its postmerger electromagnetic (EM) counterparts~\cite{LIGOScientific:2017ync}, marked the beginning of GW multimessenger astronomy, and has helped shed a new light on the EOS problem.
However, the constraints on the EOS set by GW170817 are limited by the insufficient sensitivity of current GW detectors at the frequency band of the late-inspiral and merger, where matter effects on the GW waveform become most salient.

An upgrade to the current ground-based detectors is pending for the coming years, and the next observing run of the International Gravitational-Wave Observatory Network, tentatively planned for 2028, will provide an opportunity to detect GW signals emanated during the merger from GW170817-like events.
A central question to be addressed is then: To what extent can observations of binary mergers truly reveal the composition and state of matter inside NSs?
Because a NS’s characteristic oscillation frequencies are determined by its internal structure, identifying quasiperiodic oscillations (QPOs) in some observations offers a way to probe these extreme environments. This article focuses on a subset of the stellar spectrum that may imprint observable signatures in gravitational waveforms or EM flares through tidal interactions.

Much work has focused on the tidal deformation of NSs during binary interaction in the static limit~\cite{Hinderer:2007mb,Flanagan:2007ix,Hinderer:2009ca}. 
The secular imprint on the GW signal associated to it is characterized by the so-called tidal deformability. 
The GW170817 event placed a constraint on the binary's mass-weighted average tidal deformability, disfavoring extremely stiff EOSs~\cite{Annala:2017llu,LIGOScientific:2017vwq,De:2018uhw,Zhou:2017pha,Gao:2021uus}. Combined with lower bounds on the maximum mass from radio pulsar timing~\cite{Demorest:2010bx,Antoniadis:2013pzd,Fonseca:2021wxt,NANOGrav:2019jur} and joint mass–radius constraints from NICER X-ray timing~\cite{Miller:2021qha,Miller:2019cac,Riley:2019yda}, these observations form the current picture of EOS constraints from the astrophysical side. 

The tidal response characterized by the tidal deformability is only the leading order effect obtained under the assumption that the tidal interaction is adiabatic.
As the binary approaches the merger, this assumption gradually breaks down, and dynamical, frequency-dependent contributions in the tidal response function must be taken into account~\cite{Hinderer:2016eia,Steinhoff:2016rfi,Pitre:2023xsr,HegadeKR:2024agt}.
In particular, NSs respond dynamically to the tidal field through 
the excitation of various oscillation modes
as the orbital frequency sweeps through the detector band~\cite{Shibata:1993qc,Lai:1993di,Kokkotas:1995xe,Tsang:2011ad,Chakrabarti:2013lua,Kuan:2021sin}.
Therefore, dynamical tidal interactions provide a realization 
of NS asteroseismology~\cite{Andersson:1996pn,Andersson:1997rn,Andersson:2017iav}. That is, the study of the interior composition the star through its oscillation spectrum.

Dynamical tides transfer energy and angular momentum from the orbit into oscillation modes of the star, resulting in a GW phase shift. 
Many studies (see, for instance,~\cite{Steinhoff:2016rfi,Hinderer:2016eia,Andersson:2019dwg,Pratten:2021pro}) have shown that the fundamental ($f$-) mode --- acoustic oscillations driven by pressure gradients --- can be appreciably excited even without reaching exact resonance during the inspiral. 
Although the resulting impact on the GW signal is relatively modest, incorporating dynamical tides has been demonstrated to improve waveform accuracy~\cite{Hinderer:2016eia}. 
In rotating NSs, the $f$-mode spectrum splits into prograde and retrograde ones~\cite{Kruger:2019zuz}.
Of particular interest is the retrograde $f$-mode, which appears at lower frequencies in the inertial frame of the observer and can enter into resonance with the orbital frequency {when the spin of the NS is antialigned with respect to the orbital angular momentum.
This resonance is expected to produce a much larger GW phase shift~\cite{Ho:1998hq,Ma:2020rak,Steinhoff:2021dsn,Kuan:2023qxo}, and was recently captured
in high-precision numerical simulations for the first time~\cite{Kuan:2024jnw}.

The interior of NSs is more complicated than a prescribed density-pressure relation. 
Matter densities span more than 14 orders of magnitude from the surface to the core, accompanied by changes in composition, from ordinary baryons to possibly exotic particles such as hyperons or quarks, as well as by distinct phase structures, such as solid and superfluid. 
Many low-frequency modes (compared to $f$-mode) restored by different physical mechanism may become tidally excited. Examples include gravity ($g$-) modes that arise from compositional gradients~\cite{ Shibata:1993qc,Reisenegger:1994a,Yu:2016ltf,Li:2023ijg,Counsell:2024pua,Zhao:2025pgx,Zhao:2022tcw,Zhao:2022toc}, interface ($i$-) modes associated with discontinuities in density~\cite{Miao:2023jqe,Counsell:2025hcv,Pereira:2025xsi,Zhao:2022toc,Lau:2020bfq} or shear modulus~\cite{Tsang:2011ad,Pan:2020tht,Zhu:2022pja,Lau:2020bfq}, and shear ($s$-) modes restored by shear stresses of the solid crust~\cite{Kruger:2014pva}. 

The tidal coupling of the foregoing modes is typically much smaller than that of the $f$-mode, but their lower frequencies make resonance more accessible earlier in the inspiral, and thus potentially detectable with current GW detectors. 
Several works showed that the GW phase shift induced by compositional $g$-modes is generally smaller than $10^{-3}$~radians;
see, e.g., Refs.~\cite{Yu:2016ltf,Counsell:2024pua,Kuan:2021jmk}. 
However, Refs.~\cite{Kwon:2025zbc,Kwon:2024zyg} suggested that nonlinear effects could lock the $g$-mode resonance resulting in several radians of dephasing.   
Moreover, Refs.~\cite{Tsang:2011ad,Pan:2020tht,Zhu:2022pja,Miao:2023jqe,Counsell:2025hcv,Pereira:2025xsi,Zhu:2022pja} suggested that the $i$-mode associated with the crust-core interface or a first-order phase transition may produce approximately $1$ to $10$ radians of GW dephasing. This offers a promising probe of the existence of phase transitions to quark matter, as well as into nuclear parameters around the saturation density.

Mode excitation may still yield observable signatures even if the associated GW dephasing is negligible, provided that the mode's energy is efficiently converted into coherent EM emission. 
In particular, if a mode acquires a large amplitude during resonance, the resulting stresses can break the star's crust, excite high-frequency crustal oscillations, and trigger bursts of EM waves radiation through coupling with the magnetosphere. This mechanism, first proposed for $i$-modes~\cite{Tsang:2011ad,Neill:2021lat}, has been invoked to explain the {\it precursor} emission observed before some short gamma-ray bursts (sGRB)~\cite{Troja:2010zm,Koshut:1995ApJ,Troja:2010zm,Minaev:2016gck,Zhong:2019shb,Coppin:2020stp,Wang:2020vvr,Li:2020nhs,Wang:2021Galax,Xiao:2022quv,Troja:2022yya}. 
It has since been extended to other modes, such as $g$-~\cite{Kuan:2021jmk,Kuan:2021sin,Kuan:2023kif,Kuan:2022bhu} and $f$-modes; see Ref.~\cite{Suvorov:2024cff} for a review.
Although the overstressing of the crust is triggered by resonant tidal excitation rather than slow magnetic-field evolution, the subsequent magnetic coupling and EM energy release are analogous to the mechanisms proposed for magnetar giant flares~\cite{Thompson:1995MNRAS}. 
The recent detection of a 22~Hz QPO~\cite{Xiao:2022quv} in a sGRB may have originated from shear modes~\cite{Suvorov:2022ldw} or Alfven modes excited following a crust breaking, echoing similar interpretations proposed for the QPOs in the giant-flare tails of magnetars~\cite{Israel:2005av,Strohmayer:2005ks}.

In this paper, we will investigate the dynamical tides of nonrotating NSs and their potential observational consequences. Our work improves upon the previous literature in three aspects. First, we use a unified EOS that includes both a solid crust and compositional stratification, while earlier works often treated them separately or introduced stratification phenomenologically. Second, we employ the framework of relativistic linear perturbation theory, without approximations such as the Cowling approximation or the use of Newtonian perturbations on relativistic backgrounds. Third, we analyze how resonant and nonresonant modes contribute to the breaking of the crust. Together, these improvements give a more realistic description of dynamical tides in compact binaries.

The paper is organized as follows. In Sec.~\ref{sec.osc}, we describe the EOSs and the formalism to calculate the full set of quasi-normal modes (QNMs) in relativistic linear perturbation theory. The properties of both high frequency (fundamental and shear) and low frequency (gravity and interface) modes are discussed in Secs.~\ref{sec:high} and~\ref{sec:low}, respectively. We then investigate the tidal excitation of different modes in Sec.~\ref{sec:resonance}. Therein, we include a discussion of energy transfer and GW dephasing in Sec.~\ref{sec:energy_transfer}, and a detailed study of crust breaking in Sec.~\ref{sec:crust_breaking}. 
We summarize our finding in Sec.~\ref{sec:summary}. In Appendix~\ref{sec:appendixA}, we present the master equations and describe the numerical methods we used to compute the mode frequencies.
Throughout this paper, we use geometrical units $c = G = 1$, where $G$ is the gravitational constant and $c$ is the speed of light.

\section{Neutron star model and its oscillation spectrum}
\label{sec.osc}

\subsection{The equation of state}\label{sec.eos}

We employ a unified EOS predicted by the relativistic density functional TW99, that adopts density-dependent couplings~\cite{Niu:2025tvd}. 
The EOS describes cold $npe\mu$ matter (i.e., consisting of neutrons, protons, electrons, and muons) which fulfills the (global) charge neutrality condition. 
The adopted relativistic-mean-field theory parameters and the corresponding nuclear saturation properties are given in Tables I and II of Refs.~\cite{Niu:2025tvd,Xia:2022dvw,Xia:2022pja}. 
The EOS is formulated in a two-parameter form, 
\begin{equation}
p = p(n_{\mathrm{b}}, Y_p),
\end{equation}
where $p$ is the pressure, 
$n_{\rm b}$ is the baryon number density, and $Y_p=n_p/n_{\mathrm{b}}$ is the proton fraction with $n_p$ the proton number density. 
Similarly, we also express the energy density $\epsilon = \epsilon(n_{\mathrm{b}}, Y_p)$ and chemical potential $\mu_i = \mu_i(n_{\mathrm{b}}, Y_p)$, for a fermion $i$, in a two-parameter form.
Charge neutrality requires that $Y_p=Y_e+Y_\mu$, while the chemical equilibrium of leptons requires that $\mu_e=\mu_\mu$.
For $\beta$-stable NS matter, that is, matter fulfilling $\mu_n -\mu_p = \mu_e = \mu_\mu$, the thermodynamic quantities such as the pressure and energy density become functions of density alone.

\subsubsection{Adiabatic index and stratification}
\label{sec:eos:adiabatic}

The adiabatic index $\Gamma_{0}$ of NS matter is defined as 
\begin{equation}
    \Gamma_{0} \equiv \left( \frac{\partial \ln p}{\partial \ln n_{\mathrm{b}}} \right)_\mathrm{\beta-stable} = \frac{\epsilon + p}{p} \frac{\dd p}{\dd \epsilon}\,,
\end{equation}  
where the derivative is taken along the $\beta$-stability.
This adiabatic index can be deemed as a measure of an EOS's stiffness as a function of $n_{\rm b}$ (larger $\Gamma_0$ is stiffer), and is thus closely related to the bulk properties of an NS in equilibrium, such as its mass $M$ and radius $R$.
In \cref{fig:gamma}, we show $\Gamma_0$ as a function of the rest-mass density $\rho = m_u n_{\mathrm{b}}$, where $m_u=931.494$~MeV is the atomic mass unit, for the TW99 EOS. 
We see that at the outer crust, where densities are smaller than the neutron drip density $\rho_{\rm nd} \simeq 3.0 \times 10^{11}\,\mathrm{g/cm^{3}}$, the adiabatic index is roughly constant $\Gamma_{0}\simeq 4/3$ as the pressure is dominated by the relativistic electron Fermi gas.\footnote{The density discontinuity arising from the phase transition between different species of nuclei~\cite{Baym:1971ApJ} has been smoothed out in this region since the nucleon numbers in the nuclei vary smoothly with density and take noninteger values in our EOS.}
As the density increases toward the neutron drip point, the EOS softens because of the emergence of free neutron gas, leading to a sharp drop in $\Gamma_0$.
Beyond this point, the matter stiffens as the neutron gas contributes to increase the Fermi pressure with additional contributions from nuclear force and Coulomb interactions. 
At the crust-core interface where $\rho_{\rm cc}\simeq 1.03\times 10^{14}\,\rm g/cm^3$, the adiabatic index $\Gamma_0$ first drops abruptly and then rises sharply as nuclei dissolve. 
The adiabatic index increases to $\Gamma_{0}\simeq3$ at densities $\rho \gtrsim 5 \times 10^{14}\,\mathrm{g/cm^{3}} $ as repulsive nucleon interactions become progressively stronger, before gradually decreasing at higher densities wherein the EOS softens due to the density dependence of nuclear interactions. 

\begin{figure}
    \centering
    \includegraphics[width=\linewidth]{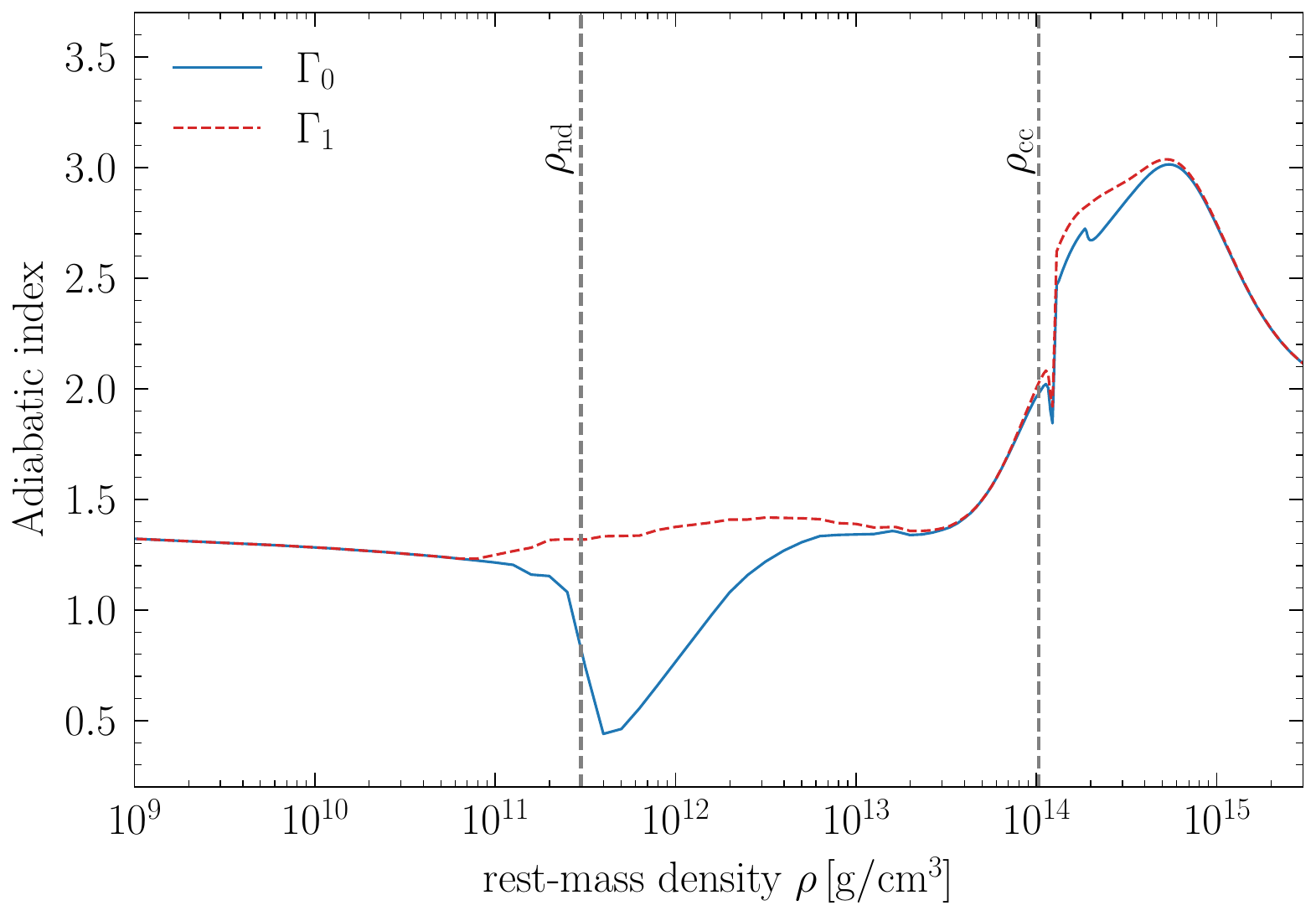}
    \caption{The adiabatic indices $\Gamma_0$ and $\Gamma_1$ as functions of rest-mass density for the TW99 EOS.The vertical dashed lines indicate the neutron drip density $\rho_{\rm nd}$ and the crust-core interface density $\rho_{\rm cc}$, respectively. }
    \label{fig:gamma}
\end{figure}

If the nuclear reactions occur rapidly enough to maintain chemical equilibrium as a fluid element is perturbed, $\Gamma_0$ can also be used to describe the perturbations. 
However, current studies
suggest that the equilibration times for nonsuperfluid matter are of the order of
\begin{equation}
    \label{eq:time_scale_with_t9}
    \tau_{\mathrm{M}} \sim \frac{2 \text { month }}{T_9^6}\,, 
    \quad \textrm{and} \quad 
    \tau_{\mathrm{D}} \sim \frac{20 \mathrm{~s}}{T_9^4}\,,
\end{equation}
for the modified and direct Urca reactions, respectively~\cite{Reisenegger:1992ApJ,Haensel:2001mw,Haensel:2002qw}.
Here, $T_9$ is the temperature in units of $10^9$~K. 
Inspiralling NSs are old and cold, typically with $T_9 \lesssim 0.01$, which makes both {$\tau_{\mathrm{M}}$} and {$\tau_{\mathrm{D}}$} much longer, 
and suggests that the adiabatic index for perturbations is different from that for the background, unperturbed matter.
Quantitatively, the typical timescale $\tau_{\rm osc}$ of the oscillations considered in this work ranges from $10^{-4}$ to~$10^{-1}$~s.
%
%
Comparing these timescales with Eq.~\eqref{eq:time_scale_with_t9} we conclude that the weak reactions are not fast enough to maintain $\beta$-equilibrium between the perturbed fluid element and its surroundings. 
For our purposes, it is therefore reasonable to assume that the composition of the fluid element remains effectively frozen during oscillations.
The adiabatic index for the perturbed matter $\Gamma_1$ is then: 
\begin{align}
    \Gamma_1 &\equiv \left( \frac{\partial \ln p}{\partial \ln n_{\mathrm{b}}} \right)_{Y_i} = \Gamma_0 + \frac{n_{\mathrm{b}}^3}{p} \left[  
    \left. \frac{\partial\, ( \mu_n - \mu_e - \mu_p )}{\partial n_{\mathrm{b}}} \right|_{Y_i} \,
    \frac{\mathrm{d} Y_p}{\mathrm{d} n_{\mathrm{b}}}\nonumber \right.\\ 
    & \left.+ \left. \frac{\partial \, ( \mu_e - \mu_\mu )}{\partial n_{\mathrm{b}}} \right|_{Y_i} \,
    \frac{\mathrm{d} Y_\mu}{\mathrm{d} n_{\mathrm{b}}} \right]\,.
\end{align}
The derivatives are taken at fixed particle fractions with $Y_i=n_i/n_\mathrm{b}$, and the second equality assumes $npe\mu$ matter~\cite{Jaikumar:2021jbw,Sun:2025zpj}.

As shown in \cref{fig:gamma}, $\Gamma_1$ generally follows the behavior of $\Gamma_0$ across most densities.
However, it notably deviates from the latter near the neutron drip point and outer core close to the crust–core interface, when free neutrons and muons emerge, respectively.
Around the neutron drip, the difference $\Gamma_1 - \Gamma_0$ can be of order unity,
while near the crust–core boundary it remains approximately $0.2$.

The difference between the adiabatic index of the background matter and that of perturbed fluid elements results in a displaced fluid element with a density different from its surroundings.
The magnitude of this difference sets the strength of the buoyancy and determines the spectrum of compositional $g$-modes; see~Sec.~\ref{sec:g_mode}.

\subsubsection{Elasticity and shear modulus}

\begin{figure}
    \centering
    \includegraphics[width=\linewidth]{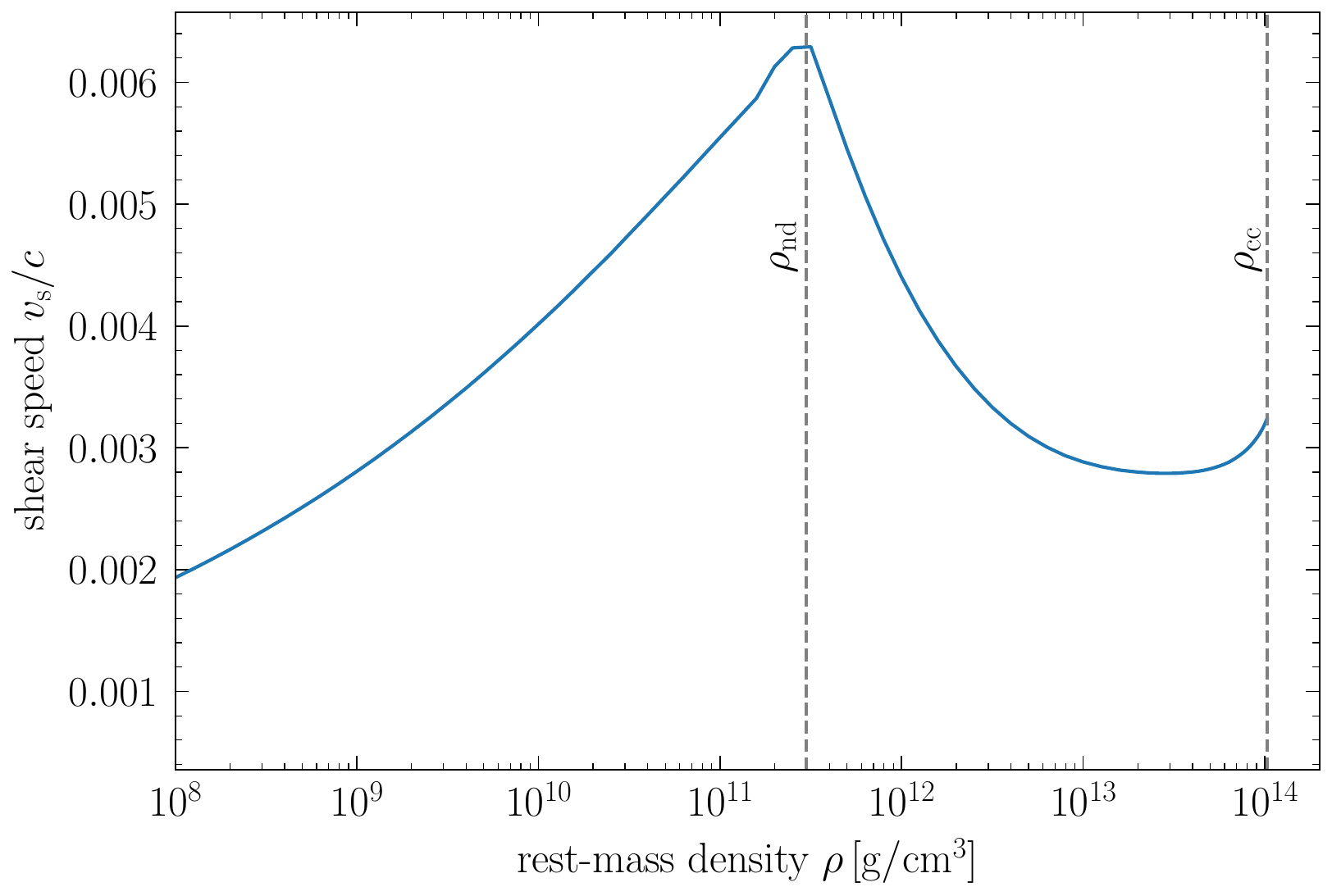}
    \caption{The shear speed $v_{\rm s}$ in the solid crust (normalized by the sound speed $c$) as a function of rest-mass density $\rho$. The vertical dashed lines indicate the neutron drip density $\rho_{\rm nd}$ and the crust-core interface density $\rho_{\rm cc}$, respectively.}
    \label{fig:shear_speed}
\end{figure}

The crust of NSs behaves as a solid due to the presence of a Coulomb lattice. Assuming a body-centered cubic lattice, the shear modulus of a zero-temperature crust can be estimated as
\begin{equation}
    \mu=0.1194\left(\frac{4 \pi}{3}\right)^{1 / 3}(Z e)^2\left(\frac{1-X_{\rm n}}{A}\right)^{4 / 3} n_{\mathrm{b}}^{4 / 3}\,,
\end{equation}
where \(Z\) is the mean charge number of the nuclei, \(A\) is the mean nucleus number, and \(X_{\rm n}\) denotes the fraction of ``free'' neutrons~\cite{Strohmayer:1991ads,Steiner:2009yg}. 

The shear speed $v_{\rm s}$ characterizes the velocity of elastic waves in the crust; its relativistic value can be computed as~\cite{Carter:1973zz}  
\begin{equation}
    v_{\rm s}=\left(\frac{\mu}{\epsilon + p}\right)^{1/2}\,.
\end{equation}
As illustrated in \cref{fig:shear_speed}, the shear speed is of the order of $10^{-3}$--$10^{-2}$ of the speed of light for densities $\sim 10^{8}$--$10^{14}\,\rm g/cm^{3}$. 
Below the neutron drip density, the shear speed gradually increases with $\rho$ because of the increasing mean charge number as well as the density.
However, the fraction of free neutrons grows rapidly when $\rho$ exceeds $\rho_{\rm nd}$, which progressively suppresses the shear speed.

The solid crust is important to NS phenomenology for several reasons. Its ability to sustain large elastic stresses allows the formation of nonaxisymmetric ``mountains"~\cite{Johnson-McDaniel:2012wbj,Gittins:2020cvx,Gittins:2024zbg}, which can both modulate pulse profiles through free precession~\cite{Jones:2000ud,Gao:2020zcd,Gao:2022hzd,Desvignes:2024vle} and act as sources of continuous GWs~\cite{Gao:2020zcd,Jones:2001yg}. Because the crust supports shear motion, it hosts oscillation families that cannot exist in a perfect fluid, most notably shear ($s$-) and torsional ($t$-) modes~\cite{McDermott:1988ApJ,Schumaker:1983ads,Strohmayer:2005ks,Samuelsson:2006tt,Sotani:2006at,Silva:2014ora}. Moreover, the sharp transition between the solid crust and the underlying fluid core gives rise to additional $i$-modes~\cite{McDermott:1988ApJ,Tsang:2011ad,Piro:2004bh}.

\subsection{Polar perturbation and oscillation spectrum}\label{sec.II.B}

Having described the EOS we will use and presented some of its properties, we now explain how we construct our equilibrium NS models and how we compute their oscillation spectra.

The spacetime line element of a static spherically symmetric NS in hydrostatic equilibrium can be expressed as
\begin{equation}
    \label{eq:metric}
    \mathrm{d} s^2=-e^\nu \, \mathrm{d} t^2+e^\lambda \, \mathrm{d} r^2+r^2 \mathrm{~d} \theta^2+r^2 \sin ^2 \theta \mathrm{~d} \phi^2 \,,
\end{equation}
where $\nu$ and $\lambda$ are the metric functions 
that depend only on the radial coordinate $r$.
We assume that the matter content of star can be described as a perfect fluid, with energy-momentum tensor
\begin{equation}
    T^{\rm fluid}_{a b} = \left(\epsilon+p\right) u_a u_b + p g_{ab}\,,
\end{equation}
where $u^a$ is the four-velocity and $g_{ab}$ the spacetime metric.

Through Einstein's equations, 
\begin{equation}
    G_{ab} = 8 \pi T_{ab}^{\rm fluid},
\end{equation}
where $G_{ab}$ is the Einstein tensor with $u^a=u^t \delta^a_{~t}$, these 
assumptions lead to the Tolman--Oppenheimer--Volkoff (TOV) equations that describe an NS in equilibrium~\cite{Tolman:1939jz,Oppenheimer:1939ne}. These equations can be readily integrated once an EOS has been specified.
We note that strictly speaking, the perfect fluid assumption is inconsistent with the presence of a solid component.
Nevertheless, {\it it still makes sense to assume the background is in a relaxed state}, which may be reasonable for NSs close to merger.

Nonrotating NSs
modeled with the TW99 EOS have a maximum mass of $M_{\rm TOV} = 2.08\,M_{\odot}$, consistent with the most massive pulsars observed to date~\cite{Fonseca:2021wxt,Antoniadis:2013pzd,Demorest:2010bx,NANOGrav:2019jur}. 
Besides, for an NS with mass $1.35\,M_{\odot}$, the corresponding radius is $R_{1.35} = 12.29\,\mathrm{km}$, and the dimensionless tidal deformability is $\Lambda_{1.4} = 510.0$, in agreement with observations X-ray timing of NSs from NICER~\cite{Miller:2019cac,Riley:2019yda,Vinciguerra:2023qxq,Salmi:2024aum} and from the binary NS inspiral GW170817~\cite{LIGOScientific:2017vwq}.

The oscillations of nonrotating NSs are typically studied by linearizing Einstein's equations on a background equilibrium solution.
We employ the relativistic Lagrangian perturbation formalism developed in Refs.~\cite{Friedman:1975ApJ,Andersson:2020phh,Andersson:2004nv,Friedman:2013xza}. 
The key dynamical quantities in this framework include the metric perturbation, $\delta g_{ab}$, and the fluid displacement vector, $\xi^{a}$. 
For any spacetime quantity $Q(t,x)$, the relationship between the Lagrangian perturbation, $\Delta Q(t,x)$, and the Eulerian perturbation, $\delta Q(t,x)$, is given by  
\begin{equation}
    \label{eq:lagrangian_perturbation}
    \Delta Q = \delta Q + \mathcal{L}_{\boldsymbol{\xi}} Q\,,
\end{equation}  
where $\mathcal{L}_{\boldsymbol{\xi}}$ denotes the Lie derivative along the displacement vector $\xi^{a}$.
Because of the spherical symmetry of the background, we decompose the perturbations into spherical harmonic and assume they have an harmonic time dependence, that is, $Y_{\ell}^{{m}} e^{i \omega t}$, where $\omega$ is the mode's oscillation angular frequency and $Y_{\ell}^{{m}}$ is the spherical harmonics with quantum numbers $\ell$ and $m$.
By adopting the Regge--Wheeler gauge~\cite{Regge:1957td}, we can express the metric perturbation of polar parity as
\begin{equation}
    \label{eq:metric_perturbation}
    \delta g_{a b}=-r^\ell \, \left(\begin{array}{cccc}
    H_0 e^\nu & i \omega r H_1 & 0 & 0 \\
    i \omega r H_1 & H_2 e^\lambda & 0 & 0 \\
    0 & 0 & r^2 K & 0 \\
    0 & 0 & 0 & r^2 \sin ^2 \theta K
    \end{array}\right) Y_{\ell}^{{m}} e^{i \omega t},
\end{equation}
where $H_0$, $H_1$, $H_2$, and $K$ are functions of $r$ only, and that dictate the perturbations in the metric.
The associated displacement vector has the form
\begin{equation} \label{eq:displacement_vector}
    \xi^a=r^{\ell} \, \left(\begin{array}{c}
    0 \\
    W r^{-1} e^{-\lambda / 2} \\
    -V r^{-2} \partial_\theta \\
    -\displaystyle\frac{V}{r^2 \sin ^2 \theta} \partial_\phi
    \end{array}\right) Y_{\ell}^{{m}} e^{i \omega t},
\end{equation}
where $W$ and $V$ are functions of $r$. 

For a perfect fluid, the master equations for $\{H_0$, $H_1$, $H_2$, $K\}$ and $\{W$, $V\}$ can be derived by solving the perturbed Einstein's equations,
\begin{equation}
    \delta G_{ab} = 8\pi \delta T_{ab}^{\rm fluid}.
\end{equation}
The perturbed Einstein tensor $\delta G_{ab}$ can be obtained from the background metric in~\cref{eq:metric} and the perturbed metric~\cref{eq:metric_perturbation}. 
The perturbed energy-momentum tensor $\delta T_{ab}^{\text{fluid}}$ can be written in terms of the background four-velocity, pressure, energy density, and their corresponding perturbations as 
\begin{align}
        \delta T_{a}{}^{b{\, \text{fluid} }} &=  (\delta \epsilon+\delta p) u_a u^b+\delta p \, \delta_{a}{}^{b} \nonumber \\
        &\quad +(\epsilon + p)\left(u_a \delta u^b+\delta u_a u^b\right)\,.
\end{align}    
The Lagrangian perturbation of the metric, given by,
\begin{equation}
    \Delta g_{a b} = \delta g_{a b} + \mathcal{L}_{\boldsymbol{\xi}} g_{a b} = \delta g_{a b} + \nabla_a \xi_b + \nabla_b \xi_a\,,
\end{equation}
is related to the Lagrangian perturbation of the four-velocity as~\cite{Friedman:1975ApJ} 
\begin{equation}
    \Delta u^a = \frac{1}{2} u^a u^b u^c \Delta g_{b c}\,.
\end{equation}  
Using~\cref{eq:lagrangian_perturbation}, we obtain 
\begin{equation}
    \delta u^{a}=q_{b}{}^{a} \mathcal{L}_{\boldsymbol{u}} \xi^b+\frac{1}{2} u^a u^b u^c \delta g_{b c}\,,
\end{equation}
where 
\begin{equation}
    q^{a b}=g^{a b}+u^a u^b\,,
\end{equation}
is the projection tensor orthogonal to the four-velocity {$u^{a}$}. 

To obtain the expressions for $\delta \epsilon$ and $\delta p$, a useful relation is the Lagrangian perturbation of the rest-mass density~\cite{Thorne:1967ApJ},
\begin{align}
    \frac{\Delta \rho}{\rho} =& -\frac{1}{2} q^{a b} \Delta g_{a b}= r^{\ell-2}\Bigg[r^2\left(K+\frac{1}{2} H_0\right)-\ell(\ell+1) V  \nonumber\\
    &-(\ell+1) e^{-\lambda / 2} W-r e^{-\lambda / 2} W^{\prime}\Bigg] Y_{\ell}^{{m}}e^{\ii \omega t}\,.
\end{align}
where a prime indicates differentiation with respect to $r$.
With this expression, the Lagrangian and Eulerian perturbations of the energy density are  
\begin{equation}
    \Delta \epsilon = (\epsilon+p) \frac{\Delta \rho}{\rho}\,, \quad \delta \epsilon = {\Delta \epsilon}-\rho'\xi^{r}\,,
\end{equation}
respectively.
The pressure perturbation is related to the the energy density perturbation by the adiabatic index. 
However, which adiabatic index, $\Gamma_0$ or $\Gamma_1$ should we use?
If we assume that matter is in $\beta$-equilibrium during the perturbation, the pressure perturbation is then given by
\begin{equation}
    \frac{\Delta p}{p}=\Gamma_0 \, \frac{\Delta \varepsilon}{\varepsilon+p}\,,\quad \delta p = \Delta p -p'\xi^{r}\,.
\end{equation}
On the other hand, for frozen composition, the pressure perturbation can be obtained by replacing $\Gamma_0$ with $\Gamma_1$ in the equation above.
{In reality, the adiabatic index should be somewhere in between because of the finite weak reaction timescale. 
However, as we argued in Sec.~\ref{sec:eos:adiabatic}, the oscillation periods of the modes under consideration are likely to be much shorter than the typical weak reaction times in inspiralling NSs for our EOS.
We will therefore focus on the QNM spectrum computed with $\Gamma_1$ for the perturbations, and use the spectrum with $\Gamma_0$ as a reference when investigating the impact of stratification.

What about the perturbation of solid crust? The foundations of elasticity and its perturbation in GR were formulated by Carter and Quintana in the 1970s~\cite{Carter:1972,Carter:1973zz}; see also Refs.~\cite{Carter:1975,Karlovini:2002fc,Beig:2002pk,Beig:2023pka} for subsequent developments and Ref.~\cite{Andersson:2020phh} for a review. 
For a perturbed spherical NS, the Lagrangian perturbation of the strain tensor is given by 
\begin{equation}
    \label{eq:strain}
    \Delta s_{a}{}^{b}=\frac{1}{2}\left(q_{a}{}^{c} \, q^{d b}-\frac{1}{3}q_{a}{}^{b} \, q^{c d}\right) \Delta g_{c d}  \,.
\end{equation}
Assuming a Hookian elastic solid, the stress tensor is given by
\begin{equation}
    \label{eq:stress}
    \delta \pi_{a}{}^{b}=-2{\mu} \delta s_{a}{}^{b}\,,
\end{equation}
where $\delta s_{a}{}^{b}$ is the Eulerian perturbation of the strain tensor. The Eulerian perturbation of the total energy-momentum tensor is then:
\begin{equation}
    \delta T_a^{~b}=\delta T_a^{~b\, \text{fluid}}+\delta \pi_a^{~b}\,.
\end{equation}
We follow Refs.~\cite{Finn:1990ads,Kruger:2014pva,McDermott:1988ApJ} to further define the radial traction $T_1$ and tangential traction $T_2$ as
\begin{align} \label{eq:tractions}
    T_1 Y_{\ell}^{m}e^{i\omega t}   \equiv \delta \pi_r^{~r} \,,
    \quad \textrm{and} \quad    
    \frac{1}{r}T_2 \partial_{\theta}Y_{\ell}^{m} e^{i\omega t}  \equiv \delta \pi_r^{~\theta} \,,
\end{align}
respectively. Both of them vanish in the fluid and are useful for the determination of the junction conditions at the solid-fluid interface. The remaining components of $\delta \pi_{a}{}^{b}$ can be expressed in terms of $\{T_1$, $T_2$, $V\}$, the shear modulus $\mu$, and the background metric functions.
In summary, we have two additional variables $\{T_1$, $T_2\}$ describing the perturbations of the solid crust besides the six variables $\{H_0$, $H_1$, $H_2$, $K$, $W$, $V\}$ describing the perturbation of the fluid core.
To avoid diverting further from the main topic, we present the master equations for both fluid and solid perturbations, the junction conditions at the solid–fluid interface, and the numerical methods we used to compute the oscillation frequencies $\omega$ in \cref{sec:appendixA}. These frequencies are complex valued, known as QNMs. The real part of the frequencies are related to the oscillations of the system, and the imaginary part of the frequencies are related with the energy dissipation through GW emission. In our problem, this damping timescale is much longer than the star's oscillation and binary-inspiral timescales. For this reason, we only consider the real part of the QNMs in this work.

\begin{figure*}
    \centering
    \includegraphics[width=0.9\linewidth]{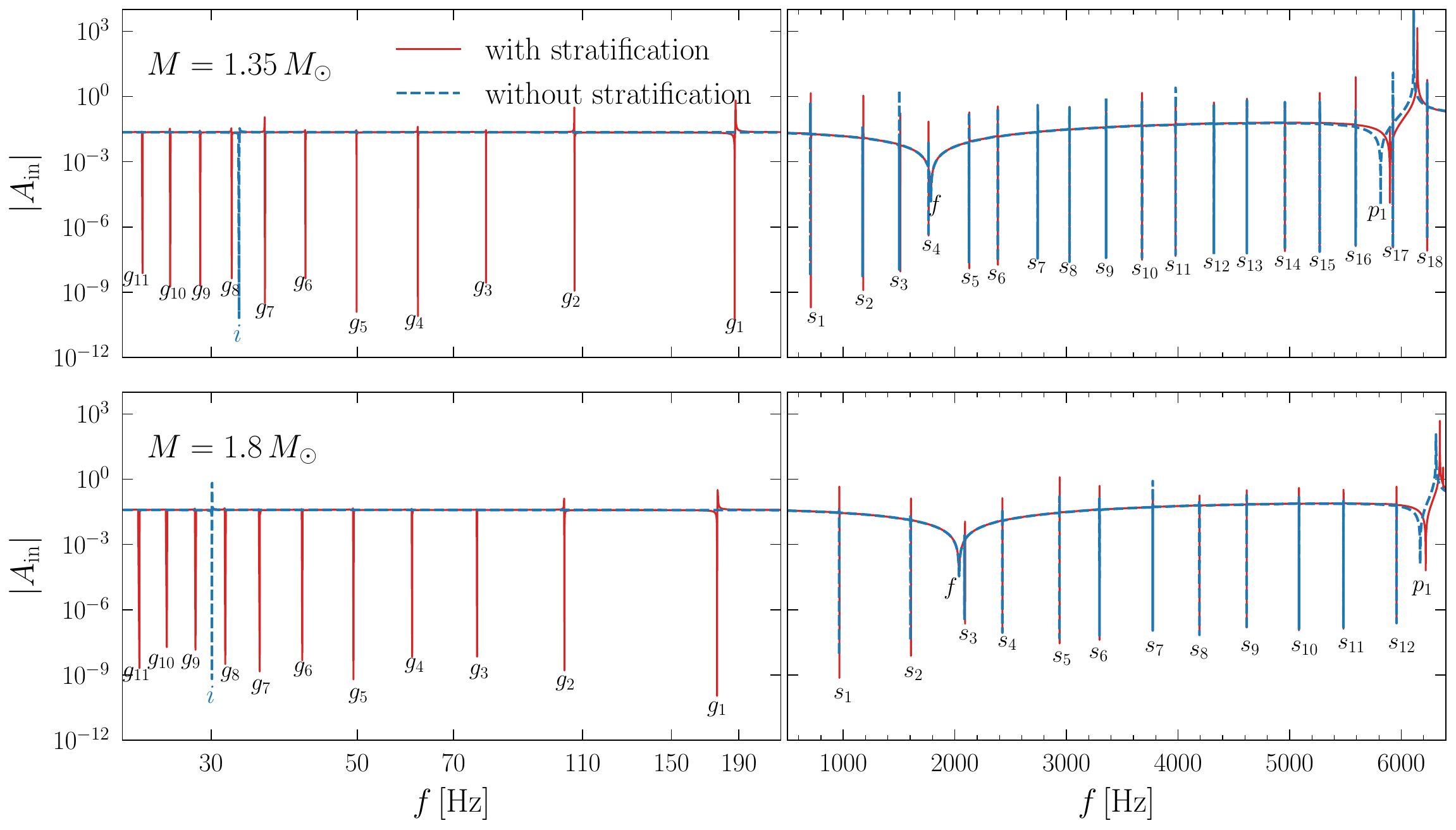}
    \caption{Real parts of the oscillation spectra, computed with and without compositional stratification, for the canonical star ($M = 1.35\,M_\odot$, top panel) and the higher-mass star ($M = 1.8\,M_\odot$, bottom panel). See the text for details and \cref{tab:model} for the background parameters.
    }
    \label{fig:spectral}
\end{figure*}

\begin{table}
\centering
\setlength{\tabcolsep}{6pt}
\begin{threeparttable}
\caption{Parameters for the canonical model ($M = 1.35\,M_\odot$) and a higher-mass model ($M = 1.8\,M_\odot$). The columns list the mass $M$, central rest-mass density $\rho_{\rm c}$, stellar radius $R$, crust–core interface radius $R_{\rm cc}$, and neutron-drip radius $R_{\rm nd}$.}
\begin{tabular}{lcccc}
\arrayrulecolor{gray}
\hline\hline
 $M\,[M_{\odot}]$ & $\rho_{\rm c}\,[10^{14}\,\rm g/cm^3]$ & $R\,[\rm km]$ &  $R_{\rm cc}\,[\rm km]$ & $R_{\rm nd}\,[\rm km]$\\
\hline
 1.35 & $7.59$ & $12.29$ & $11.30$ & 11.86 \\
 1.80 & $10.5 $ & $11.89$ & $11.31$ &  11.65\\
\hline\hline
\end{tabular}
\label{tab:model}
\end{threeparttable}
\end{table}

In \cref{fig:spectral} we plot the real parts of the mode spectra for NSs with $M = 1.35\,M_\odot$ (top) and $M = 1.8\,M_\odot$ (bottom), both with a solid crust. The $1.35\,M_\odot$ case will be referred as the canonical model, and the background parameters for both stars are listed in \cref{tab:model}.
We adopt the adiabatic index $\Gamma_1$ for the perturbations of stratified NSs. 
The vertical axis shows the amplitude $A_{\rm in}$ of the ingoing GWs at the stellar surface. As we explain in Appendix~\ref{sec:appendixA}, a frequency $f$ for which $|A_{\rm in}| = 0$ 
represents a QNM of the star. These are easily visible in the figure as narrow kinks.
We separately display the low-frequency band ($\lesssim 300\,\rm Hz$), which consists of $g$-modes, and the high-frequency band ($\gtrsim 800\,\rm Hz$), which includes acoustic waves ($f$- and $p$-modes) and $s$-modes.
To examine the influence of stratification on the spectrum, we overlay our results for a nonstratified (i.e., $\Gamma = \Gamma_0$) NS of the same mass.
We observe that stratification has little impact on the high-frequency band as the restoring forces for these modes are much stronger than buoyancy.
Unsurprisingly, in the low frequency band, $g$-modes, that have buoyancy as restoring their force, cease to exist.
However, an $i$-mode is revealed in the nonstratified scenario.
We explore in detail the interplay between $i$- and $g$-modes in \cref{sec:ig_mode}.

For comparison, we also show the results for a more massive NS with $M=1.8\,M_{\odot}$ in the lower panel of \cref{fig:spectral}.
The overall picture still applies, though the high-frequency part of the spectrum shifts to higher frequencies, while the low-frequency part shifts to lower frequencies. We also observe that the $s$-mode spectrum is noticeably sparser in the heavier NS than in the lighter one as we discuss in \cref{sec:high}. 

In the next two sections, we will discuss in detail the high and low-frequency modes separately.

\section{Characteristics of high-frequency modes}
\label{sec:high}

The high-frequency band in the right panel of Fig.~\ref{fig:spectral} accommodates acoustic modes (i.e., the $f$-mode and $p_1$-mode) that are restored by pressure gradient as well as the shear modes that are restored by the shear force of the solid crust.
In the short-wavelength approximation, the dispersion relation of acoustic waves in Newtonian gravity is given by~\cite{McDermott:1988ApJ}
\begin{equation}
    \omega^2 = k^2 c_{ l}^2 =k^2 \left(c_{\rm s}^2 + \frac{4}{3}v_{\rm s}^2\right)= k^2 \left(\frac{\Gamma p}{\epsilon + p} + \frac{4}{3}v_{\rm s}^2\right) \,.
\end{equation}
Here, $k$ is the wave number, $c_{ l}$ is the wave propagation speed, and $c_{\rm s}$ is the adiabatic sound speed where $\Gamma$ refers either to $\Gamma_0$ or $\Gamma_1$ for nonstratified and stratified cases, respectively.
On average, the shear velocity $v_{\rm s}$ is about 5\% of the adiabatic sound speed $c_{\rm s}$, so the solid crust has little effect on the acoustic modes.
{Indeed, we verified that the frequencies of the $f$- and $p_1$-modes are altered by less than 0.01\% for the canonical model when we artificially set the shear modulus $\mu$ to zero. }
Stratification also has a negligible effect on the $f$-mode, but changes the $p_1$-mode frequency by about 1\%. 
In the upper panel of \cref{fig:f_mode}, we show the $f$-mode frequency as a function of the NS mass, which increases monotonically as heavier NSs are more compact and have a higher average sound speed.
In the lower panel of \cref{fig:f_mode}, we show the real parts of the radial, $\xi^{r}$, and tangential, $\xi^{\theta}$, eigenfunctions for the canonical model and the model with $M=1.8\,M_{\odot}$.
The radial eigenfunction grows from the center of the star located at the right end of the plot and slightly decreases in the outer crust. 
On the other hand, the tangential eigenfunction is nearly constant in the fluid core, and its absolute value increases sharply in the outer crust.
This behavior is important to study the breaking the outer crust because the shear strain is proportional to the gradient of the eigenfunction with respect to the radial coordinate, as we will discuss in~\cref{sec:crust_breaking}.

\begin{figure}
    \includegraphics[width=\columnwidth]{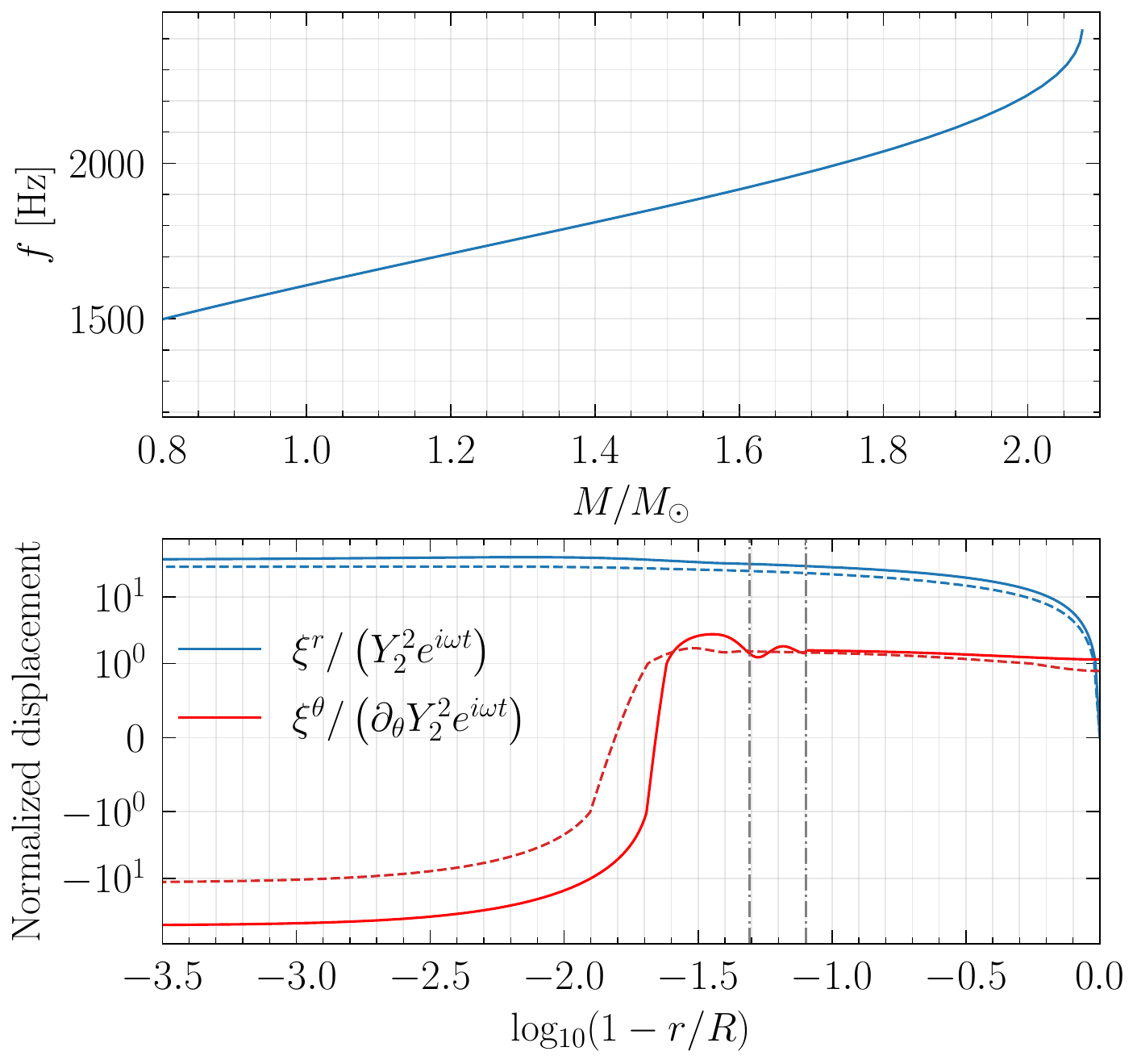}
    \caption{Top: Relation between the frequency of the $f$-mode and the mass.
    Bottom: Normalized radial and tangential eigenfunctions of the $f$-mode for the canonical model with $M=1.35\,M_{\odot}$. The eigenfunctions for the $M = 1.8\,M_{\odot}$ model are also shown as dashed lines. The two vertical dashed lines (from left to right) indicate the crust-core boundaries of the $1.8\,M_{\odot}$ and $1.35\,M_{\odot}$ models, respectively.}
    \label{fig:f_mode}
\end{figure}

\begin{figure}
    \includegraphics[width=\columnwidth]{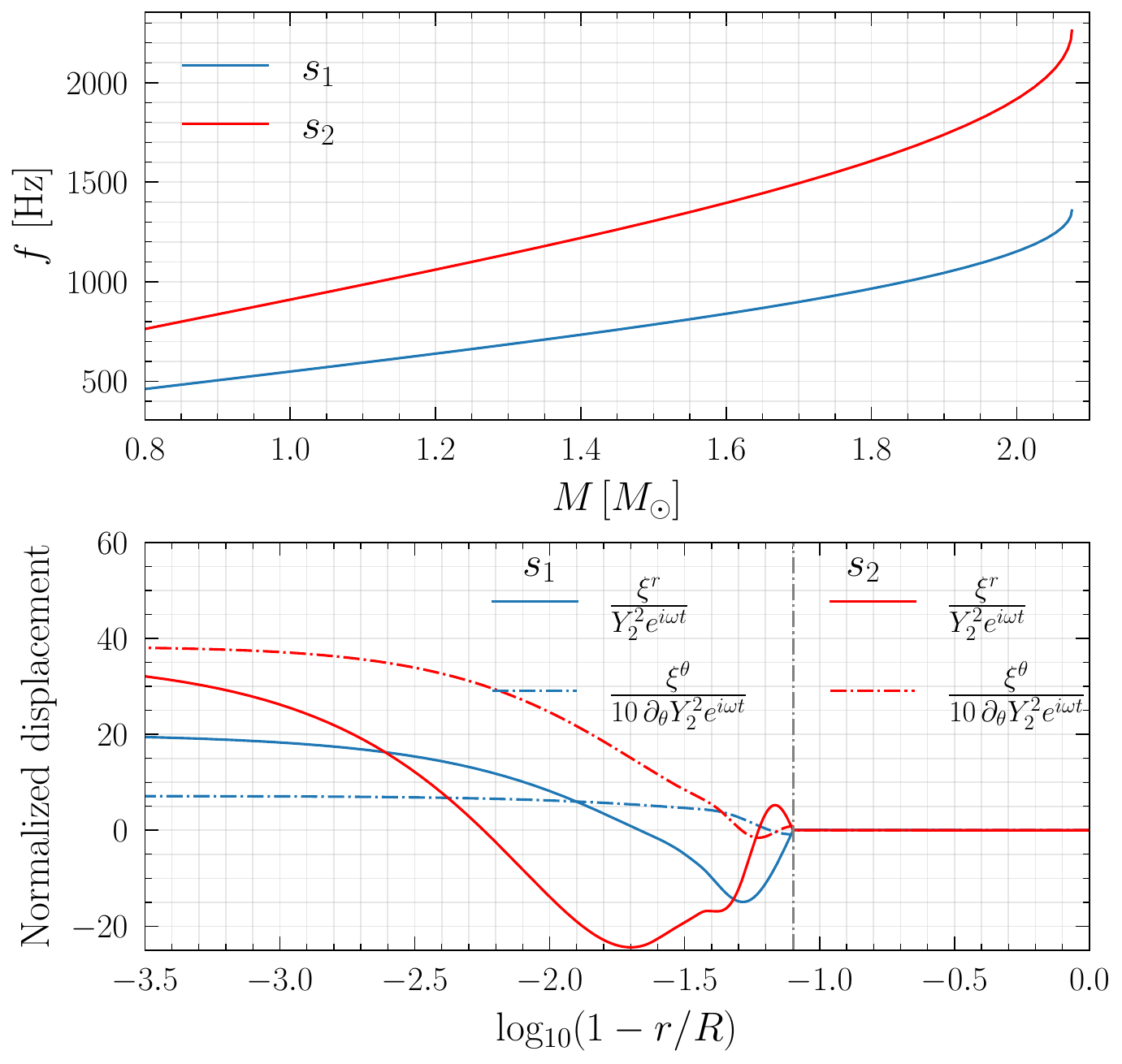}
    \caption{{Shear modes $s_1$ and $s_2$ for the canonical model. The format of the figure follows that of \cref{fig:f_mode}, except that only the results for the canonical model are shown. The tangential eigenfunctions have been divided by a factor of 10 for better visibility.}}
    \label{fig:s_mode}
\end{figure}

As we saw in \cref{fig:spectral}, the $s$-modes are approximately evenly-spaced in frequency. Their dispersion relation in Newtonian gravity is given by
\begin{equation}
\omega^2 = k^2 v_{\rm s}^2 \,,
\end{equation}
where $v_{\rm s}$ is the shear wave speed~\cite{McDermott:1988ApJ}. The corresponding frequency can be estimated as
\begin{equation}
f \simeq 10^3 \left(\frac{R - R_{\rm cc}}{\lambda} \right)\left(\frac{1 \,\mathrm{km}}{R - R_{\rm cc}}\right)\left(\frac{v_{\rm s}}{10^8\,\mathrm{cm/s}}\right)\,\mathrm{Hz} \,,
\end{equation}
where $R-R_{\rm cc}$ is the the thickness of the crust, that is, the difference between the star's radius $R$ and the location of the crust-core interface $R_{\rm cc}$. 
{For low overtones of $s$-modes, the wavelength $\lambda$ scales with the crust thickness, resulting in a characteristic frequency of approximately 1$\,$kHz.
Because of this scaling, the frequency of a given $s$-mode is higher in the heavier NS, as its crust is thinner. 
Moreover, the s-mode spectrum is sparser for the more massive star than in the lighter one. This is expected because the spacing between neighboring $s$-modes scales approximately as $1/(R - R_{\rm cc})$.

In the upper panel of \cref{fig:s_mode}, we show the frequencies of the $s_1$ and $s_2$ modes as functions of the mass. The mode frequencies increase with mass due to the progressive thinning of the crust, which reduces the wavelength and thus increases the frequency. As shown in \cref{fig:spectral}, stratification has a negligible effect on the $s$-mode spectrum, as the shear stresses in the solid crust are typically an order of magnitude larger than the local buoyancy forces in this region. In the lower panel of \cref{fig:s_mode}, we present the real parts of the radial and tangential eigenfunctions of the $s_1$ and $s_2$ modes for the canonical model. The oscillations are confined almost completely within the crust because only the solid can sustain shear stresses. The tangential displacement is one order of magnitude larger than the radial displacement. For this reason we divide it by a factor of 10 for for the purpose of presentation. 

\section{Characteristics of low-frequency modes}\label{sec:low}

We now turn our attention to the low-frequency band in the left panel of \cref{fig:spectral} that includes the $i$-mode and a subset of $g$-modes (viz.,~$g_1$ to $g_{11}$).
We examine the spectral properties and eigenfunctions of these modes in Secs.~\ref{sec:i_mode} and~\ref{sec:g_mode}, respectively.
As we will see, an interesting feature of the low-frequency modes is that the $i$-mode appears only in nonstratified models, while it is eliminated by stratification.
However, its eigenfunction characteristics are imprinted on the $g$-modes, particularly at higher overtones.
This phenomenon is further discussed in Sec.~\ref{sec:ig_mode}.

\subsection{ $i$-mode of nonstratified NSs}\label{sec:i_mode}

The $i$-modes were first studied in geophysics where they are known as Scholte waves at the interfaces between elastic and fluid media~\cite{scholte1942stoneley}, and have since been investigated in the context of NSs (e.g., Refs.~\cite{McDermott:1988ApJ,Tsang:2011ad,Tsang:2013mca,Yoshida:2002vd,Kruger:2024fxn,Sotani:2023ypt,Passamonti:2020fur,Zhu:2022pja}).
The basic picture of the crust-core interfacial mode is a surface wave propagating along the core-crust interface. 
As a pressure perturbation travels across the crust-core boundary, it compresses the crust, leading to a pronounced radial displacement at the interface. 
This interficial mode of oscillation is eliminated by buoyancy when NS stratification is taken into account.} 
Nevertheless, its characteristics manifest in the overtones of $g$-modes as will be further demonstrated later.
Therefore, all results we present about the $i$-modes 
are obtained by setting the buoyancy to zero, in other words, 
using $\Gamma = \Gamma_0$.

\begin{figure}
    \includegraphics[width=\columnwidth]{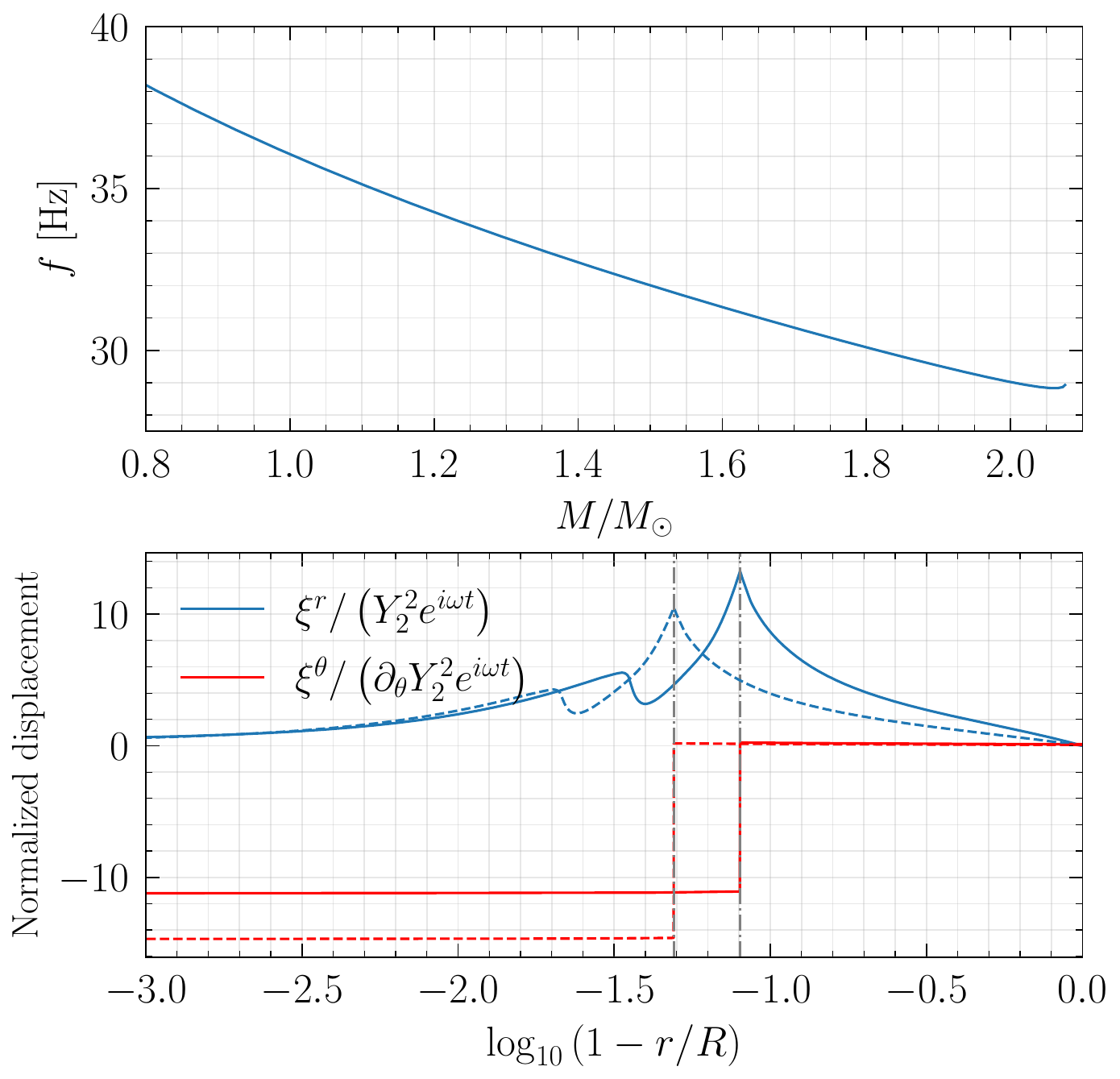}
    \caption{Same as \cref{fig:f_mode} but for the $i$-mode in the nonstratified NS.}
    \label{fig:i_mode_eigen}
\end{figure}

We depict $i$-mode's eigenfunctions in the bottom panel of \cref{fig:i_mode_eigen} for the canonical model.
The radial displacement $\xi^{r}$ decays into both the core and the crust, as has a characteristic kink-like structure at the core-crust interface. 
The tangential displacement $\xi^{\theta}$ is significantly excited in the crust but is highly suppressed in the core due to the abrupt change in shear modulus, resulting in a discontinuity at the interface. 
Since the stress is related to the spatial derivative of displacements, the large displacement gradient at the the base of the crust can induce large shear stresses there. 
Some literature (e.g.,~Refs.~\cite{Tsang:2011ad,Pan:2020tht})
thus suggests that the resonantly excited $i$-mode can potentially yield the crust breaking if \reply{the amplitude is large enough}.

We also present the $i$-mode frequency as a function of mass in the upper panel of \cref{fig:i_mode_eigen}. 
The frequency decreases with increasing mass, primarily due to crust thickening. 
For NSs with $M \gtrsim 1.0\,M_{\odot}$, the $i$-mode frequency remains below approximately 40\,Hz. 
These frequencies are of the same order of magnitude as those reported in Refs.~\cite{Zhu:2022pja,Kruger:2024fxn,Sotani:2023ypt}, but are in tension with Refs.~\cite{Neill:2020szr,Neill:2022psd}, which report $i$-mode frequencies in the range of 100--200$\,\rm Hz$ for a broad set of nuclear parameters. 
The inconsistencies may arise from the EOS prescriptions or the Newtonian calculations used in these latter works.
Understanding this inconsistency is left for future. A difference of an order of magnitude in these mode frequencies would significantly impact the resonance onset during binary inspirals, and would thus render very different observational implications.
Should an observation be made, the difference could also result in biased Bayesian inferences about the underlying nuclear parameters, which determine the onset density of the crust-core transition that the $i$-mode frequency is sensitive to. Therefore, this discrepancy warrants further investigation. A systematic study of the $i$-mode frequency band across a broad range of EOSs will be carried out elsewhere.

\subsection{$g$-mode of stratified NSs}\label{sec:g_mode}

Once stratification is included, a perturbed fluid element experiences buoyancy due to the composition gradient, giving rise to a spectrum of $g$-modes. The characteristic angular frequency of the oscillating fluid parcel, known as the Brunt–V{\"a}is{\"a}l{\"a} frequency $N$, is given by~\cite{Ipser:1992ApJ}
\begin{equation}
    \label{eq:BV}
   N^2 =e^{\nu-\lambda} \, \frac{p^\prime}{\epsilon+p}A \,, \quad A = \frac{p'}{p}\left(\frac{1}{\Gamma_0}-\frac{1}{\Gamma_1}\right)\,,
\end{equation}
where $A$ is the relativistic Schwarzschild criterion~\cite{Detweiler:1973ApJ,Kuan:2021jmk,Gao:2025nfj}. 
If $\Gamma_1 > \Gamma_0$ (i.e., $A < 0$ and $N^2 > 0$), the fluid is stable against convection and supports the propagation of $g$-modes.

\begin{figure}
    \centering
    \includegraphics[width=\linewidth]{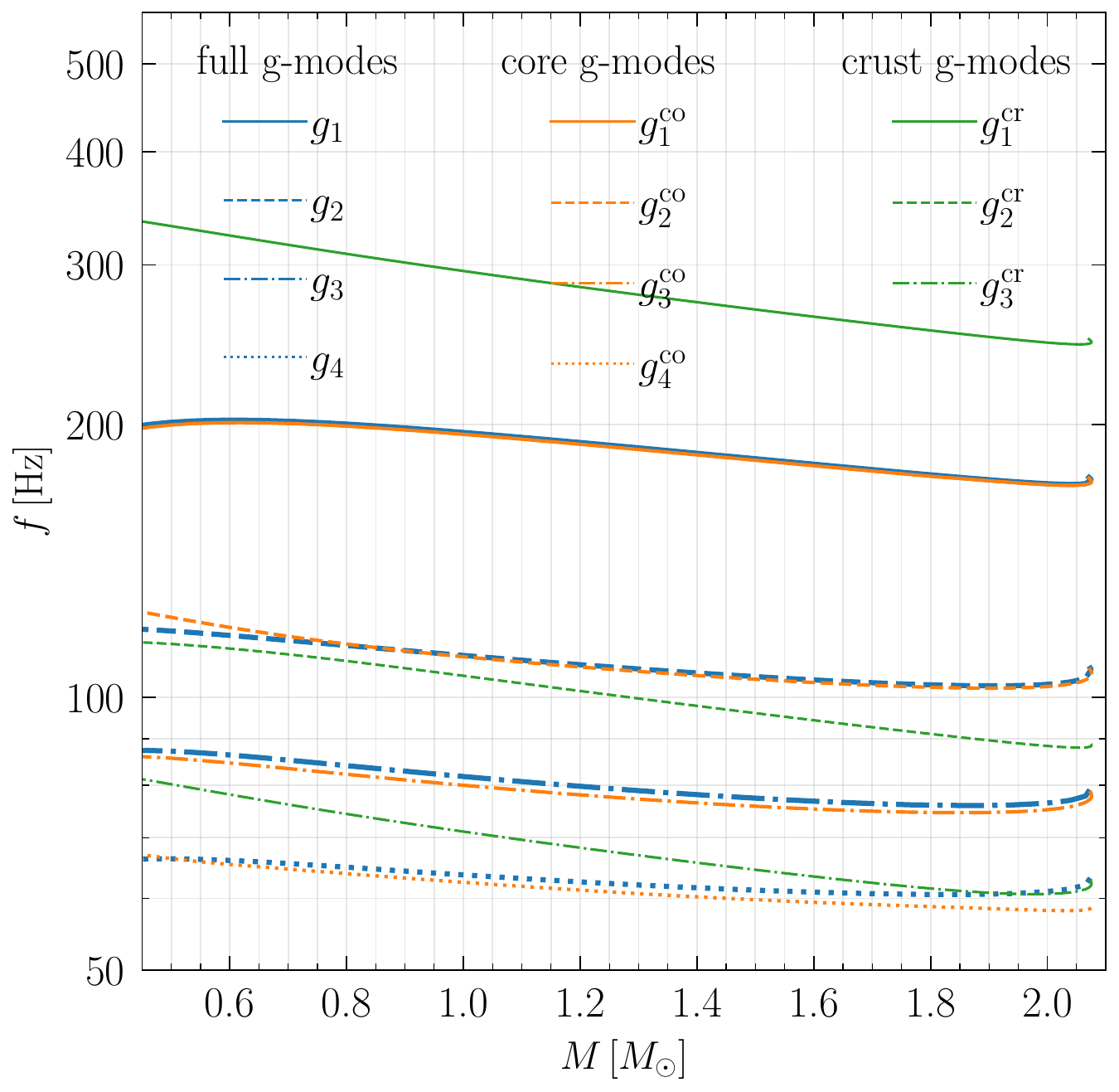}
    \caption{The frequency of the $g$-modes as a function of mass. The ``full $g$-modes'' branch corresponds to NSs with a solid crust. In contrast, the ``core $g$-modes'' and ``crust $g$-modes'' branches refer to models where the elasticity of the crust has been removed. See the text for details. 
    }
    \label{fig:g_mode_spec}
\end{figure}

As shown in \cref{fig:gamma}, the TW99 EOS satisfies the condition $\Gamma_1 - \Gamma_0 \geqslant 0$, with the difference becoming particularly pronounced near the neutron drip point and in the outer core around $\rho \sim 2 \times 10^{14}\,\mathrm{g/cm^{3}}$. Neglecting the solid phase, we expect two distinct classes of $g$-modes propagating in the crust and core, respectively, as previously discussed by \citet{Counsell:2024pua} using the BSK family of EOSs~\cite{Goriely:2010bm,Pearson:2012hz,Potekhin:2013qqa,Fantina:2013A&A,Shchechilin:2023erz}. We perform a similar analysis for the TW99 EOS by setting the shear modulus to zero. The resulting mode spectrum is shown in \cref{fig:g_mode_spec}, where the core $g$-modes and crust $g$-modes are labeled as $g^{\,\rm co}_{n}$ and $g^{\,\rm cr}_{n}$, respectively. {To distinguish between the two families, we artificially suppress the buoyancy in a selected region by setting $\Gamma_1=\Gamma_0$. For instance, when the buoyancy around the neutron-drip layer is removed, the crustal $g$-modes disappear from the spectrum while the core $g$-modes persist. Conversely, eliminating the buoyancy in the outer core leaves only the crustal $g$-modes.} The subscript $n$ labels the overtone number of the mode. The frequencies of the $g$-modes decrease monotonically with respect to the mass, except near the maximal mass, but the dependence on the mass is weak.
This weak dependence can be understood by noting that $g$-mode frequencies scale approximately with the Brunt–V{\"a}is{\"a}l{\"a} frequency $N$ defined in \cref{eq:BV}, which itself scales as $N^2 \sim g A$, where $g$ is the local gravitational acceleration and $A$ characterizes the stratification. Since $N$ is primarily determined by the composition gradient in the outer core, and $g$ does not vary significantly across models with different masses, the $g$-mode frequencies remain relatively insensitive to the stellar mass.

The eigenfunctions of the two lowest-order modes for the same background model as in \cref{fig:spectral} are displayed in \cref{fig:fluid_eigen}. For the crust $g$-modes, the amplitudes of the radial and tangential eigenfunctions are primarily concentrated in the outer crust, especially for the $g^{\rm cr}_{1}$-mode. The morphology of the eigenfunctions near the neutron drip point resembles that expected for $i$-modes, with a kink structure in $\xi^r$ and a sharp jump in $\xi^{\theta}$ at the neutron drip point. For the core $g$-modes, the amplitude of the radial eigenfunctions peaks in the core and becomes small in the crust, while the amplitude of the tangential eigenfunctions resides in both the core and crust, indicating that the buoyancy force in the core is sufficiently strong to penetrate into the crust region if we neglect the elasticity.

\begin{figure}
    \centering
    \includegraphics[width=\linewidth]{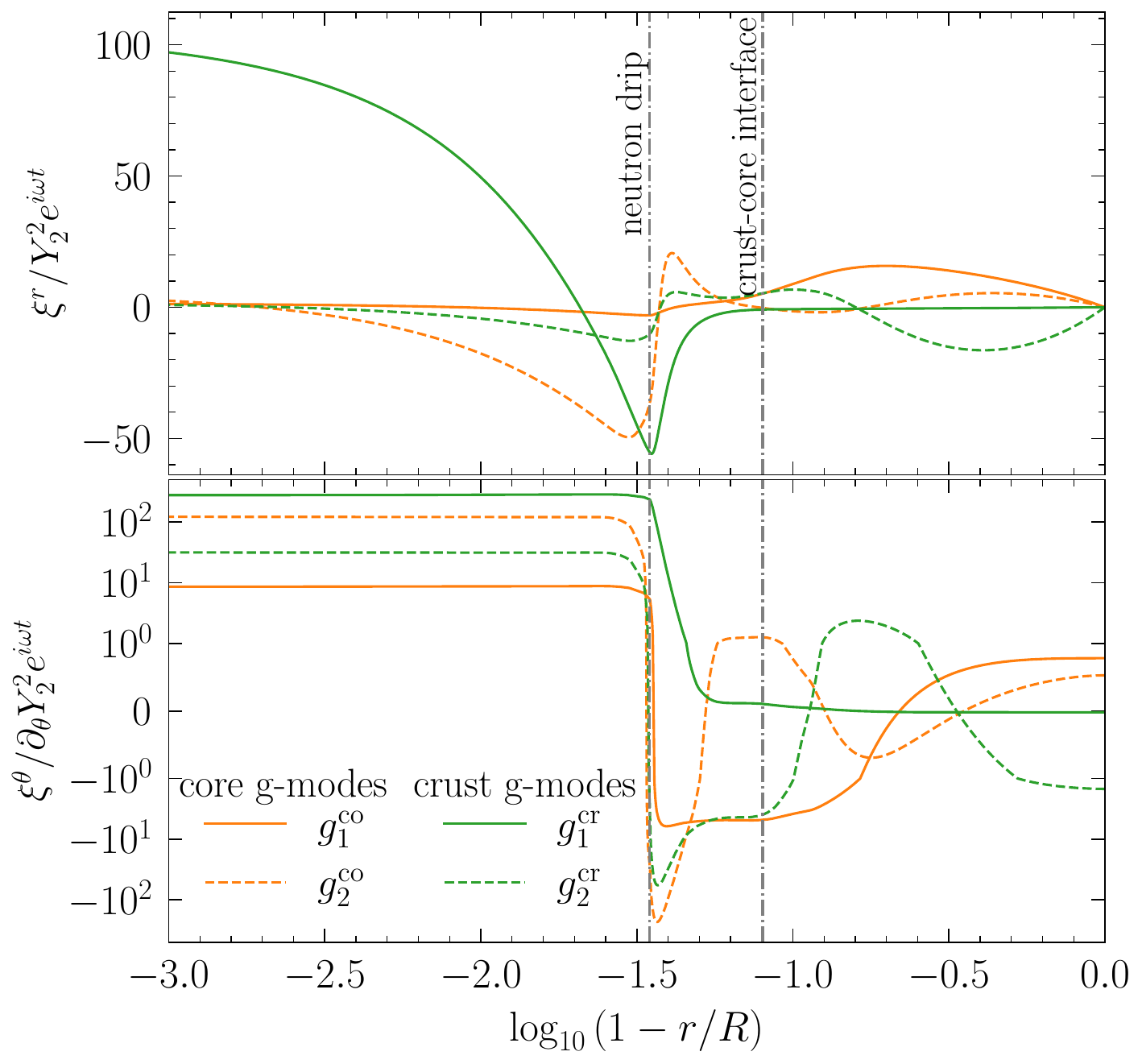}
    \caption{The radial eigenfunctions of the core $g$-modes ($g^{\rm co}_1$--$g^{\rm co}_2$) and the crust $g$-modes ($g^{\rm cr}_1$--$g^{\rm cr}_2$) for {the canonical model where the elasticity of the crust has been removed}.}
    \label{fig:fluid_eigen}
\end{figure}

Once elasticity is included in the crust, the crust $g$-modes disappear from the spectrum, and the core $g$-modes dominate the entire $g$-mode spectrum. To distinguish this case from the perfect-fluid model, we refer to the $g$-modes in NSs with a solid crust as ``full $g$-modes'' and denote them as $g_n$. As shown in \cref{fig:g_mode_spec}, the frequencies of the full $g$-modes closely follow those of the core $g$-modes, with increasing deviations for overtones.
We further illustrate the eigenfunctions of the $g_1$–$g_6$ modes in \cref{fig:g_mode_eigen}. 
Compared to the $g^{\rm co}_1$ and $g^{\rm co}_2$ modes shown in \cref{fig:fluid_eigen} for the fluid case, the amplitudes of both $\xi^r$ and $\xi^\theta$ are heavily suppressed in the crust region.}
The amplitude within the crust is further quenched for higher overtones (corresponding to lower mode frequencies), and the eigenfunctions near the crust-core interface begin to exhibit features of $i$-modes.

\begin{figure}
    \centering
    \includegraphics[width=\linewidth]{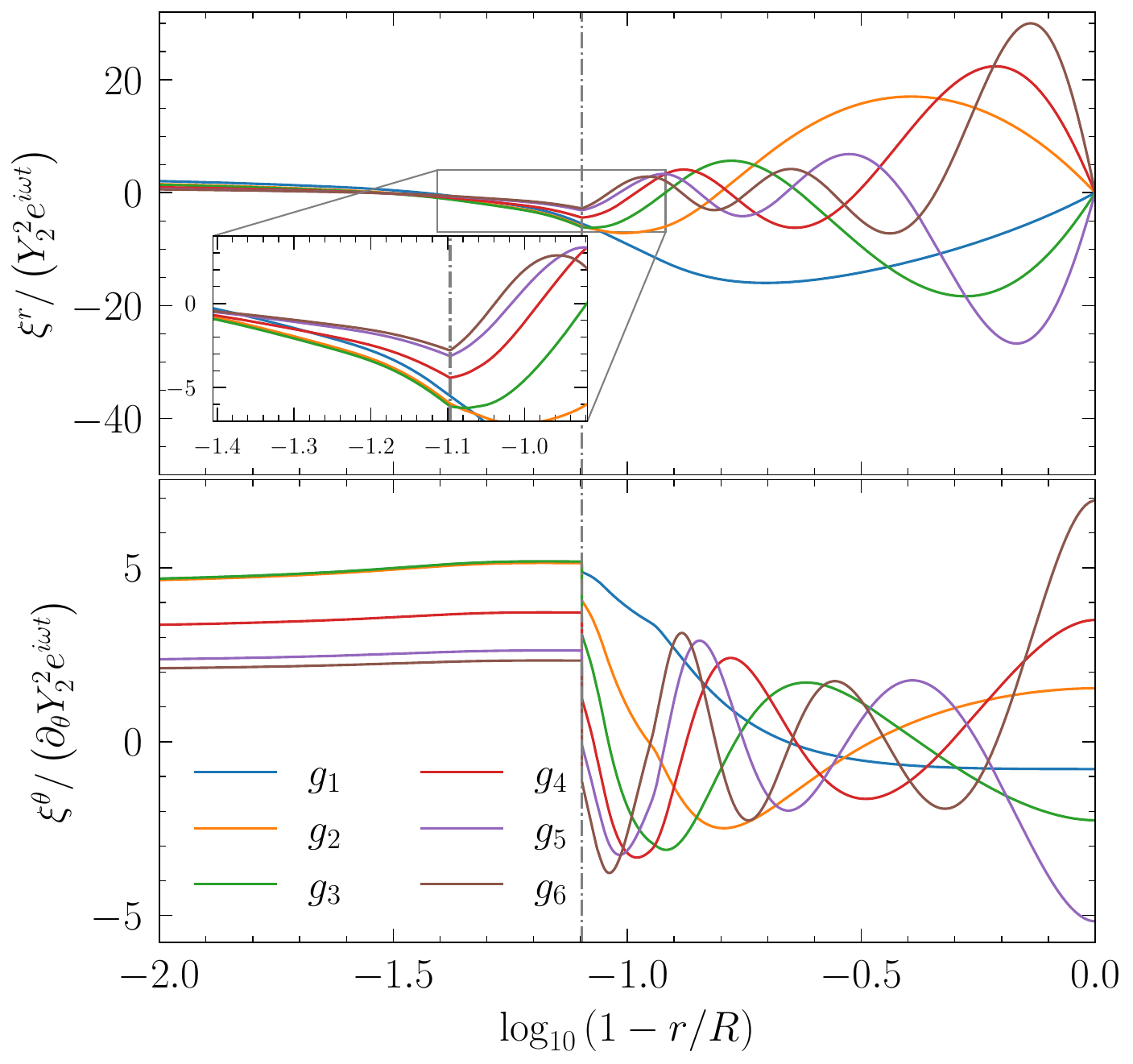}
    \caption{The radial ({\it top}) and tangential ({\it bottom}) eigenfunctions for $g_1$--$g_6$ modes for the canonical model. The vertical dashed line denotes the crust-core interface.}
    \label{fig:g_mode_eigen}
\end{figure}

To understand the behavior of the $g$-mode eigenfunctions in the solid crust, we compare the mode's transverse momentum with the contribution from crustal shear stresses. For simplicity, we adopt the short-wavelength approximation, in which the eigenfunction varies as $\exp(ikr)$. According to Eqs.~(\ref{eq:strain}) and (\ref{eq:stress}), the dominant tangential contribution from the solid can be approximated as
\begin{equation}
    \label{eq:criterion}
    \frac{\dd \, \delta \pi_{r}^{\theta}}{\dd r} \sim \frac{\dd \, (r T_2)}{\dd r} \sim \mu \frac{\dd^2 \xi^{\theta}}{\dd r^2} \sim -\mu k^2 \xi^{\theta}\,,
\end{equation}
where $\mu$ is the shear modulus. This term strongly modifies the eigenfunctions when it becomes comparable to the transverse acceleration, which scales as $(\epsilon + p)\omega^2 \xi^\theta$. This yields a critical frequency,
\begin{equation}
    \label{eq:fcrit}
    f_{\rm crit} = \left( \frac{\mu}{\epsilon + p} \right)^{1/2} k \approx 100\,\left( \frac{v_{\rm s}}{10^8\,\mathrm{cm\,s^{-1}}} \right)\left( \frac{R}{\lambda} \right)\,\mathrm{Hz}\,,
\end{equation}
where $\lambda$ is the mode's wavelength and $v_{\rm s}$ is the shear speed. If the mode frequency $f$ is comparable to $f_{\rm crit}$, the mode can still propagate into the crust but is strongly affected by elasticity. On the other hand, for $f \ll f_{\rm crit}$, the mode is effectively excluded from the crust, or its frequency band is substantially altered.

The shear velocity in the crust is typically of the order of $10^{8}\,\mathrm{cm\,s^{-1}}$. For the lowest-order core $g$-modes, the frequency is around $200\,\mathrm{Hz}$, with a characteristic wavelength $\lambda \sim R$, leading to a critical frequency $f_{\rm crit} \sim 100\,\mathrm{Hz}$. In this case, the mode can still penetrate into the crust, but its amplitude is suppressed by a factor of $\sim 2$ due to the elasticity. For higher-order core $g$-modes, the frequency becomes increasingly small relative to $f_{\rm crit}$, resulting in stronger suppression of the eigenfunction in the crust.
In contrast, the lowest-order crust $g$-modes have frequencies on the order of $300\,\mathrm{Hz}$, but much shorter wavelengths, $\lambda \sim 0.05R$, yielding a much higher critical frequency, $f_{\rm crit} \sim 2000\,\mathrm{Hz}$. As a result, these modes are unlikely to propagate in the crust once elasticity is included. Indeed, we find that crust $g$-modes vanish from the spectrum in the elastic case. However, due to numerical noise dominating the spectrum below $10\,\mathrm{Hz}$, we cannot rule out the possibility that they survive with very low, but finite, frequencies.

\subsection{Interplay between $g$- and $i$-mode}\label{sec:ig_mode}

\begin{figure}
    \includegraphics[width=\columnwidth]{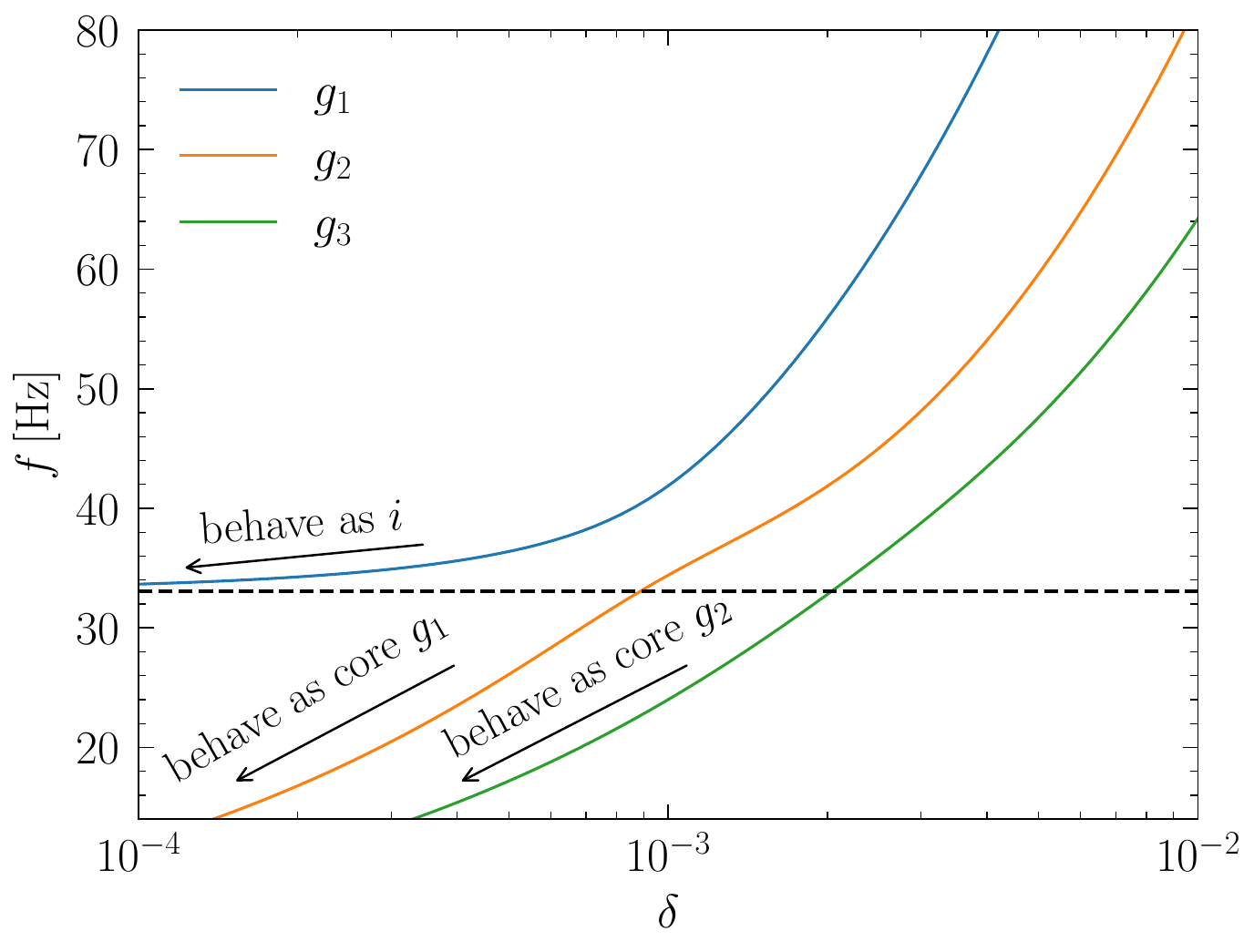}
    \caption{Frequencies of $g$-modes for an NS with $1.35\,M_{\odot}$ under a phenomenological stratification $\Gamma_1 = \Gamma_0(1 + \delta)$. The frequency of the $i$-mode in the absence of stratification is represented by the horizontal dashed line.}
    \label{fig:ig}
\end{figure}

As discussed earlier, the $i$-mode seemingly ``disappears'' when stratification is included. At the same time, the eigenfunctions of high-order $g$-modes --- whose frequencies are close to that of the $i$-mode --- begin to resemble the characteristic behavior of $i$-modes near the crust-core interface. 
This interesting interplay between $i$- and $g$-modes arises from the competition between their respective restoring forces. 
A simple way to assess the relative strength of these forces is to compare the characteristic frequencies of the mode families they support. 
To this end, we introduce a phenomenological modification to the adiabatic index, 
\begin{equation}
\Gamma_1 = \Gamma_0 \, (1 + \delta)\,,
\end{equation}
and examine how the strength of the buoyancy force, and consequently the $g$-mode frequency, changes with varying $\delta$.

In \cref{fig:ig}, we present the relationship between the frequencies of the first three $g$-modes and the control parameter $\delta$. As $\delta$ decreases from $10^{-2}$ to $10^{-4}$, the buoyancy force weakens, leading to a corresponding decrease in the frequencies of the $g$-modes. The $g_1$-mode, which has no nodes in the interior of the NS, gradually transitions into an $i$-mode, with its frequency approaching the $i$-mode frequency of a nonstratified star, $f = 32.9\,\rm Hz$. This transition is illustrated in \cref{fig:ig_eigen} for three representative values of $\delta$. 

When $\delta = 10^{-2}$, the buoyancy force of the $g_1$-mode is much stronger than the restoring force of the $i$-mode, and the eigenfunction clearly exhibits the characteristics of a $g_1$-mode. For $\delta = 9.4 \times 10^{-4}$, the frequency of the $g_1$-mode is slightly higher than that of the $i$-mode, and the eigenfunction displays a mixed character, with a dominant contribution from the $i$-mode. At $\delta = 10^{-4}$, the frequency of the $g_1$-mode is nearly equal to that of the $i$-mode, and the eigenfunction behaves as a pure $i$-mode; compare with \cref{fig:i_mode_eigen}. Since the frequency of $i$-modes is in the range of $30$--$40\,\rm Hz$, comparable to $f_{\rm crit}\sim 100\,\rm Hz$, the mode can penetrate into the crust. As the difference between $\Gamma_1$ and $\Gamma_0$ in the outer core for the TW99 EOS is much larger than $10^{-2}$, the $i$-mode naturally disappears and only leaves imprints near the crust-core interface in the higher-order $g$-modes.

The frequencies of the $g_2$ and $g_3$ modes drop well below the $i$-mode frequency as $\delta \sim 10^{-4}$, and these modes gradually transition into the core $g_1$ and core $g_2$ modes, respectively. As shown in \cref{fig:ig_eigen}, the $g_2$ mode becomes nearly confined to the core and exhibits the characteristic structure of a core $g_1$ mode. This is consistent with expectations, as the mode frequency of the $g_2$-mode, $f = 12.7\,\mathrm{Hz}$, is much smaller than the critical frequency $f_{\rm crit} = 100\,\mathrm{Hz}$ (for $\lambda \sim R$), and thus the mode is effectively excluded from the crust by the restoring shear forces.

It is worth emphasizing here that compositional $g$-modes are different from thermal $g$-modes that are supported by temperature gradients. For an old NS with a typical temperature of $\sim 10^7\,\mathrm{K}$, the frequencies of low-order thermal $g$-modes are of the order of $0.001$–$0.01\,\mathrm{Hz}$, well below $f_{\rm crit}$, and are therefore completely excluded from the solid crust. These modes are confined to the fluid core or the ocean layer, and are referred to as core $g$-modes and ocean $g$-modes in \citet{McDermott:1988ApJ}.

\begin{figure}
    \includegraphics[width=\columnwidth]{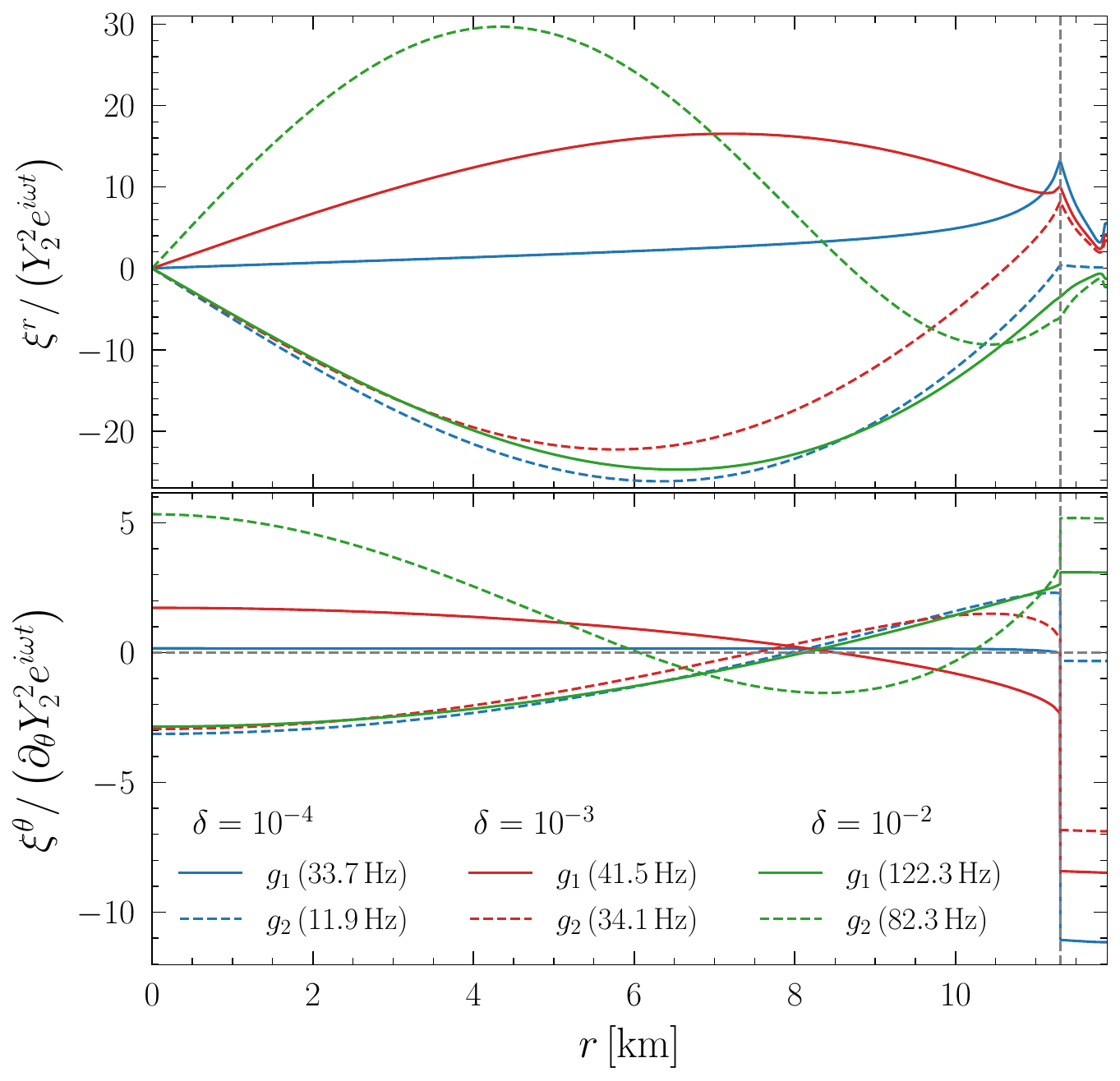}
    \caption{The radial ({\it top}) and tangential ({\it bottom}) eigenfunctions of the first two $g$-modes for three representative $\delta$ with NS mass $M=1.35\,M_{\odot}$.}
    \label{fig:ig_eigen}
\end{figure}

\section{Tidal resonance kinetics and orbital energy transfer}
\label{sec:resonance}

The spectrum of an NS \emph{almost} completely dictates its internal fluid motion to the linear order (though see Refs.~\cite{Pitre:2023xsr,Pitre:2025qdf,Kuan:2024jnw,Reboul-Salze:2025gyi,Kwon:2024zyg,Kwon:2025zbc} for progress in nonlinearity).
In particular, the perturbative flows $\xi^{i}$ inside an inspiraling NS driven by the tidal field of its companion can be approximately modeled as a sum of harmonic oscillators,
\begin{equation}
    \boldsymbol{\xi}(\boldsymbol{r}, t)=\sum_\alpha q^\alpha(t) \, \boldsymbol{\xi}_\alpha(\boldsymbol{r})\,,
\end{equation}
where $q^{\alpha}$ is the amplitude of a specific mode with $\alpha$ denoting the quantum number $\{n,\,\ell,\,m\}$. 
Hereafter we use the indices $\alpha$ and $\{n,\,\ell,\,m\}$ interchangeably. 
How these oscillators are forced and excited as the binary orbiting around at a chirping frequency is governed by the Hamiltonian
\begin{align}
    H = H_{\rm orb+gw} + H_{\rm osc} + H_{\rm tid} \,.
\end{align}
Here $H_{\rm orb+gw}$ captures the inspiral motion for point particles, which includes the conservative orbital dynamics and the dissipative part caused by back reaction of GWs. They can be well described by post-Newtonian dynamics via the effective-one-body formalism~\cite{Buonanno:1998gg}. 
The Hamiltonian for the harmonic oscillators 
is denote by $H_{\rm osc}$, 
and $H_{\rm tid}$ denotes the tidal work done on the NS through the excitation of its internal modes of oscillation. 

In Newtonian gravity, $\xi_{\alpha}^i$ satisfies a self-adjoint eigenvalue equation (see, e.g., Refs.~\cite{Lai:1993di,Press:1977ApJ}), 
\begin{equation}
   \frac{1}{2} \mathcal{\rho} \,\omega_\alpha^2\, \xi_\alpha =  \mathcal{C} \, \xi_\alpha\,,
\end{equation}
where $\mathcal C$ is appropriate potential operator. Different modes are orthogonal to each other  
\begin{equation}
     \int \rho \, (\xi_{\alpha})_{i}  \,  {(\bar{\xi}_{\alpha'})^{i}}\dd x^3= \mathcal{A}_\alpha^2 \, \delta_{\alpha \alpha^{\prime}}\,,
\end{equation}  
where $\mathcal{A}_\alpha^2$ is the normalization factor, and overhead bar denotes complex conjugation. 
The Hamiltonian $H_{\rm osc}$ can be written as~\cite{Alexander:1987MNRAS,Kokkotas:1995xe}
\begin{equation}
    \label{eq:H_osc}
    H_{\mathrm{osc}}=\frac{1}{2} \sum_\alpha\left(\frac{{p^\alpha \bar{p}^{\alpha} }}{{\mathcal{A}_\alpha^2}}+{\mathcal{A}_\alpha^2}\omega_\alpha^2 q^\alpha \bar{q}^{\alpha}\right)\,,
\end{equation}
where $p^{\alpha}$ is the canonical momenta associated with $q_{\alpha}$.

In general relativity, the eigenvalue problem for stellar oscillations is generally non-self-adjoint due to the presence of GW emission, which leads to complex mode frequencies. 
However, for certain modes such as $g$-modes and interface $i$-modes, the GW damping timescales are extremely long~\cite{Kruger:2014pva,Kruger:2024fxn}. 
For these modes, the imaginary part of the frequency can be safely neglected with their frequencies and eigenfunctions essentially unchanged.
Assuming a real spectrum leads to treating the eigenvalue problem as Hermitian system when computing mode energies and couplings. 
By adopting the variational principle used in Refs.~\cite{Detweiler:1973ApJ,Schumaker:1983ads,Finn:1990ads}, and retaining only the conservative contributions, we arrive at
\begin{align}
    &\frac{1}{2}\omega^2 \int \bigg[e^{-\nu} (\epsilon+p) \, \xi_{i} \, \bar{\xi}^{i}  -\frac{r^{2\ell} e^{-\nu}}{16 \pi}\left(K \bar{K} + K \bar{H_2}+H_2 \bar{K}\right) \nonumber \\ & -\frac{\ell(\ell+1) e^{-\lambda-\nu} r^{2\ell}}{16 \pi} H_1 \bar{H_1}\bigg] \sqrt{-g} \, \dd^3x = \text{Potential terms}
     \,,
\end{align} 
with $g$ being the determinant of the metric $g_{\mu\nu}$.
Here, on the left-hand side, the first term represents the kinetic energy of the perturbed fluid while 
the terms involving metric perturbations have no Newtonian counterpart and may be broadly interpreted as the kinetic energy of the gravitational field.
The right-hand side consists of potential terms with lengthy expressions that are not needed in this paper; we therefore refer the reader to Refs.~\cite{Detweiler:1973ApJ,Finn:1990ads} for more details.
By setting the normalization factor
\begin{align}
    \label{eq:norm}
    \mathcal{A}_\alpha^2 =& \int \bigg[e^{-\nu}(\epsilon+p)   (\xi_{\alpha})_{i}   {(\bar{\xi}_{\alpha})^{i}} -\frac{\ell(\ell+1) e^{-\lambda-\nu} r^{2\ell}}{16 \pi} H_1 \bar{H_1} \nonumber \\
    & -\frac{e^{-\nu}r^{2\ell}}{16 \pi}\left(K \bar{K} + K \bar{H_2}+H_2 \bar{K}\right)\bigg] \sqrt{-g} \, \dd^3x \,,
\end{align}
the corresponding Hamiltonian retains the canonical oscillator forms as given in \cref{eq:H_osc}. We omit the subscript “$\alpha$” on the metric functions for simplicity.
For the literature adopting the Cowling approximation (e.g., Ref.~\cite{Counsell:2024pua}), the metric perturbations are omitted completely throughout (not only restricted to the real frequencies).
In this limit, the terms associated with the metric perturbation are absent from the normalization factor.
Within the relativistic framework, an alternative choice of the normalization factor used in Ref.~\cite{Kuan:2021sin} normalizes eigenfunction by its components in fluid motion while the contributions of metric functions are suppressed since their contributions are considerably less important than the former.
In this work, we take the normalization factor in \cref{eq:norm} to be 
\begin{equation}
    \mathcal{A}_\alpha^2=MR^2 \,.
\end{equation}
 
The Hamiltonian $H_{\rm tid}$ characterizes the gravitational energy absorbed by the NS due to its coupling to the companion's tidal field. 
A full treatment requires solving for the dynamical tidal response in the buffer zone surrounding the star~\cite{Thorne:1984mz}, which calls for further development along the lines of Refs.~\cite{HegadeKR:2024agt,Katagiri:2024wbg,Steinhoff:2016rfi}, and is beyond the scope of this work. 
Instead, we adopt a simplified approximation following~\citet{Kuan:2021sin}, in which the interaction Hamiltonian takes the form
\begin{align}
    H_{\rm tid} &= \int \delta \bar{\epsilon}\,\Phi^{\mathrm{T}}  \sqrt{-g} \, \dd^3 x 
    \nonumber \\
    &= \sum_{\alpha} \bar{q}^{\alpha} \int (\epsilon + p)\, \bar{\xi}_{\alpha}^{i} \nabla_i \Phi^{\rm T} \sqrt{-g}\, \dd^3x \nonumber \\
    &\quad + \sum_{\alpha} \bar{q}^{\alpha} \int r^{\ell}\left(\frac{1}{2}\bar{H}_2+\bar{K}\right)\bar{Y}_{\ell}^{m} \, \Phi^{\rm T} \sqrt{-g}\, \dd^3x\,,
\end{align}
where $\delta \bar{\epsilon}$ is the conjugation of the Eulerian perturbation of the energy density, $\delta \epsilon$, and the external tidal potential is given by the Newtonian one~\cite{Press:1977ApJ}
\begin{equation}
    \Phi^{\rm T}=-G M^{\prime} \sum_{\ell=2}^{\infty} \sum_{m=-\ell}^\ell \frac{W_{\ell m} \, r^\ell}{a^{\ell+1}} Y_{\ell}^{m} e^{-i m \Phi(t)}\,.
\end{equation}
Here, $M'$ is the mass of the companion, $a$ is the orbital separation, $\Phi(t)$ is the orbital phase, and $W_{lm}$ is zero for odd $\ell+m$, and otherwise 
\begin{align}
    W_{\ell m}= &\, (-1)^{(\ell+m) / 2}\left[\frac{4 \pi}{2 \ell+1}(\ell+m)!(\ell-m)!\right]^{1 / 2} \nonumber \\
    & \times\left[2^\ell\left(\frac{\ell+m}{2}\right)!\left(\frac{\ell-m}{2}\right)!\right]^{-1}\,.
\end{align}
For quadrupolar perturbations, we have
\begin{equation}
    W_{20}=-\sqrt{\frac{\pi}{5}}, \quad W_{2 \pm 1}=0, \quad W_{2 \pm 2}=\sqrt{\frac{3 \pi}{10}}\,.
\end{equation}
The Hamiltonian $H_{\rm tid}$ can then be recasted into 
\begin{equation}
    H_{\rm tid} = -\frac{G M' M}{R} \sum_{\alpha}  \bar{q}_{\alpha}{W_{\ell m}}\left(\frac{R}{a}\right)^{\ell+1} Q_{n \ell} \, e^{-i m \Phi(t)} \,,
\end{equation}
with
\begin{align}
    \label{eq:tidal_overlap}
    Q_{n \ell} =& \frac{1}{MR^{\ell}}\int \bigg[{\left(\epsilon+p\right)} \, \bar{\xi}_\alpha^{i} \, \nabla_i\left(r^\ell Y_{\ell}^{m}\right) \nonumber \\
    &+ r^{2\ell}\left(\frac{1}{2}\bar{H}_2+\bar{K}\right)\bar{Y}_{\ell}^{m} {Y}_{\ell}^{m}\bigg]\sqrt{-g}\, \dd^3 x \,,
\end{align}
the relativistic analogue of the tidal overlap integral.

\begin{figure}
    \centering
    \includegraphics[width=0.9\linewidth]{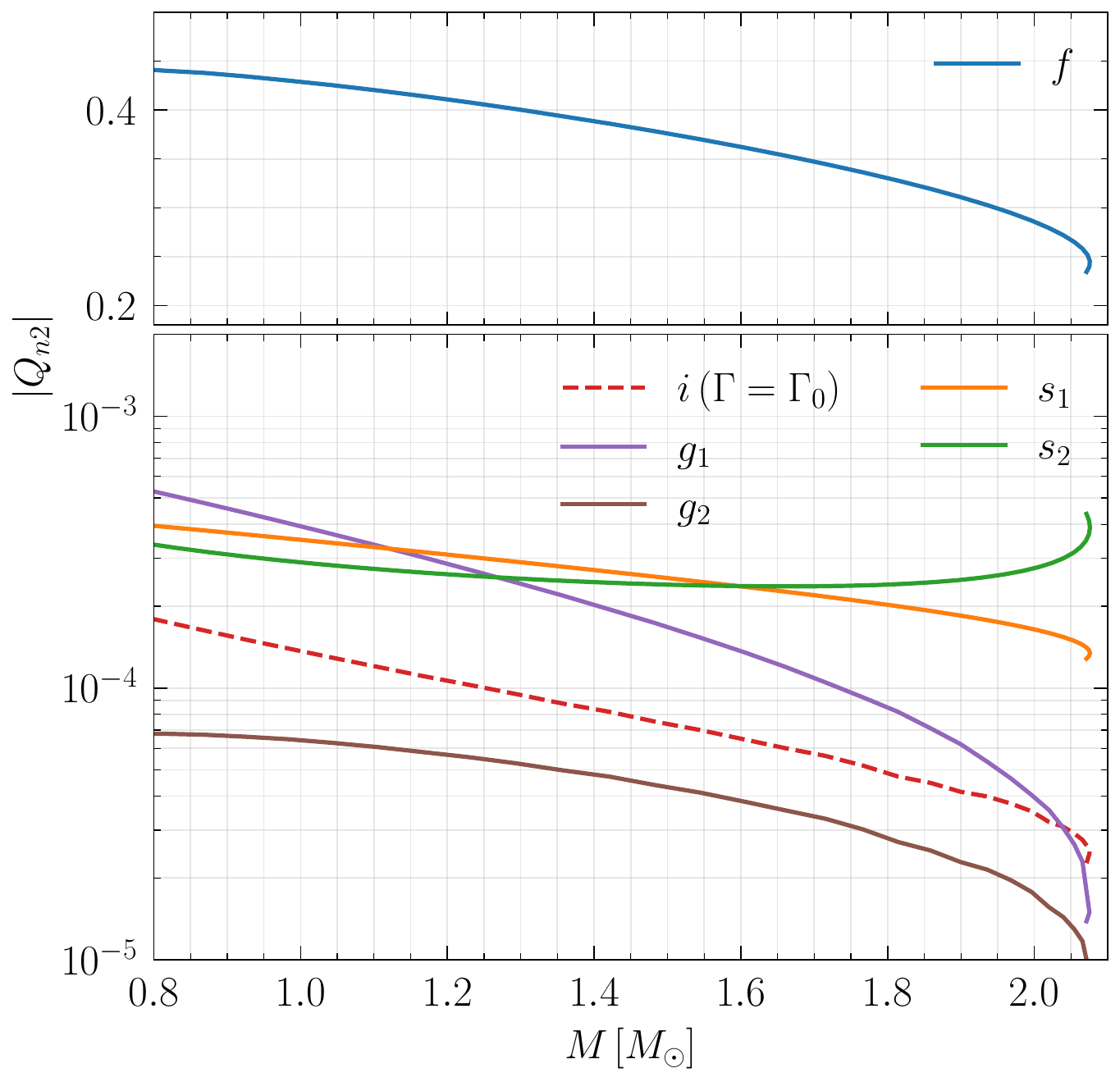}
    \caption{Tidal overlap integrals for different modes as functions of the NS mass. 
    }
    \label{fig:Qnl}
\end{figure}

\begin{table*}[]
    \centering
    \setlength{\tabcolsep}{4.5pt}
    \begin{threeparttable}
    \caption{The spectral and resonance parameters of the $f$, $s_1$–$s_2$, $g_1$–$g_2$, and $i$ ($\Gamma = \Gamma_0$) modes for a canonical NS with mass $M = 1.35\,M_{\odot}$ in an equal-mass binary system. The columns list the mode frequency $f_{\alpha}$, the tidal overlap integral $|Q_{n\ell}|$, the maximum mode amplitude $|q^{\alpha}_{\rm max}|$ attained after resonance by integrating \cref{eq:q_evolution}, and its approximation from \cref{eq:q_alpha}. Also included are the mode energy after resonance $E_{\rm mode}^{\rm max}$, the resulting GW dephasing $\Delta\Phi$, the energy deposited in the crust $E_{\rm crust}$, and the post-resonance elastic energy $E_{\rm ela}$. }
    \begin{tabular}{cccccccccc}
    \arrayrulecolor{gray}
    \hline\hline
    mode & $f_{\alpha}\,[\rm Hz]$& $|Q_{n2}|$ & $|q^{\alpha}_{\rm max}|
    \,[\rm \cref{eq:q_evolution}] $& $|q^{\alpha}_{\rm max}|\,[\rm \cref{eq:q_alpha}]$ & $E_{\rm mode}^{\rm max}\,[\rm erg]$  & $\Delta\Phi$ & $E_{\rm crust}\,[\rm{erg}]$ & $E_{\rm ela}\,[\rm{erg}]$ \\
    \hline 
    $f$ &$1785.4$ & 0.394 &  $0.102\tnote{a}$  & $0.337$  & $4.52\times 10^{51}\tnote{a}$ & $-0.04\tnote{a}$ & $7.58\times 10^{49}\tnote{a}$ & $1.99\times 10^{47}\tnote{a}$ \\
    $s_2$ & $1179.5$&  $2.48\times 10^{-4}$ & $3.12\times 10^{-4}$  & $3.00\times 10^{-4}$ & $2.96\times 10^{46}$ & $-1.16\times 10^{-5}$ & $2.76\times 10^{46}$ & $2.67\times 10^{45}$ \\
    $s_1$ & $709.8$&  $2.81\times 10^{-4}$  & $5.32\times 10^{-4}$  & $5.17\times 10^{-4}$ & $2.65\times 10^{46}$ & $-4.10\times 10^{-5}$ & $2.21\times 10^{46}$ & $2.24\times 10^{45}$\\
    $g_1$ & $187.2$&  $2.21\times 10^{-4}$ & $1.25\times 10^{-3}$  & $1.23\times 10^{-3}$ & $8.42\times 10^{45}$ & $-3.64\times 10^{-4}$ & $1.83\times 10^{45}$ & $6.70\times 10^{42}$\\
    $g_2$ & $106.9$ &  $5.12\times 10^{-5}$ &  $4.55\times 10^{-4} $ &  $4.56\times 10^{-4}$ &  $3.98\times 10^{44}$ & $-6.00\times 10^{-5}$ & $7.93\times 10^{43}$ & $8.90\times 10^{41}$\\
    $i\,(\Gamma=\Gamma_0)$ &$33.1$ & $9.60\times 10^{-5}$  & $2.28\times 10^{-3} $ &  $2.28\times 10^{-3}$  &  $9.18\times 10^{44}$ & $-2.21\times 10^{-3}$ & $1.95\times 10^{43}$ & $3.95\times 10^{42}$\\
    \hline\hline
\end{tabular}
\begin{tablenotes}
\footnotesize
\item[a] The values are evaluated at the time of merger, as the $f$-mode resonance does not occur during the inspiral phase in this model.
\end{tablenotes}
\label{tab:m1}
\end{threeparttable}
\end{table*}

In \cref{fig:Qnl}, we present $Q_{n\ell}$ as a function of the NS mass for several representative modes: the $g_1$-, $g_2$-modes, the $i$-mode in the nonstratified case, the $s_1$-, $s_2$-modes, and the $f$-mode. The results corresponding to the canonical model is shown in \cref{tab:m1}. 
Among these, the $f$-mode exhibits the strongest coupling, with $Q_{n\ell} \sim 0.2$ -- $0.5$~\cite{Shibata:1993qc,Reisenegger:1994a,Lai:1993di,Kokkotas:1995xe,Kuan:2021sin,Counsell:2024pua}. 
In contrast, the $g$-modes possess significantly weaker couplings (3 to 5 orders of magnitude smaller than that of the $f$-mode) that is smaller for higher overtones. 
The $s_1$- and $s_2$-modes yield overlap integrals of the order of $10^{-4}$. 
Unlike the $g$-modes, their values do not show a clear trend of decreasing with increasing radial order. The $s_1$-mode has larger tidal overlap than $s_2$-mode for small masses and their dominance switches when $M \approx 1.5\,M_{\odot}$. The $i$-mode in the nonstratified case has a tidal overlap comparable to that of the $g_1$-mode. 
Neglecting the metric perturbation terms in the normalization [\cref{eq:norm}] and in the tidal overlap integral [\cref{eq:tidal_overlap}] affects the computed value of $Q_{n\ell}$ by no more than $10\%$ across most NS masses. While our simplified treatment of $H_{\rm tid}$ may introduce some quantitative deviations, it is not expected to qualitatively alter our results.

Given the mode frequency, tidal overlap integral, and binary parameters, the evolution of the orbital phase and mode amplitude can be determined by numerically integrating the Hamiltonian equations of motion.

\subsection{Energy transfer and estimates of GWs}
\label{sec:energy_transfer}

To have a semi-quantitative estimation on the evolution of modes in the late inspiral phase, we treat the orbit as Newtonian and adopt the quadrupole formula of GW dissipation, which provides a reasonable approximation for our purposes. 
The corresponding Hamiltonian can be found in Ref.~\cite{Kokkotas:1995xe}. 
The orbital separation, $a$, of GW-driven quasi-circular inspirals decreases at the rate 
\begin{equation}
\frac{\dot{a}}{a}=-\frac{2}{3}\frac{\dot \Omega}{\Omega}=-\frac{64}{5} \frac{G^3}{c^5} \frac{M^{\prime} M\left(M+M^{\prime}\right)}{a^4}\,,
\end{equation}
where the dot means the derivative with respect to time $t$.
The orbital angular frequency of the binary,
\begin{equation}
    \Omega=\dot{\Phi}(t)=\left[\frac{G\left(M+M^{\prime}\right)}{a^3}\right]^{1 / 2}\,,
\end{equation}
increases with time.
A resonance between a given mode with $l=|m| =2$ and tidal force occurs when $\omega_\alpha \simeq 2 \Omega = 2\pi f_{\rm gw}$, where $f_{\rm gw}$ is the GW frequency. 
After evolving the Hamiltonian equations of motion, the energy transferred from the orbit to the modes can be directly computed from the oscillator Hamiltonian $H_{\rm osc}$, which includes both kinetic and potential energy contributions. 
From the Hamiltonian, one can readily derive the amplitude evolution equation for nonrotating NSs,
\begin{equation}
    \label{eq:q_evolution}
\ddot{q}^\alpha + \omega_{\alpha}^2 \, q^\alpha = \frac{G M'}{R^3} \, Q_{n\ell} \, \left( \frac{R}{a} \right)^{\ell+1} \,W_{\ell m} \, e^{-i m \Phi(t)}\,.
\end{equation}
This equation describes a forced harmonic oscillator, and is formally identical to the Newtonian result; see, e.g., Refs.~\cite{Lai:1993di,Press:1977ApJ,Yu:2016ltf}.

In \cref{fig:resonance_bns}, we show the evolution of the mode energy $E_{\rm mode}$ computed from \cref{eq:H_osc} during the inspiral for a binary NS system with equal mass $M'=M=1.35\,M_{\odot}$ in the unit of $E_{\rm b}=GM^2/R$. 
The modes are gradually excited and the amplitude grows as the separation of the binary decreases.  
Before reaching the resonance, i.e., when $\omega_{\alpha}^2\gg (m\Omega)^2$, the solution of $q^{\alpha}$ scales as 
\begin{equation}
|q^{\alpha}| \sim |Q_{n\ell}| {\Omega^2}/{\omega_{\alpha}^2}\sim  |Q_{n\ell}| {f_{\rm gw}^2}/{\omega_{\alpha}^2}\,,
\end{equation}
and consequently, the mode energy scales as 
\begin{equation}
E_{\rm mode}\sim \omega_{\alpha}^2 q^{\alpha}\bar{q}^{\alpha}\sim |Q_{n\ell}|^2{f_{\rm gw}^4}/{\omega_{\alpha}^2}\,. 
\end{equation}
The $f_{\rm gw}^4$ scaling of the $f$-mode energy is clearly visible. We also see how tidal resonances excite different oscillation modes of the NS at different values of $f_{\rm gw}$.
After the orbital frequency goes off resonance with a given oscillation mode, the mode's energy nearly remains constant with minor modulations.
This is because the tidal force continues to act upon the oscillation, although the net energy transfer is negligible. For the stratified NS, the $f$-mode always dominates over the $g$-modes, even though it does not enter into the resonance regime during the inspiral in this model. This result is consistent with the Newtonian study in Ref.~\cite{Passamonti:2020fur}. We also show the energy evolution of the $i$-mode for the nonstratified model. Since its frequency is approximately one order of magnitude smaller than the frequency of the $g_1$-mode, the $i$-mode has more time to be in resonance with binary's orbital frequency and hence to accumulate energy. Around the resonance point of the $i$-mode, its mode amplitude is larger than that of $f$-mode. The ``post-resonance'' energy of the $i$-mode lies between that of the $g_1$ and $g_2$ modes.

To give a better physical interpretation, we employ approximate methods to quantify the post-resonance amplitude and energy of the modes in the following.
The orbital decay timescale can be estimated as
\begin{align}
    t_D & \equiv \frac{a}{|\dot{a}|}=\frac{5 c^5}{64 G^3} \frac{a^4}{M M'\left(M+M'\right)} \nonumber\\
    & =9.2\,\mathrm{s} \ M_{1.35}^{-5 / 3}\left(\frac{1+q}{2 q^3}\right)^{1 / 3}\left(\frac{f_{\alpha}}{100 \mathrm{~Hz}}\right)^{-8 / 3}\,,
\end{align} 
where $q = M'/M$ is the mass ratio of the binary. Since this timescale is much longer than the resonance timescale, we can determine $q^{\alpha}$ with the stationary-phase approximation. In particular, the Green-function solution of \cref{eq:q_evolution} gives the following estimate for the peak amplitude of the excited modes:
\begin{equation}
\label{eq:q_alpha}
    |q^\alpha_{\rm max}| \simeq \frac{\pi}{32} \, \omega_\alpha^{-5/6} \, |Q_{n\ell}| \, M^{-5/6} \, q^{1/2} \left( \frac{2}{1 + q} \right)^{5/6}\,.
\end{equation} 
Note that both $m = \pm 2$ modes contribute equally to the tidal response.
In \cref{tab:m1}, we present the maximum amplitudes computed using Eq.~\eqref{eq:q_alpha}, alongside the exact results obtained from direct numerical evolution of Hamilton's equations. The discrepancy between the two approaches is less than 5\%.

\begin{figure}
    \centering
    \includegraphics[width=\linewidth]{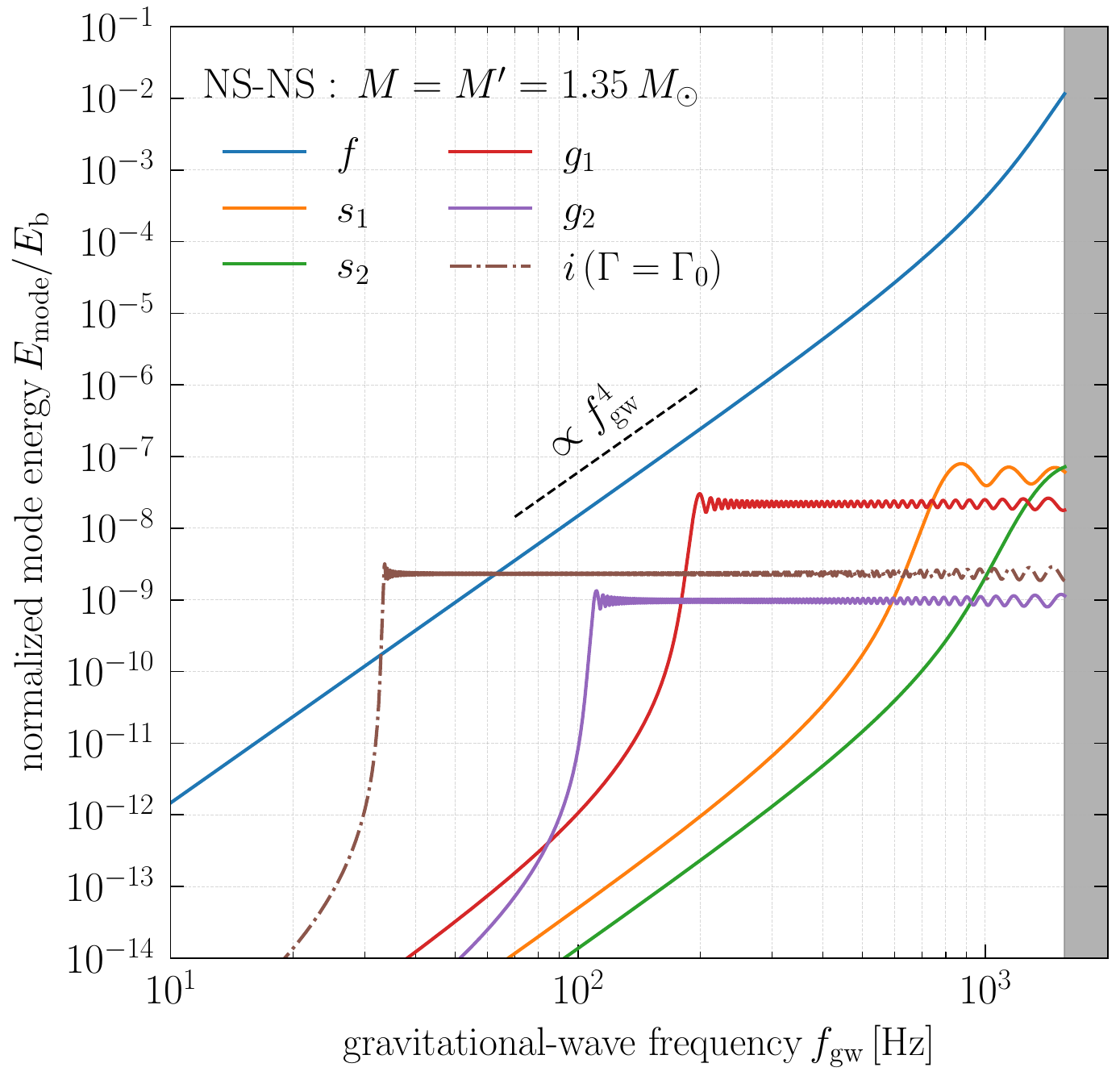}
    \caption{The evolution of the mode energy as a function of the GW frequency for a $1.35$+$1.35\,M_{\odot}$ equal-mass binary NS. The energy is normalized by $E_{\rm b}=GM^2/R$. 
    The hypothetical evolution of $i$-mode that exists when the buoyancy is artificially turned off is represented by the dashed curve.
    }
    \label{fig:resonance_bns}
\end{figure}

\begin{figure}
    \centering
    \includegraphics[width=\linewidth]{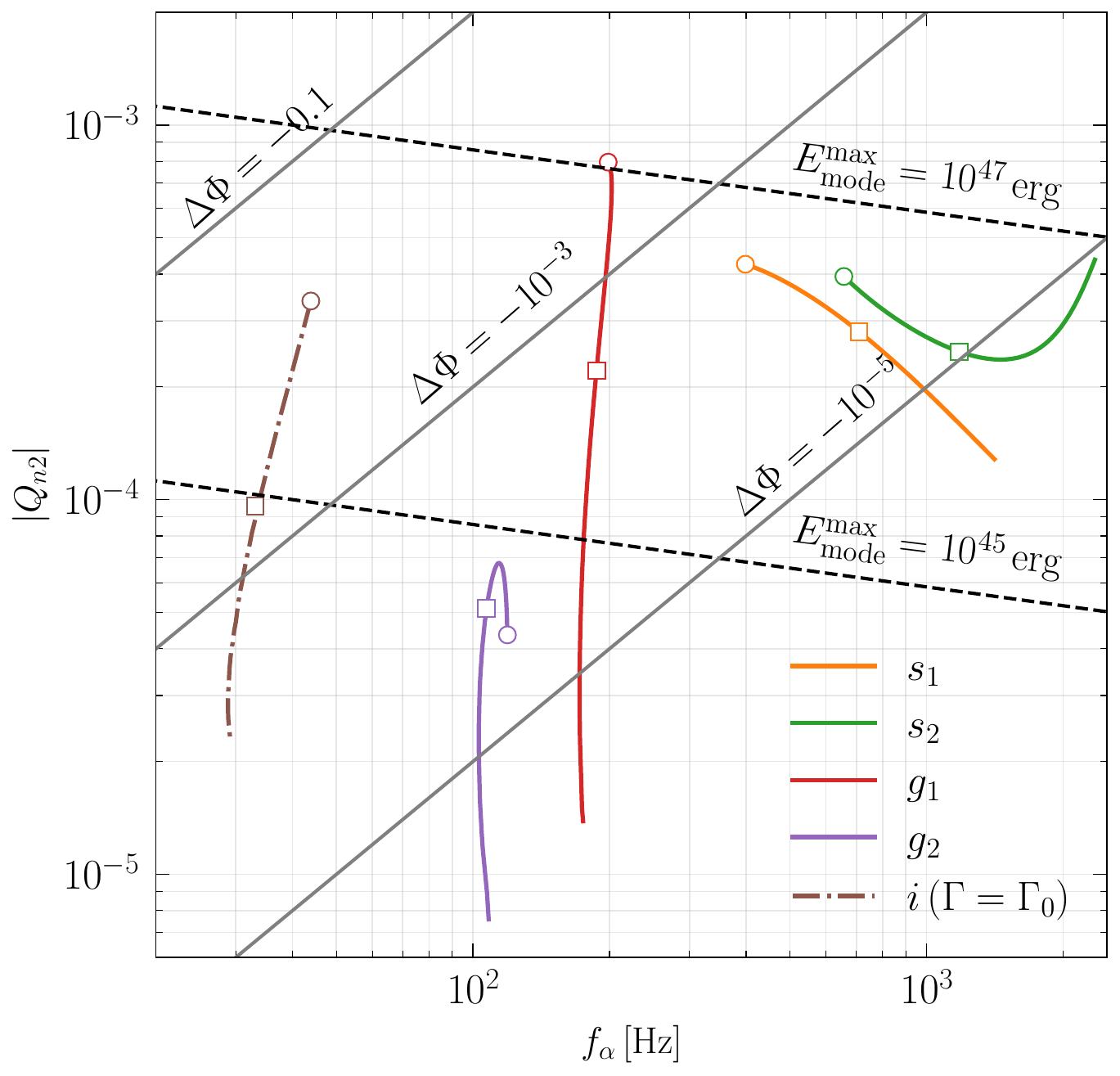}
    \caption{The relation between the mode frequency $f_{\alpha}$ and the tidal overlap integral $Q_{n2}$ for models with masses from $0.4\,M_{\odot}$ up to the maximum mass of $2.08\,M_{\odot}$. The circles and the stars denote the cases for $0.4\,M_{\odot}$ and $1.35\,M_{\odot}$, respectively. 
    Also plotted are some representative contour lines of $E_{\rm mode}^{\rm max}$ [\cref{eq:E_mode}] and $\Delta \Phi$ [\cref{eq:phase_shift}].}
    \label{fig:Qnl_Emode}
\end{figure}

The kinetic energy and the potential energy are equal after the resonance, and the total energy of the mode can be estimated as
\begin{align}
    \label{eq:E_mode}
    E_{\rm mode}^{\rm max} &\simeq \mathcal{A}_\alpha^2 \, {{\omega}_\alpha}^{2} \mid q^{\alpha}_{\rm max} \mid^2 \nonumber\\
    &= \frac{\pi^2}{1024} \left(\frac{{\omega}_{\alpha}}{\Omega_{0}}\right)^{\frac{1}{3}}\left|Q_{n l}\right|^2 \left(\frac{Rc^2}{GM}\right)^{\frac{5}{2}} q\left(\frac{2}{1+q}\right)^{\frac{5}{3}} E_{\rm b}\nonumber \\
    & = 6.8 \times 10^{45}\times  q\left(\frac{2}{1+q}\right)^{5/3}  \nonumber \\  
    &\quad \times R_{12}^2 M_{1.35}^{-2/3} \left(\frac{f_{\alpha}}{200\,\rm Hz}\right)^{\frac{1}{3}} \left(\frac{\left|Q_{n l}\right| }{0.0002}\right)^2\,{\rm erg} \,,
\end{align}
where $\Omega_0 = (GM/R^3)^{1/2}$, $R_{12}=R/12\,\rm km$, $M_{1.35}=M/1.35\,M_{\odot}$, and $E_{\rm b}=GM^2/R$.
The dependence on binary's component masses in the expression for the mode's maximal energy enters primarily through the factor $q\left(2/(1 + q)\right)^{-5/3}$, which is relatively weak compared to the dependence on the frequency and the tidal overlap integral. We will discuss this below in this section. 

\begin{figure}
    \centering
    \includegraphics[width=\linewidth]{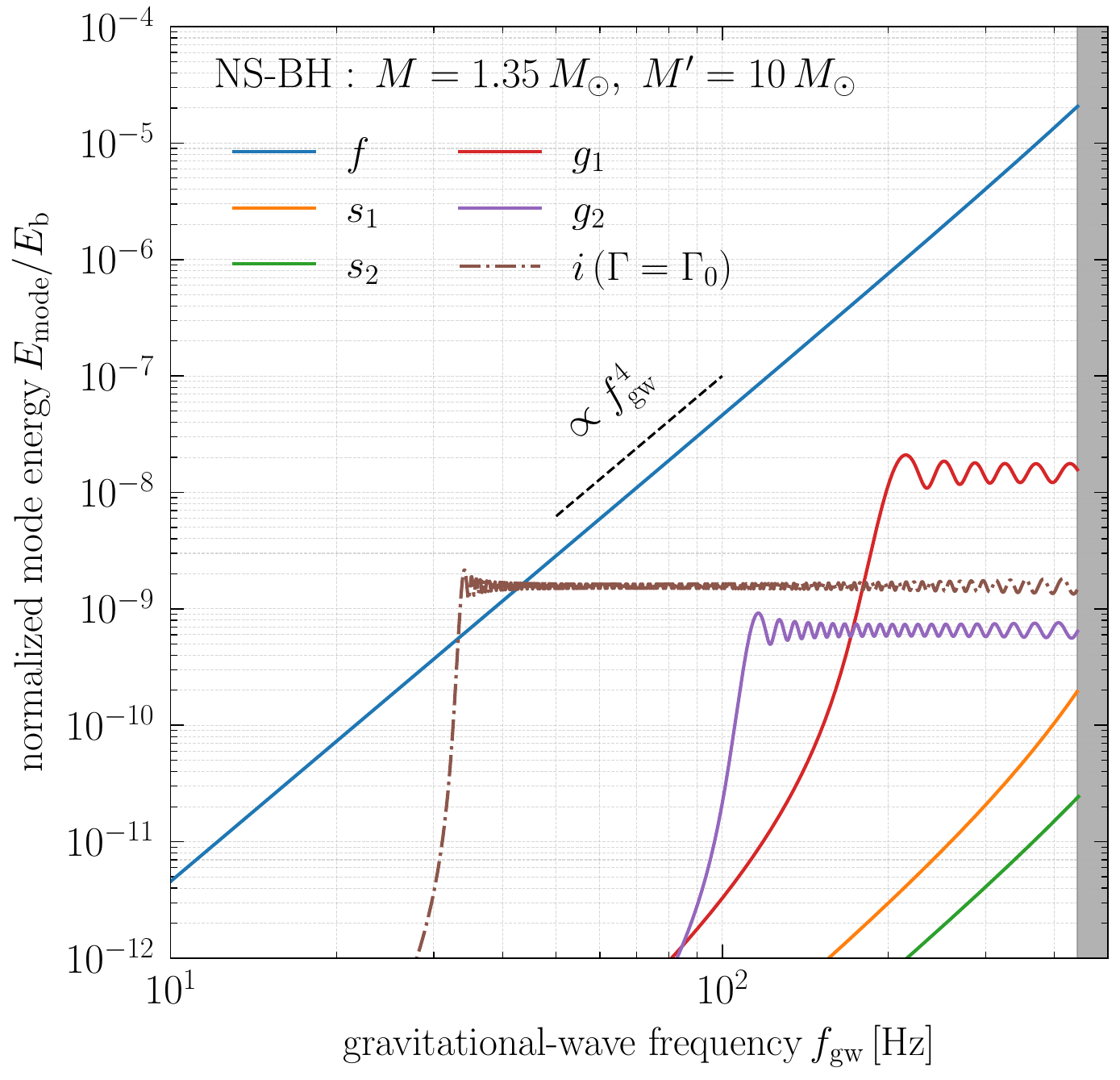}
    \caption{Same as \cref{fig:resonance_bns}, but for a NS-BH binary with the NS mass $M=1.35\,M_{\odot}$ and the BH mass $M'=10\,M_{\odot}$. Relative to the binary NS case in \cref{fig:resonance_bns}, the mode energy is noticeably lower, owing to the smaller merger frequency and the additional suppression from the mass-ratio factor.}
    \label{fig:resonance_bhns}
\end{figure}

Because a fraction of the orbital energy is absorbed by stellar oscillations, the binary’s orbital phase will be shifted. The characteristic energy scale of the binary is
\begin{align}
    E_{\mathrm{orb}} & =-\frac{G M M'}{2 a} \nonumber \\
    & =-1.6 \times 10^{52} \, M_{1.35}^{5 / 3}\, q\left(\frac{2}{1+q}\right)^{1 / 3}\left(\frac{f_{\rm gw}}{100 \mathrm{~Hz}}\right)^{2 / 3}\,\rm{erg}\,.
\end{align}
An estimate of the shift in the orbital phase $\Delta \Phi$ due to the energy transfer $E_{\rm mode}^{\rm max}$ can be estimated as~\cite{Lai:1993di,Ho:2023shr}
\begin{align}
    \label{eq:phase_shift}
    \frac{\Delta \Phi}{2 \pi} & \approx-\frac{t_D}{t_{\mathrm{orb}}} \frac{ E_{\rm mode}^{\rm max}}{\left|E_{\mathrm{orb}}\right|} \nonumber \\
    & =-145 M_{1.35}^{-5 / 3}\left(\frac{1+q}{2 q^3}\right)^{1 / 3}\left(\frac{f}{200 \mathrm{~Hz}}\right)^{-5 / 3} \frac{E_{\rm mode}^{\rm max}}{\left|E_{\mathrm{orb}}\right|}\nonumber \\
    & =-\frac{5 \pi}{4096}\left(\frac{c^2 R}{G M}\right)^5 \frac{2}{q(1+q)} \left(\frac{{\omega}_\alpha}{{\Omega_0}}\right)^{-2} Q_{nl}^2\nonumber \\
    & =-4\times 10^{-5} M_{1.35}^{-4} R_{12}^2 \frac{2}{q(1+q)} \left(\frac{f_{\alpha}}{200 \mathrm{~Hz}}\right)^{-2}\left(\frac{Q_{nl}}{0.0002}\right)^{2}\,,
\end{align}
where $t_{\rm orb}=2\pi/\Omega$ is the orbital period. In \cref{fig:Qnl_Emode}, we plot the relation between mode frequency and the tidal overlap integral for a range of NS masses from $0.4\,M_{\odot}$ up to the maximum mass of $2.08\,M_{\odot}$. As visual representations of the analytic expressions given in \cref{eq:E_mode} and \cref{eq:phase_shift}, we overlay contour lines of the mode energy $E_{\rm mode}^{\rm max}$ and the GW phase shift $\Delta\Phi$ for an NS with $M = 1.35\,M_{\odot}$ and $R = 12\,\rm km$ in an equal-mass binary. Note that the $s$-modes do not reach resonance in all models, and the contour lines represent an upper limit. We also find that the mode energy remains below $\sim10^{47}\,\rm erg$ for the $s$-modes, $g$-modes, and $i$-modes in NSs with masses $M\gtrsim 1.1\,M_{\odot}$. Correspondingly, the GW phase shift induced by resonant excitation of these modes stays below $10^{-3}\,\rm rad$.
This extent of dephasing is quite challenging to be observable even with the next generation ground-based GW observatories~\cite{Read:2023hkv}.

So far we have focused on equal-mass binaries. Here, we briefly examine how the maximal mode energy depends on the mass ratio.
The mass ratio $q$ enters into $E_{\rm mode}^{\rm max}$ in \cref{eq:E_mode} through the factor $q[2/(1 + q)]^{-5/3}$, which equals to unity for equal-mass binaries, reaches a maximum at $q = 1.5$, and decreases to zero for large values of $q$.
For asymmetric NS–NS binaries, the typical mass ratio is smaller than 1.5, and thus the mode energy deposited through a resonance is comparable to that in equal-mass binaries. 
However, for NS–BH binaries, the mass ratio is typically much greater than $1.5$~\cite{LIGOScientific:2021qlt}.
We therefore expect a significant suppression of the amount of energy transferred into mode energy after resonance. We confirm this expectation in \cref{fig:resonance_bhns}, for a representative NS–BH system with masses $M = 1.35\,M_{\odot}$ for the NS and $M' = 10\,M_{\odot}$ for the BH.
Comparing to the equal-mass NS–NS scenario, the maximum mode energy of the resonantly excited mode is reduced by approximately $21\%$.
Another key effect is that NS–BH binaries merge at lower merger frequency compared to NS–NS systems.
As shown in \cref{fig:resonance_bhns}, the $s$-modes can no longer reach resonance, and the energy of the nonresonantly excited $f$-mode is suppressed by three orders of magnitude relative to the NS–NS case shown in \cref{fig:resonance_bns}.

\begin{figure*}
    \includegraphics[width=0.9\linewidth]{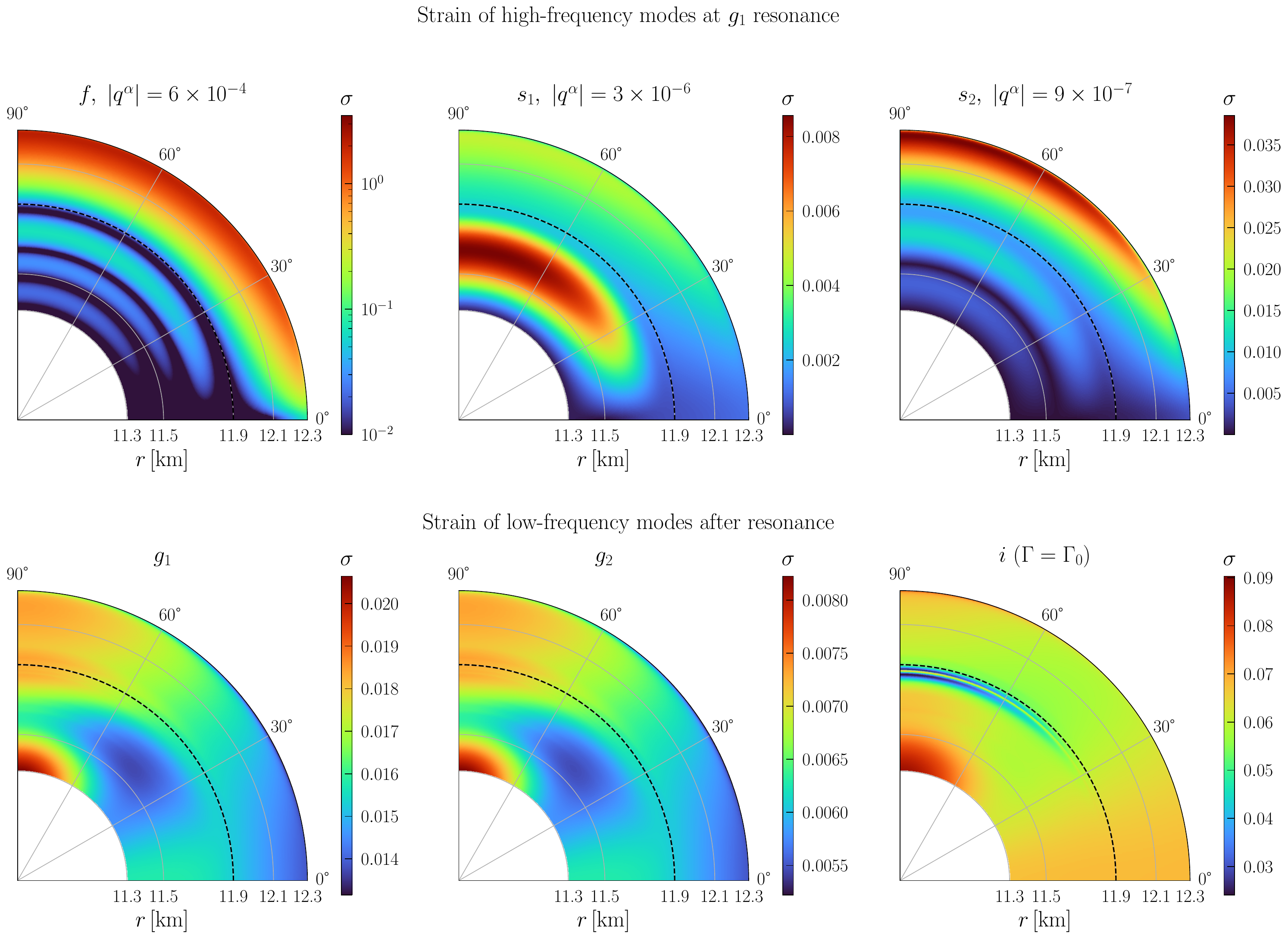}
    \caption{The distribution of the strain $\sigma$ for different modes based on the canonical model. The upper panel presents the results for high-frequency modes—$f$, $s_1$, and $s_2$ at the time when $g_1$-mode reaches resonance.  The lower panel displays the results for low-frequency modes, including the $g_1$ and $g_2$ modes, as well as the $i$-mode in the $\Gamma = \Gamma_0$ case. The black dashed lines indicate the location of the neutron drip. The companion is located at the direction of $\theta=90^{\circ}$. \reply{An animation of the strain evolution as a function of the inspiral phase can be found at \cite{strain}.}}
    \label{fig:von_Mises}
\end{figure*}

\subsection{Crust breaking and the von Mises criterion}
\label{sec:crust_breaking}

The resonance-induced oscillations generate stresses in the crust, which may cause it to break. For a relaxed background star, the local maximum strain is given by
\begin{equation}
\label{eq:mises}
\sigma = \sqrt{2} \, |q^{\alpha}| \, \sqrt{\tfrac{3}{2} \, \Delta s_{ab} \, \Delta \bar{s}^{ab}} \,.
\end{equation}
Remember that $q^{\alpha}$ is the amplitude of an oscillation mode and $\Delta s_{ab}$ is the Lagrangian perturbation of the strain tensor; cf.~Eq.~\eqref{eq:strain}.
The prefactor $\sqrt{2}$ accounts for contributions from both $m = 2$ and $m = -2$ components. For perturbations with $\ell = |m| = 2$, this expression takes the explicit form
\begin{align}
    \label{eq:von_Mises}
    \sigma^2 =&\frac{45 |q^{\alpha}|^2}{256\, \pi} \Big[ 
    \; 3 \sin^4\theta\, \left(\frac{|T_{1}|^2}{\mu^2}\right) + 8\, e^{-\lambda} \sin^2\theta \left( 3 + \cos 2\theta \right) \left(\frac{|T_{2}|^2}{\mu^2}\right) \nonumber \\
    & + 16 \left( 8 - 8\sin^2\theta + \sin^4\theta \right) |V|^2 
        \Big]\,.
\end{align}
Remember again that $T_1$ and $T_2$ are the radial and tangential tractions, defined in Eq.~\eqref{eq:tractions}; $V$ arises in the angular components of the displacement vector, defined in Eq.~\eqref{eq:displacement_vector}; $\lambda$ is a background metric function; and $\mu$ the shear modulus.
We note that our expression for the strain differs from that in Ref.\cite{Passamonti:2020fur} (Eq.(58)), although the source of this discrepancy is not yet understood. For reference, we have made a Mathematica notebook with our derivation available online~\cite{von_Mises}.

In \cref{fig:von_Mises}, we illustrate the strain amplitude for different oscillation modes of the canonical model. Since the $f$-mode does not undergo resonance, and the $s_1$- and $s_2$-modes reach the resonance point very close to merger, we present the results for these high-frequency modes at the amplitude $|q^{\alpha}|$ corresponding to the orbital phase at which the $g_1$-mode is in resonance. In contrast, the results for the low-frequency modes ($g_1$-, $g_2$-, and $i$-modes) are shown at their respective resonance points, with amplitudes evaluated at resonance. Except the $i$-mode, a general feature is that the strain is largest along the direction to the companion ($\theta=90^{\circ}$) compared to other $\theta$ values. Owing to the linear nature of the problem, the strain amplitude at any other orbital phase can be obtained by rescaling with the value of $|q^{\alpha}|$ provided in \cref{fig:resonance_amplitude}. \reply{The time before merger is related to the frequency of GW as  
\begin{equation}
    t-t_{\rm merger} \simeq -2.29 \mathrm{~s}\left(\frac{1.175\,M_{\odot}}{M_c}\right)^{5 / 3}\left(\frac{100 \,\mathrm{Hz}}{f_{\mathrm{gw}}}\right)^{8 / 3}\,.
\end{equation}
Here $t_{\rm merger}$ denotes the merger time, and $M_c$ is the chirp mass of the binary, which equals to $1.175\,M_{\odot}$ for an equal-mass binary NS with $M' = M = 1.35\,M_{\odot}$.}

\begin{figure}
    \includegraphics[width=\columnwidth]{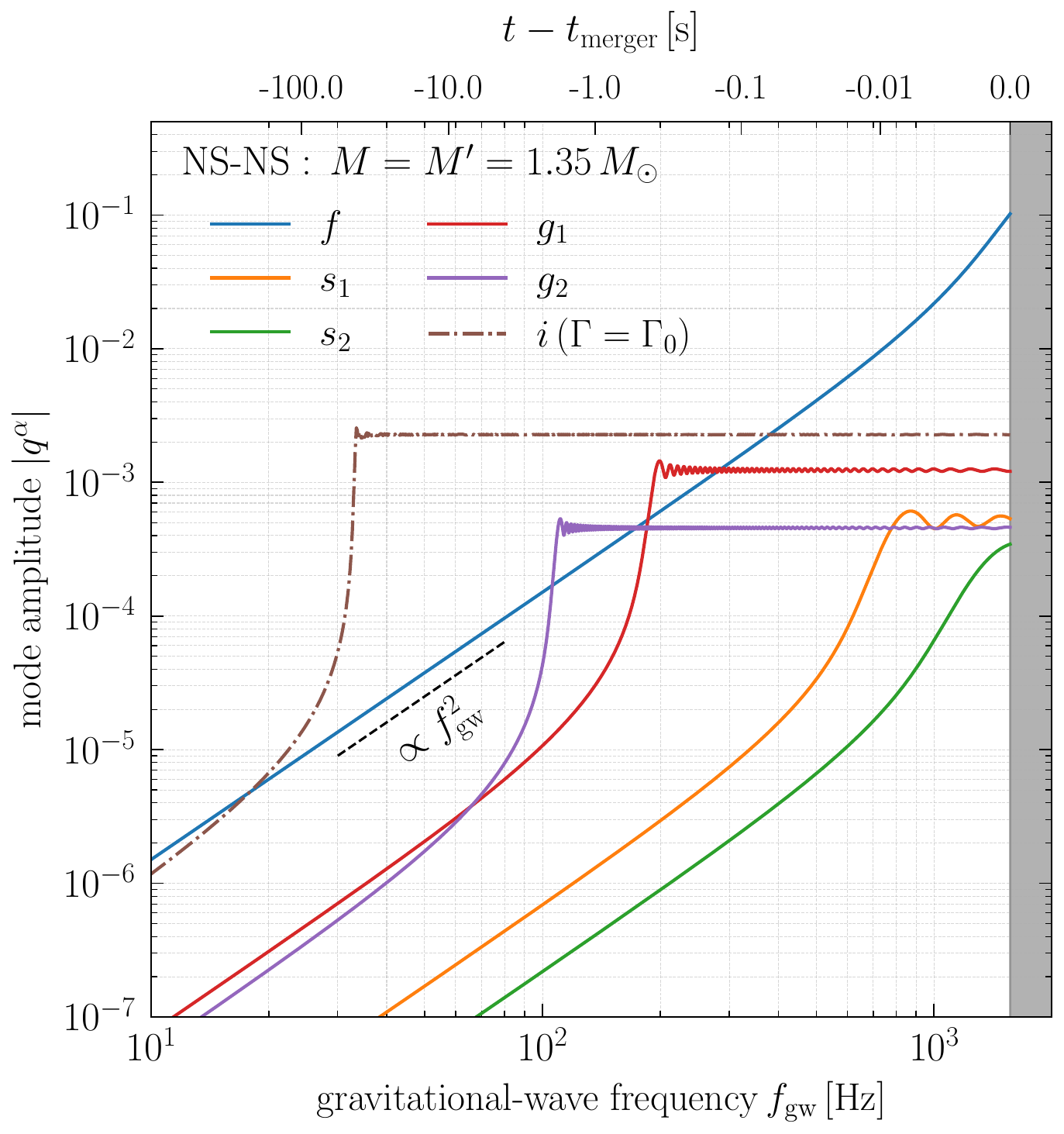}
    \caption{The evolution of the mode amplitude as a function of the GW frequency for a $1.35$+$1.35\,M_{\odot}$ equal-mass binary NS. \reply{The time before merger, $t - t_{\rm merger}$, is indicated on the upper x-axis.}
    {This is another measure of mode excitation, complementary to the mode energy shown in \cref{fig:resonance_bns}.}
    }
    \label{fig:resonance_amplitude}
\end{figure}

We also observe in \cref{fig:von_Mises} that the $f$-mode's strain amplitude exhibits several peaks and valleys in the inner crust, and increases nearly monotonically in the outer crust toward the star's surface. This behavior is driven by the characteristics of the horizontal displacement, which oscillates in the inner crust and grows steadily toward the surface in the outer crust, as shown in \cref{fig:f_mode}.

The crust is broken when the strain amplitude exceeds a critical threshold, i.e., $\sigma > \sigma_{\rm break}$, which is the so-called von Mises criterion. The breaking strain is commonly taken to be \(\sigma_{\rm break} = 0.1\), based on molecular dynamics simulations of the NS crust~\cite{Horowitz:2009ya}. 
As shown in the contour plot, the outer crust has already been broken at the position where $g_1$-mode reaches resonance. If we consider a more conservative estimate of \(\sigma_{\rm break} = 0.04\), as suggested by Ref.~\cite{Baiko:2018jax}, a part of the inner crust will also be broken.
To our knowledge, we have thus demonstrated for the first time that the nonresonant excitation of the $f$-mode can break the crust at $f_{\rm gw}\sim 100\text{--}200\,\rm Hz$.
A Newtonian study in \cite{Passamonti:2020fur} also indicated that the $f$-mode can break the crust, but it only occurs very close to the merger. 
This inconsistency may be caused by the simplified EOS and the different strain expression [Eq.~(89) in \cite{Passamonti:2020fur}] adopted therein.

To obtain a clearer understanding of crustal breaking induced by the nonresonant $f$-mode, we analyze the contributions of the three terms appearing in \cref{eq:von_Mises}. We find that the term involving the horizontal traction, $T_2$, is almost always dominant—typically 1–2 orders of magnitude larger than the other two terms for the $f$-mode. Consequently, the radial gradient of the horizontal displacement is the primary driver of crust breaking, and the leading-order contribution to the von Mises strain is given by
\begin{align}
    \label{eq:von_Mises_estimate}
    \sigma & \sim \sqrt{\frac{45}{32\pi}}\,|q^{\alpha}|e^{-\lambda/2}\sin\theta(3+\cos2\theta)^\frac{1}{2}\frac{|T_2|}{\mu}\nonumber \,,\\
    & \sim  \sqrt{\frac{45}{32\pi}}\,|q^{\alpha}|e^{-\lambda/2}\sin\theta(3+\cos2\theta)^\frac{1}{2}r|V'|\nonumber \,,\\
    & \sim 0.2\, \left(\frac{|q^{\alpha}|}{6\times 10^{-4}}\right)\left(\frac{r|V'|}{5\times 10^3}\right)\sin\theta(3+\cos2\theta)^\frac{1}{2}e^{-\lambda/2}\,.
\end{align}
Here, the amplitude of the $f$-mode is set to its value at the $g_1$-mode resonance. In \cref{fig:rdv}, we plot the radial profile of $r V'$ within the crust for a $1.35\,M_{\odot}$ NS. Combined with the estimate in \cref{eq:von_Mises_estimate}, this analysis indicates that the outer crust is already broken in the vicinity of the $g_1$ resonance. Meanwhile, the peak-valley structure in the inner crust region observed in \cref{fig:von_Mises} arises from the radial variation of $V'$, as illustrated in \cref{fig:rdv}. It is worth noting that static tides are generally ineffective at fracturing the crust, as demonstrated in~\cite{Penner:2011br,Gittins:2020mll}. This is primarily because, under static tidal forces, the solid crust deforms much like a fluid. As a result, the star can experience significant global deformation while the internal shear strain remains small, making static tides inefficient at breaking the crust.

\begin{figure}
    \includegraphics[width=\columnwidth]{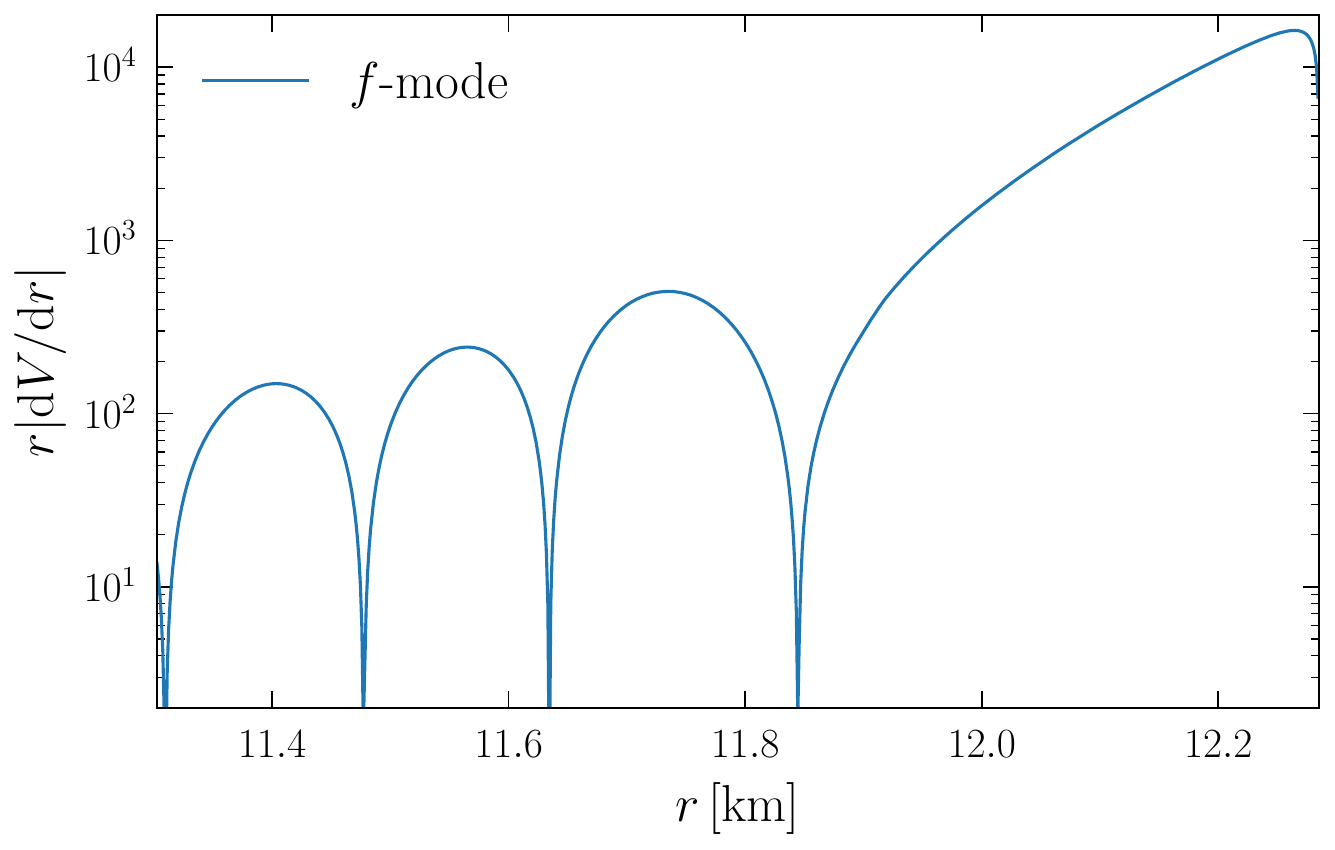}
    \caption{{Main contribution to the crustal strain ($r\,\dd V/\dd r$; see main text) as a function of $r$ for the $f$-mode of a $1.35\,M_{\odot}$ NS.}}
    \label{fig:rdv}
\end{figure}

For the $s_1$-mode, the strain amplitude peaks in the inner crust; however, the maximum amplitude at the $g_1$ resonance is insufficient to break the crust. Based on the mode amplitude evolution shown in \cref{fig:resonance_amplitude}, the inner crust is expected to break at GW frequencies of approximately $f_{\rm gw} \sim 400\,\mathrm{Hz}$ and $\sim 600\,\mathrm{Hz}$ for breaking strains of $\sigma_{\rm break} = 0.04$~\cite{Baiko:2018jax} and $\sigma_{\rm break} = 0.1$~\cite{Horowitz:2009ya}, respectively. In contrast to the $s_1$-mode, the $s_2$-mode exhibits its maximum strain in the outer crust, with an overall pattern resembling that of the $f$-mode. As shown in the contour plot, at the orbital phase corresponding to the $g_1$-mode resonance, only a small region of the outer crust approaches the breaking strain $\sigma_{\rm break} = 0.04$. The GW frequency at which the $s_2$-mode breaks the inner crust is approximately similar to that of the $s_1$-mode.

For the $g_1$- and $g_2$-modes, the strain distribution is similar, as their eigenfunctions share a comparable structure throughout the star. The strain is clearly larger along the direction of the companion relative to other directions, with the largest values occurring near the crust-core boundary. This is consistent with the fact that the buoyancy force peaks at these locations. The maximum strain of the $g_1$-mode reaches approximately $0.025$, which is close to the typical breaking strain. However, we cannot conclude that $g$-modes are incapable of breaking the crust, as their frequencies and tidal overlap integrals are highly sensitive to the nuclear physics near the crust-core interface. A more detailed investigation into the possibility of crust breaking by $g$-modes across different EOSs will be presented in a forthcoming paper.

Finally, for the $i$-mode in the nonstratified case, the strain is largest at the base of the crust and concentrated near the equator. 
Although the mode energy at resonance is smaller than that of the $g_1$-mode, the $i$-mode is more efficient at inducing strain in the crust. 
It can completely break the crust if we adopt a breaking strain of $\sigma_{\rm break} = 0.04$, with the maximum strain approaching $\sigma_{\rm break} = 0.1$. 
However, as previously discussed, this mode does not exist for stratified NSs.
The characteristic motion of $i$-mode is encoded in the high overtones of $g$-modes, while these modes are only feebly susceptible to tidal pushing force and can be barely excited to leave observables.

\subsection{Implication in EM precursors}

Although the energy deposited into oscillation modes generally has a negligible effect on the GW phase evolution except for the $f$-mode, a given mode could produce observable precursor flares before the merger if it manages to break the crust and the energy thus liberated can be efficiently converted into EM radiation.
Once a mode grows to the point where the induced stress exceeds the crust’s elastic limit, the crust may break. 
This shattering process redistributes the mode energy into high-frequency oscillations, which can couple more effectively to the star’s magnetic field and generate radiation. This mechanism was first proposed by \citet{Tsang:2011ad} and has since been explored in various studies~\cite{Tsang:2011ad,Kuan:2021sin,Kuan:2023kif,Suvorov:2022ldw,Suvorov:2024cff,Most:2024eig}. 
In this context, even modes that are energetically subdominant in GW emission may play an important role in generating EM counterparts through crustal breaking and magnetic coupling.

A complete description of the nonlinear dynamics following crust breaking and the subsequent radiative processes in the magnetosphere is highly complex. Since our study focuses on a single EOS and is primarily concerned with establishing the formalism, we restrict ourselves to presenting the energy budget available $E_{\rm crust}$ (the mode energy confined within the crust) and the elastic potential energy $E_{\rm ela}$. These contributions can be expressed as
\begin{align}
    E_{\rm crust} &= q^{\alpha} \bar{q}^{\alpha} \omega^{2} \int_{\rm crust} \bigg[
    e^{-\nu}(\epsilon + p)\, (\xi_{\alpha})_{i} (\bar{\xi}_{\alpha})^{i}\nonumber \\
    &\quad - \frac{\ell(\ell + 1)\, e^{-\lambda - \nu}\, r^{2\ell}}{16\pi} H_1 \bar{H}_1 \nonumber \\
    &\quad - \frac{e^{-\nu}\, r^{2\ell}}{16\pi} \left(K \bar{K} + K \bar{H}_2 + H_2 \bar{K} \right)
    \bigg] \sqrt{-g}\, \dd^3x\,, \\
    E_{\rm ela} &= q^{\alpha} \bar{q}^{\alpha} \int_{\rm crust} \mu\, \Delta s_{ab} \Delta \bar{s}^{\,ab} \sqrt{-g}\, \dd^3x\,,
\end{align}
where the integrals are evaluated over the crust region only.

In \cref{fig:low_energy}, we present $E_{\rm crust}$ and $E_{\rm ela}$ for low-frequency modes at {their respective} resonance {as functions of} NS masses. 
Both quantities decrease with increasing stellar mass, primarily due to the reduction in the tidal overlap integral. 
For the $g_1$- and $g_2$-modes, the elastic potential energy constitutes only about $1\%$ of the mode energy confined in the crust, whereas for the $i$-mode (in the nonstratified case) this fraction increases to approximately $10\%$. 
For high-frequency modes, not all models reach resonance during inspiral; therefore, we only show results for the canonical model as a function of orbital evolution in \cref{fig:high_energy}. 
Among these, the mode energy confined in the crust for the $f$-mode remains dominant, with the elastic contribution again being only $\sim 1\%$ of $E_{\rm crust}$. 
In contrast, for the $s_1$- and $s_2$-modes, $E_{\rm crust}$ is several orders of magnitude smaller than that of the $f$-mode, and yet $E_{\rm ela}$ can still reach about $10\%$ of $E_{\rm crust}$, similar to the $i$-mode. 
This feature reflects the stronger shear character of the $s$- and $i$-modes compared to the globally coherent $f$-mode and composition-driven $g$-modes. 
The enhanced shear motion can be directly seen in the eigenfunctions or inferred from the von Mises stress criterion shown in \cref{fig:von_Mises}.

\begin{figure}
    \centering
    \includegraphics[width=\linewidth]{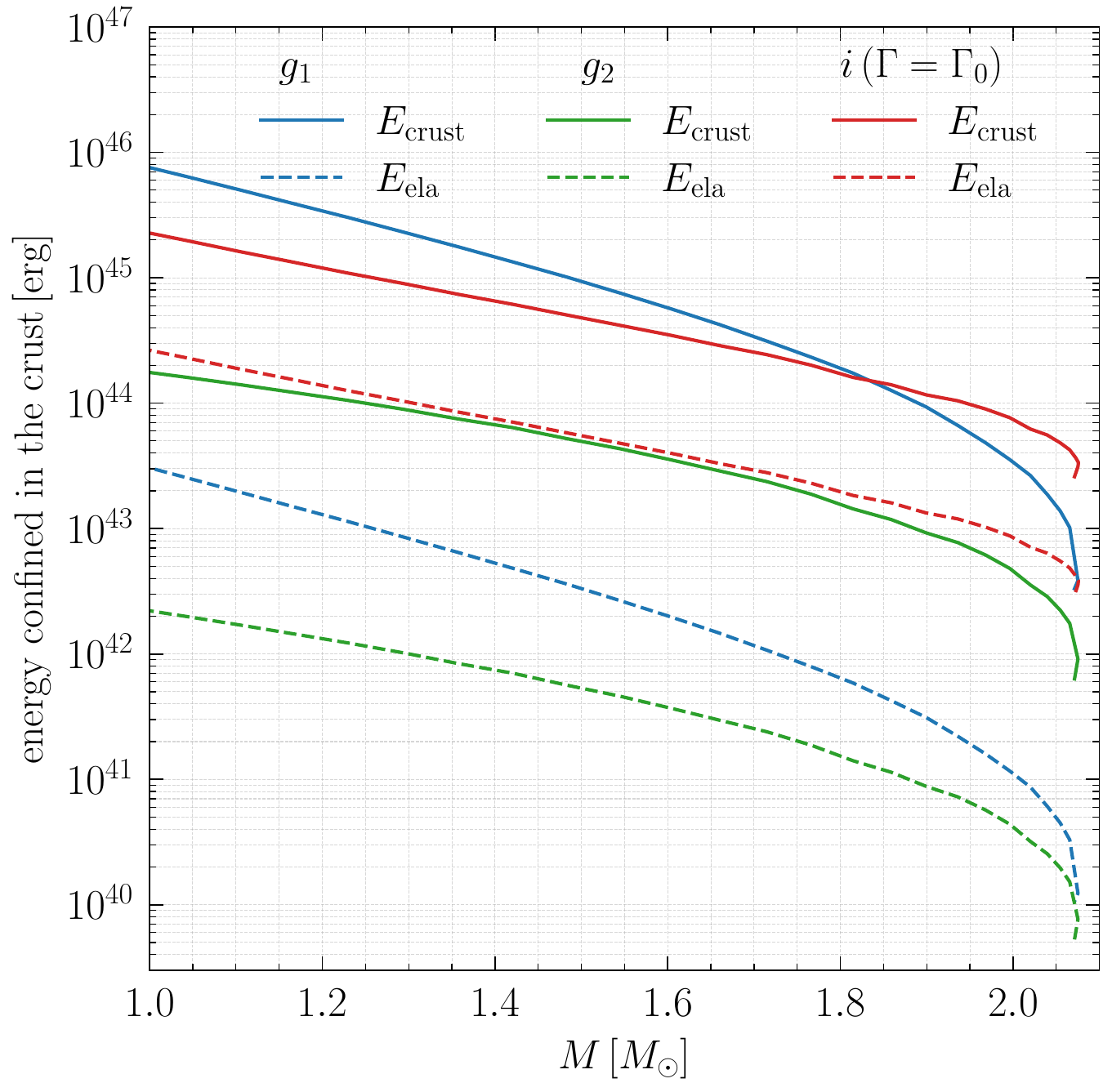}
    \caption{Mode energy in the crust ($E_{\rm crust}$) and elastic energy induced by oscillations ($E_{\rm ela}$) for low-frequency modes at resonance, shown for different stellar masses in equal-mass binaries.
    }
    \label{fig:low_energy}
\end{figure}
\begin{figure}
    \centering
    \includegraphics[width=\linewidth]{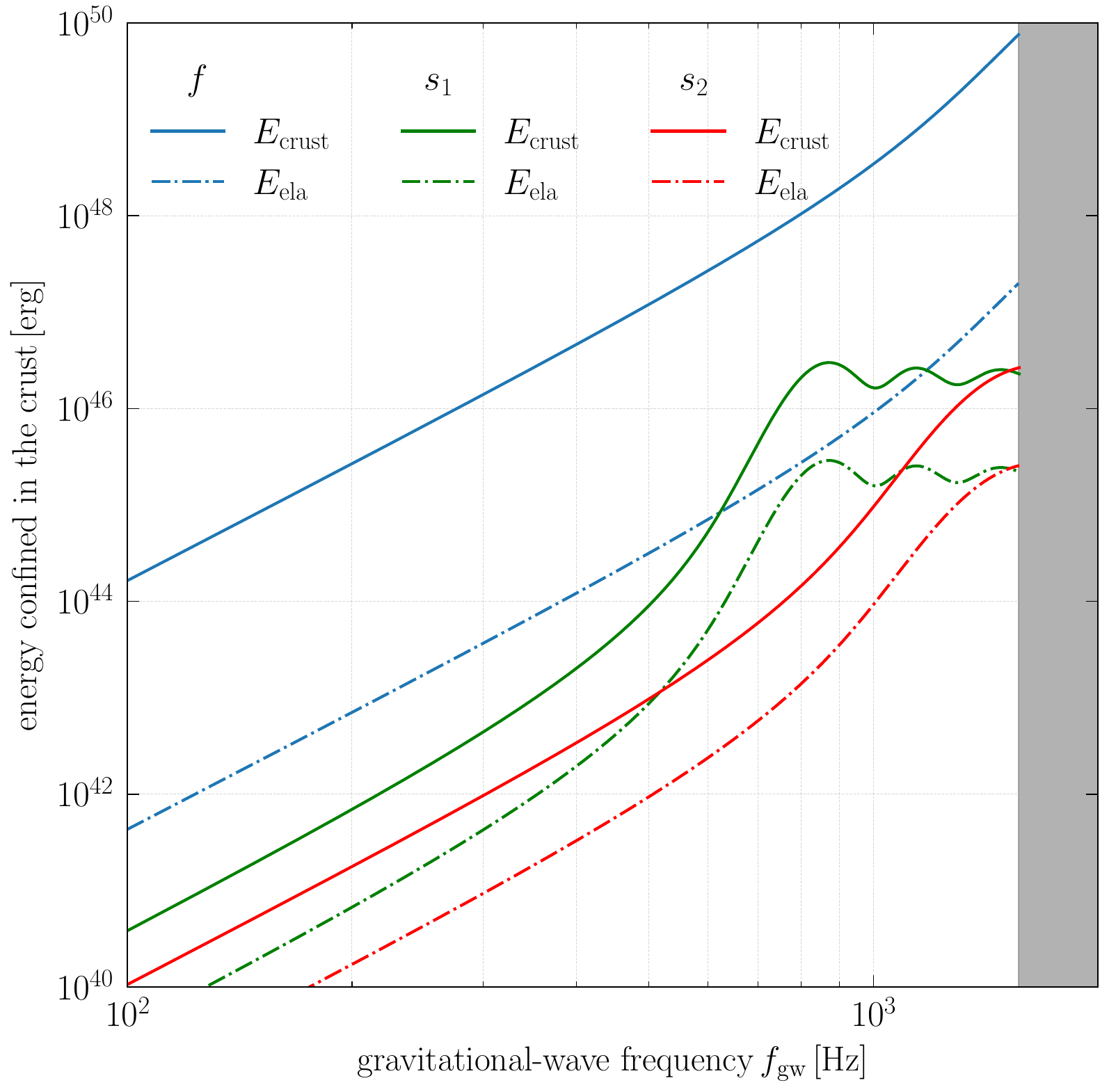}
    \caption{The mode energy confined within the crust ($E_{\rm crust}$) and the elastic energy induced by the oscillations ($E_{\rm ela}$) for high-frequency modes during the orbital evolution of a $1.35\,M_{\odot}$ equal-mass binary.
    }
    \label{fig:high_energy}
\end{figure}

Once the crust breaks, the energy stored in various oscillation modes can potentially power EM precursor emissions. 
Based on our canonical model, we propose a plausible scenario outlining this process. As the binary orbit decays, low-frequency $g$-modes are resonantly excited first, while high-frequency modes such as the $f$- and $s$-modes are nonresonantly driven, with the $f$-mode contributing to the dominant share of energy.
At the resonance of the $g_2$-mode, the crust is not significantly overstrained by any of the excited modes. The oscillation energy remains confined within the star, with $E_{\rm crust} \sim 10^{44}\,\mathrm{erg}$. We do not consider the $i$-mode here, as it is expected to mix with $g$-modes in a stratified NS, which is a more realistic scenario during the late inspiral phase~\cite{Andersson:2019ahb}.

As the orbit further shrinks, the $g_1$-mode reaches resonance. By this stage, the outer crust has already broken because of the accumulated stresses from the nonresonantly excited $f$-mode and $s_2$-mode, as indicated in \cref{fig:von_Mises}. This could initiate the onset of EM precursor emission. Shortly afterward, the system evolves to a GW frequency of $f_{\rm gw} \sim 300\,\mathrm{Hz}$ within a timescale of $\sim 0.1\,\mathrm{s}$. At this point, the crust is likely fully broken, according to Figs.~\ref{fig:resonance_amplitude} and~\ref{fig:von_Mises}. The total mode energy confined in the crust, dominated by contributions from the $f$- and $g_1$-modes, reaches $\sim 10^{46}\,\mathrm{erg}$ and becomes available to power EM emissions. If the energy release occurs on a timescale of $0.01$–$0.1\,\mathrm{s}$, the corresponding luminosity could reach up to $10^{45}$–$10^{46}\,\mathrm{erg/s}$ assuming the 1\% energy conversion efficiency~\cite{Most:2024eig}.
It may be worth mentioning that the typical luminosity of observed precursors is $10^{45\text{--}46}\,\mathrm{erg/s}$, compatible with our proposed scenario.

\section{Summary and discussions}
\label{sec:summary}

In this paper, we investigated nonradial oscillations of NSs with an elastic crust self-consistently modeled in the EOS, and a special attention has been paid of their tidal resonances during the late stage of binary inspiral. 
The QNMs of a spherically symmetric background were computed using a fully relativistic perturbation framework that incorporates both the elasticity of the solid crust and composition stratification.
We modeled the dynamical tidal response of the various modes as driven harmonic oscillators in an external tidal field, and examined the associated energy transfer, resulting GWs dephasing, and potential crust breaking.

In the low-frequency band, the polar-parity oscillations include the $i$-mode associated with the solid–liquid transition at the crust–core boundary, and composition $g$-modes. 
For the specific EOS considered here, we find that the $i$-mode exists only in nonstratified NSs with frequencies of the order of several tens of Hz, regardless of the NS masses.
In reality, composition stratification is expected to be present, which smears out the $i$-mode and withdraws the clear distinction between $i$- and the compositional $g$-modes. 
Instead, a number of modes with mixed gravity–interfacial character are revealed:
the eigenfunctions of the leading-order $g$-modes are predominantly shaped by buoyancy from the composition gradient, while the feature of $i$-mode -- a cusp in the radial motion and a discontinuous tangential motion at the crust-core boundary -- gradually manifests in overtones (\cref{sec:g_mode}).

To better understand how the transition in the motion characteristics relates to the competition between the buoyancy and shear stress, we adopted a simple phenomenological stratification model, $\Gamma_1 = \Gamma_0 (1 + \delta)$, whereby the buoyancy is smoothly controlled by varying the parameter $\delta$. (The buoyancy is completely switched off when $\delta=0$.)
As $\delta \to 0$, the global $g_1$-mode gradually transitions into an $i$-mode, while higher-order $g$-modes become more and more confined into the core (\cref{fig:ig_eigen}).
In the high-frequency regime, we analyzed the $f$-mode and $s$-modes. 
Both exhibit significant amplitude in the crust, with the latter being almost entirely confined within the crust.

The crust not only supports new families of oscillations, such as shear $s$-modes, but also exerts shear stresses that significantly influence mode propagation in this region. 
We established a penetration criterion by comparing the mode’s transverse acceleration with the restoring force from crustal shear stresses [see Eqs.~\eqref{eq:criterion} and~\eqref{eq:fcrit}]. 
We found that the crustal $g$-mode identified in Ref.~\cite{Counsell:2024pua}, associated with neutron drip at $\sim 3$--$4 \times 10^{11}\,\rm g/cm^{3}$, is absent from the QNM spectrum due to strong shear forces. 
A similar disappearance of density-discontinuity $g$-modes~\cite{Finn:1987ads}, associated with phase transitions between nuclear species, was previously reported by~\citet{Kruger:2014pva}.
This can also be accommodated by our criterion.
In contrast, the $i$-mode in the nonstratified case, core composition-gradient $g$-modes, and the $f$-mode can all penetrate the crust due to their large transverse momentum.
In particular, the noticeable permeation of the composition-gradient $g$-modes into the crust essentially removes the concern raised by~\citet{Neill:2024gfe} (and its series of works) that these $g$-modes of oscillations are unable to break the crust.
Understanding such mode penetration is essential, especially when evaluating the shear strain in the crust and the possibility of crust breaking triggered by mode excitation.

For the dynamical tidal excitation of various modes, we found that the energy of the $f$-mode is always dominant although the exact resonance is not reached.
The transferred energy into $s$-modes and $g$-modes is smaller than $\sim 1\%$ of $f$-mode for typical NSs since their tidal overlapping is by 3--4 orders of magnitude smaller than that of $f$-mode. 
Correspondingly, the GW phase shift induced by resonant excitation of these modes stays well below $10^{-3}\,\rm rad$ for typical NSs.

We also analyzed the strain induced in the crust by various oscillation modes and found that nonresonant $f$- and $s$-modes can trigger crust breaking already around the first gravity-mode resonance at $f_{\rm gw} \sim 200\,\rm Hz$ for an equal-mass binary with component masses $1.35\,M_{\odot}$. Interestingly, the $f$-mode and $s_2$-mode produce the largest strain in the outer crust with similar spatial patterns, while the $s_1$-mode peaks in the inner crust. Our results show slight tension with fully Newtonian studies~\cite{Passamonti:2020fur,Passamonti:2022yqp}, which also report crust breaking from nonresonant $f$- and $s$-modes excitation, but only very close to merger. This discrepancy may stem from different strain expressions: Eq.~(68) in~\cite{Passamonti:2020fur} versus our \cref{eq:von_Mises}. Moreover, our calculations of the modes are relativistic, which is more accurate. The $g_1$-mode resonance concentrates strain at the crust–core boundary but yields only marginal stress for breaking. These results suggest that resonant and nonresonant mode excitations can break the crust, and potentially contribute to pre-merger energy transfer into the magnetosphere via crust shattering, offering a potential mechanism for EM precursor signals. 

For asymmetric NS–NS binaries, our conclusions remain largely unaffected, as the typical mass ratio lies within the range of $\sim 1\text{--}2$, and the mode energy given in \cref{eq:E_mode} should be on the same order. 
In contrast, for NS–BH binaries with the BH mass of about $10\,M_{\odot}$, the mass ratio is much higher. This leads to the suppression of the mode energy after resonance by approximately one order of magnitude. 
Moreover, due to the substantially lower merger frequency compared to NS–NS binaries, the $s$-modes may no longer be resonantly excited, and the amplitude of the nonresonantly excited $f$-modes is reduced by several orders of magnitude. 
As a result, crust breaking is less likely to occur in NS–BH binaries, or if it does occur, it happens very close to the merger.

Our work represents a step toward more realistic modeling of dynamical tides in binary NSs. In this study, we mainly establish the formalism using a single EOS, and there remains a lot to be done. For example, as is well known, the frequencies and tidal overlap integrals of $g$-, $i$-, and $s$-modes depend sensitively on nuclear parameters near saturation density. A systematic investigation of these dependencies is crucial for clarifying the disappearance of the $i$-mode, verifying the tensions in the predicted frequencies and tidal overlaps of $i$-modes compared to Refs.~\cite{Tsang:2011ad,Pan:2020tht,Zhu:2022pja}, assessing the role of $g$-mode resonances, and evaluating crust breaking triggered by both resonant and nonresonant excitations. \reply{It is worthwhile to note that, once the crustal strain exceeds the elastic limit, plastic flow may occur, converting oscillation energy into thermal one~\cite{Baym:1971ApJ,Kochanek:1992ApJ,Pan:2020tht,Zhu:2022pja}. It will be interesting to examine how different oscillation modes contribute to such heating.}

It is also important to note that further advances are needed in both the microphysical description of matter and the relativistic formulation of the problem. 
More specifically, for matter fields, superfluidity in the outer core is expected to increase the frequencies of the gravity modes~\cite{Kantor:2014lja,Passamonti:2022yqp,Yu:2016ltf}, potentially modifying the mode spectrum and the crustal breaking behavior predicted in our work. Going beyond $npe\mu$ matter studied here, phase transitions to exotic degrees of freedom --- such as hyperons or deconfined quarks --- may lead to new phase structures, accompanied by additional $g$- and $i$-modes~\cite{Tran:2022dva,Zhao:2022tcw,Miao:2023jqe,Counsell:2025hcv,Yu:2017cxe}. 
Recent studies suggest that $i$-mode resonances associated with first-order phase transitions can produce significant phase shifts in GWs, provided that such phase transitions do exist.
In addition, bulk viscosity may be important for exotic phases of matter and could potentially result in appreciable tidal heating during resonances~\cite{Ghosh:2025wfx,Ghosh:2023vrx,Saketh:2024juq}.
Regarding the modeling of dynamical tides, our treatment of the tidal coupling is based on a relativistic extension of the Newtonian formalism. It would be interesting to repeat our calculations in a full self-consistent relativistic approach (e.g., Refs.~\cite{Steinhoff:2016rfi,HegadeKR:2024agt,HegadeKR:2024slr,Pitre:2023xsr,Miao:2025utd,HegadeKR:2025qwj} for recent progress).
Finally, nonlinear effects induced by tidal resonances may also lead to prolonged mode excitations, amplified GW phase shifts~\cite{Kwon:2025zbc,Kwon:2024zyg,Pitre:2025qdf}, and induce differential fluid flows within the NS interior~\cite{Reboul-Salze:2025gyi,Kuan:2024jnw}.
We will investigate these aspects in future work.

\section*{Acknowledgements}

\reply{We thank the anonymous referee for comments and suggestions.}
Y.~G. thanks Zexin Hu, Jan Steinhoff, and Yanbei Chen for useful discussions. 
H.~O.~S. acknowledges funding from the Deutsche Forschungsgemeinschaft
(DFG)~-~Project No.:~386119226.
Numerical computations were performed on the Sakura cluster at the Max Planck Computing and Data Facility. This work was in part supported by Grant-in-Aid for Scientific Research (grant No. 23H04900) of Japanese MEXT/JSPS.
%


\bibliography{refs}

\appendix

\section{Polar perturbation for multi-layer NSs}
\label{sec:appendixA}

In this Appendix, we present the polar-sector perturbation equations for multi-layer NSs and numerical procedures to solve the eigenvalue problem following Refs.~\cite{Finn:1990ads,Kruger:2014pva,Detweiler:1985zz,Lindblom:1983ps}. A \texttt{Mathematica} notebook that derives the perturbation equations is available at \cite{von_Mises}.

\subsection{Master equations}

\subsubsection{Perturbation equations for perfect fluid}
For the perturbations of perfect fluid, we use the form of the perturbation equations derived by Detweiler and Lindblom~\cite{Detweiler:1985zz,Lindblom:1983ps}, 
\begin{widetext}
\begingroup
\allowdisplaybreaks
\begin{align}
    { H_1'}
        & = \left[
                 \frac{1}{2} \left( \lambda' - \nu' \right)
                - \frac{\ell+1}{r}
            \right] H_1 
            + \frac{e^{\lambda}}{r}
              \left[ H_0 + K -16 \pi \left( \epsilon + p \right) V \right]\,,
                                                    \label{eq:pfV_odeH1} \\
   {K'}
        & = \frac{1}{r} H_0
            + \frac{n_{\ell}+1}{r} H_1
            + \left( \frac{1}{2} \nu' - \frac{\ell+1}{r} \right) K
            - \frac {8 \pi \, \left( \epsilon  + p \right)}{r} e^{\lambda/2} W\,,
                                                    \label{eq:pfV_odeK} \\
    W'
        & = - \left( \frac{\ell+1}{r} + \frac{p'}{\Gamma_{1} p} \right) W
            + r e^{\lambda/2}
              \left[
                \frac{ \epsilon + p }{\Gamma_{1} p}
                    \left( e^{-\nu} \omega^2 V + \frac{1}{2} H_0 \right)
                - \frac{2(n_\ell+1)}{r^2} V
                + \frac{1}{2} H_0
                + K
              \right]\,,
                                                    \label{eq:pfV_odeW} \\                                                   
    X'= & -\frac{\ell}{r} X+\frac{1}{2}(\epsilon+p) e^{\nu / 2}\left\{\left[\frac{1}{r}-\frac{\nu^{\prime}}{2}\right] H_0+\left[r \omega^2 e^{-\nu}+\frac{n_\ell+1}{r}\right] H_1\right. \nonumber\\
    & +\left[\frac{3}{2} \nu^{\prime}-\frac{1}{r}\right] K-\nu^{\prime} \frac{\ell(\ell+1)}{r^2} V-\frac{1}{r}\left[8 \pi(\epsilon+p) e^{\lambda / 2}+2 \omega^2 e^{\lambda / 2-\nu}\right.\left.\left.-\,r^2\left(r^{-2} e^{-\lambda / 2} \nu^{\prime}\right)^{\prime}\right] W\right\},
                                                       \label{eq:pfX_odeX}                                                 
\end{align}
\endgroup
where a prime indicates a derivative with respect to $r$, and $n_{\ell} = (\ell+2)(\ell-1)/2$ for convenience.  Here $X$ is related to the Lagrangian variation of the pressure $\Delta p$ as $X = -e^{\nu/2} \Delta p/r^{\ell}$, and the fluid perturbation variable $V$ can be determined by 
\begin{equation}
    \label{eq:V}
    \omega^2(\epsilon+p) V=e^{\nu / 2} X+\frac{1}{r} p^{\prime} e^{\nu-\lambda / 2} W-\frac{1}{2}(\epsilon+p) e^\nu H_0\,.
\end{equation}
The variable $H_0$ can be obtained from the following algebraic equation deriving from the $rr$ and $r\theta$ components of the perturbed Einstein equations:
\begin{align}
    \label{eq:H0}
    & {\left[(n_{\ell}+1) r-\frac{r e^{-\lambda}}{2}\left(r \lambda^{\prime}+2\right)\right] H_0=r^2 e^{-\lambda}\left[\omega^2 r e^{-\nu}-\frac{n_{\ell}+1}{2} \nu^{\prime}\right] H_1} \nonumber\\
    & +\left[n_{\ell}\, r-\omega^2 r^3 e^{-\nu}-\frac{1}{4} r^2 e^{-\lambda} \nu^{\prime}\left(r \nu^{\prime}-2\right)\right] K+4 \pi r^2(\epsilon+p)\left(e^{-\lambda / 2} \nu^{\prime} W+2 \omega^2 re^{-\nu} V\right) \,.
\end{align}
\end{widetext}
The metric perturbation function $H_2$ satisfies 
\begin{equation}
    \label{eq:H2}
    H_2=H_0\,.
\end{equation} 
These equations have been extensively employed in the analysis of high-frequency oscillation modes, such as the $f$, $p$, and $w$ modes~(see \cite{Kokkotas:1999bd} for a review).

However, the formalism is problematic for low-frequency modes (e.g., higher-order $g$-modes and $i$-modes). This is purely numerical and mainly stems from the \cref{eq:V} to calculate $V$, where numerical cancellation is inevitable and hence leads to inaccurate solutions (see \citet{Kruger:2014pva} for details). 
\citet{Kruger:2014pva} pointed out that using $V$ instead of $X$ as a ``fundamental variable" can solve this problem; the perturbation equations are essentially the same, except for changing \cref{eq:pfX_odeX} into
\begin{equation}
    V'
         = \left( -A
                + \nu' - \frac{\ell}{r} \right) V
            - \frac{A e^{\nu}}{2 \omega^2} 
                \left( H_0 + \frac{\nu'}{re^{\lambda/2}} W \right)
                                                                + r H_1
            - \frac{e^{\lambda/2}}{r} W\,.
                                                    \label{eq:pfV_odeV}
\end{equation}
As a summary, in the fluid core, we use \cref{eq:pfV_odeH1,eq:pfV_odeK,eq:pfV_odeW,eq:pfX_odeX,eq:V,eq:H0,eq:H2} if the oscillation frequency $\omega \geq 0.01/M$. Otherwise, we use \cref{eq:pfV_odeV} to replace \cref{eq:pfX_odeX}. For the fluid ocean (if added), the systematic equations \cref{eq:pfV_odeH1,eq:pfV_odeK,eq:pfV_odeW,eq:pfX_odeX,eq:V,eq:H0,eq:H2} are applied.

\subsubsection{Perturbation equations for elastic solid}

The full set of perturbation equations for elastic solid can be expressed as~\cite{Kruger:2014pva}\footnote{The perturbation equations presented in Ref.~\cite{Kruger:2014pva} miss a factor of 1/2 in the shear modulus $\mu$, which has been also corrected in Ref.~\cite{Kruger:2024fxn}.}
\begin{widetext}
\begingroup
\allowdisplaybreaks
\begin{align}
    H_1'
        & = \left[ \frac{1}{2}(\lambda' - \nu')
                     - \frac{\ell+1}{r} \right]
              H_1
            + \frac{e^{\lambda}}{r}
                \left[ H_2 + K - 16 \pi (\epsilon + p) V \right]\,,
                                                        \label{eq:odeH1} \\
    K'
        & = \frac{1}{r} H_2
            + \frac{n_{\ell}+1}{r}H_1
            + \left( \frac{1}{2} \nu' - \frac{\ell+1}{r} \right) K
            - \frac{8 \pi (\epsilon + p) e^{\lambda/2} }{r} W\,,
                                                        \label{eq:odeK} \\
   H_0'
        & =  K'
            -r e^{-\nu} \omega^2 H_1
            - \left( \frac{1}{2} \nu'
                + \frac{\ell-1}{r} \right) H_0
            - \left( \frac{1}{2} \nu'
                + \frac{1}{r}   \right) H_2
            + \frac{\ell}{r} K
            - \frac{16 \pi}{r} T_2\,,
                                                        \label{eq:odeH0} \\
    W'
        & = - \frac{\ell+1}{r} W
            + r e^{\lambda/2} \left[
                \frac{e^{-\nu/2}}{\Gamma_{1} p} X
                - \frac{2(n_{\ell}+1)}{r^2} V
                + \frac{1}{2} H_2
                + K
            \right]\,,
                                                        \label{eq:odeW} \\
    V'
        & = \frac{1}{ \mu r} T_2
            + \frac{e^{\lambda/2}}{r} W
            + \frac{2-\ell}{r} V\,,
                                                        \label{eq:odeV} \\
    T_2'
        & = - \frac{1}{2} r e^{\lambda} (\epsilon + p) H_0
            + r e^{\lambda - \nu / 2}
              \left( X 
                + \frac{1}{2r^2} e^{\nu / 2} T_1 
              \right)
            + \left[
                \frac{1}{2}(\lambda' - \nu')-\frac{\ell+1}{r}
              \right] T_2+ \left[
                    \frac{2 n_{\ell} e^{\lambda} \mu}{r}
                    -e^{\lambda-\nu} r \omega^2 (\epsilon + p) 
                  \right] V
                + e^{\lambda / 2} p' W\,.
                                                        \label{eq:odeT2}
\end{align}
\endgroup
In addition to the foregoing six ordinary differential equations, we have three algebraic relations
\begin{align}
    H_2 
        & = H_0 + 32 \pi \mu V\,,
                                                        \label{eq:alg1} \\
        \left[\frac{r e^{-\lambda}}{2}\left(r \nu^{\prime}-2\right)+(n_{\ell}+1) r\right] H_0
        &  =r^2 e^{-\lambda}\left[\omega^2 r e^{-\nu}-\frac{n_{\ell}+1}{2} \nu^{\prime}\right] H_1
                                                        \nonumber \\
        & \quad + \left[ n_{\ell} r - \omega^2 r^3 e^{-\nu} 
                     - \frac{1}{4} r^2 e^{-\lambda} \nu^{\prime}\left(r \nu^{\prime}-2\right) \right] K  +8 \pi r^3 e^{-\nu / 2} X - 8\pi r T_1 - 16 \pi r e^{-\lambda} T_2\,,
         \label{eq:alg2}\\                                          
    \frac{1}{3} e^{-\nu/2} \mu r^2 X + \frac{1}{4} \Gamma_{1} p T_1
        & = \frac{\mu \Gamma_{1} p}{2} 
            \left[ 2 e^{-\lambda/2} W - r^2 K + 2(n_{\ell}+1) V \right].
                                                        \label{eq:alg3}
\end{align}
\end{widetext}
The first algebraic equation can be used to determine $H_2$, while the last two algebraic equations return $T_1$ and $X$.

\subsection{Joint and boundary conditions}

To integrate the ODEs and obtain the interior solutions for a given oscillation angular frequency \(\omega\) and EOS, we require matching conditions at the solid-liquid interface, as well as the boundary conditions at both the center and the surface of the NS. 
The matching conditions at the solid-liquid interface arise from the requirement that the true degrees of freedom of spacetime remain continuous. We use $[Q_{r}]$ to denote the jump of a quantity $Q$ across the interface,
\begin{equation}
    [Q]_r\equiv\lim _{\varepsilon \rightarrow 0} Q(r+\varepsilon)-\lim _{\varepsilon \rightarrow 0} Q(r-\varepsilon) \,.
\end{equation}
The matching conditions at the solid-liquid interface are
\begin{equation}
    [H_0]_r = [H_1]_r = [K]_r = [W]_r = [T_2]_r = 0\,,
\end{equation}
Note that the horizontal displacement $V$ is not necessarily continuous across the fluid-solid interface. The equations of motion allow us to derive more matching conditions, though not independent. \cref{eq:alg1} indicates that $H_2$ is not continuous across the interface, and the jump of $H_2$ is given by
\begin{equation}
    \left[H_2\right]_r=32 \pi[{\mu} V]_r .
\end{equation}

The solutions must remain regular at the center of the star, necessitating a Taylor expansion of the perturbation variables near the origin to approximate the ODE solutions. It has been shown in Ref.~\cite{Detweiler:1985zz} that not all the quantities are independent. Once \( K(r=0) \) and \( W(r=0) \) are specified, all other coefficients in the Taylor expansion can be uniquely determined (see details in Ref.~\citep[pp. 27--30, 59--60]{soton384187}). The surface of the star is a matter-vacuum interface. If a two-layer NS with a solid surface is considered, the boundary conditions at the surface are 
\begin{equation}
    T_2(R) = 0\,, \quad  R^2 e^{-\nu(R) / 2} X(R)+T_1(R)=0\,,
\end{equation}
where the latter one is derived from the algebraic relation \cref{eq:alg2} by imposing $[H_0]_R = [H_1]_R= [K]_R$. While for a three-layer star with a fluid ocean, the boundary condition at the surface is simply
\begin{equation}
    X(R) = 0\,.
\end{equation}
Since the systematic equations are linear, one can still choose one overall normalization for the solutions. 
The boundary and joint conditions of the perturbation equations for the multi-layer NS, as well as the overall normalization, are summarized in \cref{fig:boundary}. It should be noted that the QNM calculations are performed with the normalization $W(R) = 1$, whereas the results reported in the main text are rescaled using $\mathcal{A}_\alpha^2 = MR^\ell$ [cf.~\cref{eq:norm}].
\begin{figure}
    \centering
    \includegraphics[width=\linewidth]{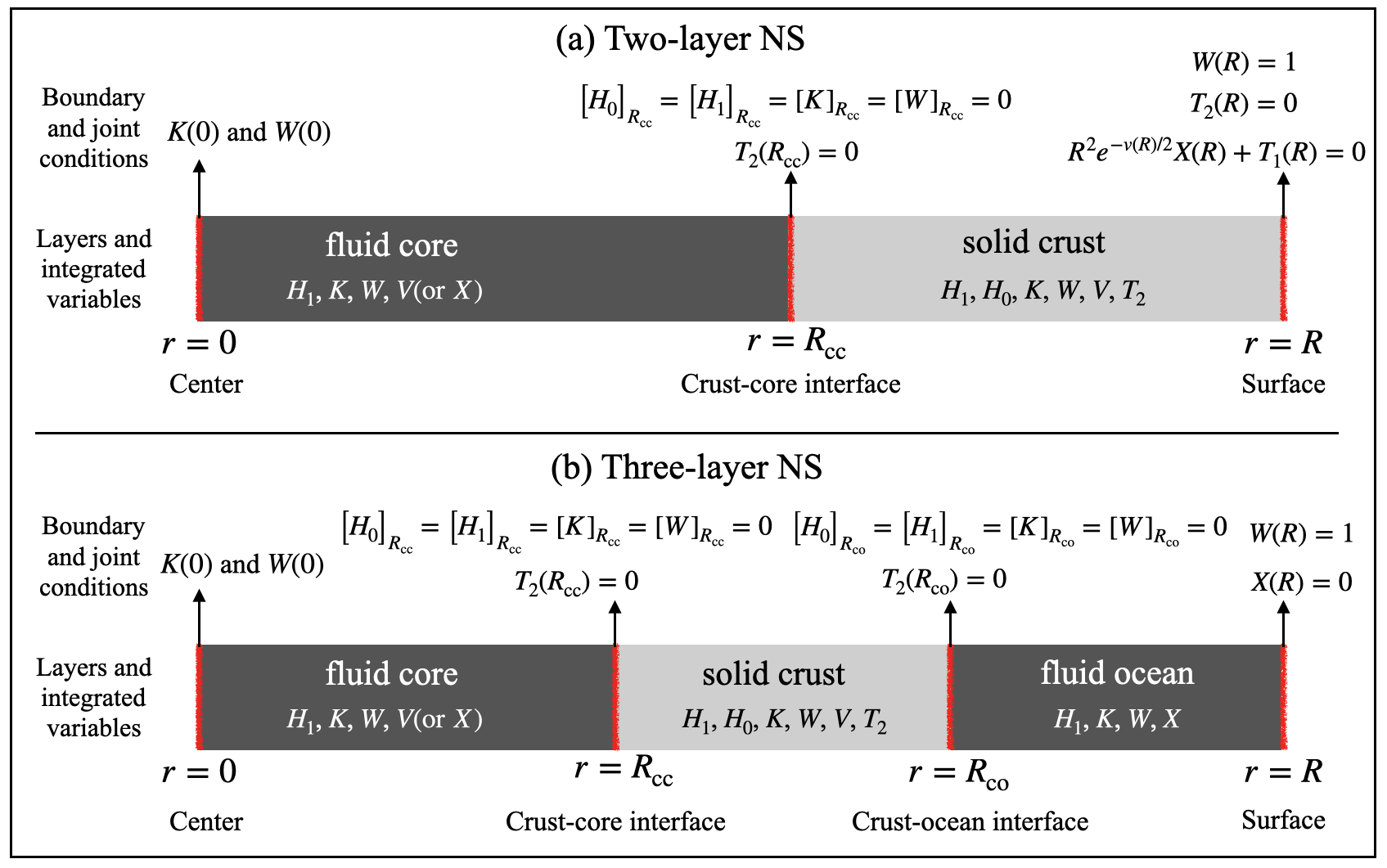}
    \caption{The boundary and joint conditions for multi-layer NSs.}
    \label{fig:boundary}
\end{figure}

\subsection{Numerical procedures}

\subsubsection{The interior solution}

Our numerical procedures to solve the perturbation equations for a multi-layer NS is similar with \citet{Lin:2007dg} and \citet{Kruger:2014pva}. Considering an NS with $n$ layers, where the interfaces including the center and the surface of the NS, are located at the radii $0=R_0<R_1<R_2<\cdots<R_{n-1}<R_n=R$. The dynamical equations that govern the perturbations at each layer depends on the nature of the matter field at this layer. For $i$-th layer, the perturbation equations can be written as
\begin{equation}
    \frac{\dd^{(i)} \mathbf{Y}}{\dd r}={ }^{(i)} \mathbf{M} \cdot{ }^{(i)} \mathbf{Y}\quad {\rm for} \quad [R_{i-1},R_{i}]\,.
\end{equation}
Here ${ }^{(i)} \mathbf{Y}=\left(y_{1}, \ldots, y_{k_i}\right)$ is the vector of perturbation variables and $k_{i}$ is the number of variables in the $i$-th layer to be integrated, ${ }^{(i)} \mathbf{M} ={ }^{(i)} \mathbf{M}(\texttt{star},r,\ell,\omega) $ is a $k_i \times k_{i}$ matrix of coefficients, with \texttt{star} represents the background fields. 

To obtain the general solution in layer $i$, we need to integrate the perturbation equations $x_{i}$ times starting from the boundary, where $x_{i}$ is the number of linearly independent solutions at each layer. The general solution in layer $i$ is the linear combination of these solutions
\begin{equation}
    { }^{(i)} \mathbf{Y}(r)=\sum_{j=1}^{x_i} c_{i, j}{ }^{(i)} \mathbf{Y}_j(r) \quad \text { for } \quad r \in\left[R_{i-1}, R_i\right],
\end{equation}
where the coefficients $c_{i,j}$ is determined by the boundary and joint conditions, with $i$ denoting the layer and $j$ counting the different solutions in the layer.

For a two-layer NS, as illustrated in \cref{fig:boundary}, the fluid interior admits two independent solutions, depending on the choices of \((K(0), W(0))\). In contrast, within the solid crust, the tangential traction at the crust-core interface vanishes \((T_2 = 0)\), reducing the number of independent solutions from six to five. Consequently, there are seven independent solutions in total, demanding seven conditions to determine the coefficients $c_{i, j}$. These conditions comprise four continuity conditions at the crust-core interface: $\left[ H_0 \right]_{R_{cc}} = \left[ H_1 \right]_{R_{cc}} = [K]_{R_{cc}} = [W]_{R_{cc}}=0$, along with three boundary conditions at the NS surface: $T_2(R) = 0, \ W(R) = 1, \ \text{and}\  R^2 e^{-\nu(R)/2} X(R) + T_1(R) = 0$.

For a three-layer model, the approach to integrating the perturbation equations in the fluid core and solid crust remains unchanged. However, the presence of a fluid ocean introduces an additional set of perturbation equations. To handle this, we integrate the system of ODEs from the surface to the crust-ocean interface, imposing the condition $X(R)=0$, which yields three independent solutions in the fluid ocean. Consequently, we obtain a total of ten independent solutions. Then we have ten independent solutions in total. Except for the four joint conditions at the crust-core interface, we have four joint conditions at the crust-ocean interface: $\left[H_0\right]_{R_{\mathrm{co}}}=\left[H_1\right]_{R_{\mathrm{co}}}=[K]_{R_{\mathrm{co}}}=[\mathrm{W}]_{R_{\mathrm{co}}}$ along with one constraint $T_2(R_{\rm co})=0$. Furthermore, at the surface, we impose additional one normalization $W(R) = 1$. With these conditions in place, the coefficients $c_{i,j}$ can be fully determined.

\subsubsection{The exterior solution}

Outside of the star, the matter field vanish and the system of perturbation equations reduces to a second-order system of equations for metric perturbation $H_1$ and $K$. These equations can be transformed into a standard Zerilli equation~\cite{Fackerell:1971ApJ,Lindblom:1983ps}, which allows two independent solutions. Far away from the star, one of the solutions can be identified as an outgoing wave, whereas the other solution describes an ingoing wave. For a given oscillation frequency $\omega$, the solution for the interior of the NS leads to a mixture of ingoing and outgoing waves. Physically, it means that a general frequency $\omega$ will not corresponds to a resonance of the stellar model, the star is forced to oscillate at this frequency, driving by the ingoing gravitational waves. The QNMs are those special frequencies at which the ingoing wave vanishes, and the star oscillates ``freely". 

We follow the phase-amplitude method \cite{Froeman:1992gp,Andersson:1995wu} to solve the exterior problem and identify the QNMs. 
Instead of solving the Zerilli function $Z$, they introduce a new independent variable $\Psi$
\begin{equation}
    Z=\left(1-\frac{2 M}{r}\right)^{-1 / 2} \Psi\,,
\end{equation}
to transform the linear wave equation to a non-linear differential equation with a slowly varying variable $q$ respect to $r$. The physically acceptable solution for the interior corresponds to a wave at spatial infinity,
\begin{equation}
    \Psi=A_{\rm {in }} \Psi^{+}+A_{\rm {out }} \Psi^{-}\,,
\end{equation}
where $A_{\rm in}$ and $A_{\rm out}$ are the amplitude of the ingoing and outgoing waves, respectively. A QNM corresponds to a solution possessing no incoming component (i.e., with $A_{\rm in}=0$).

\end{document}